\def\spose#1{\hbox to 0pt{#1\hss}}

\def\multleft#1{\hbox to size{\vbox {\halign {\lft{##}\cr #1}}\hfill}\par}
\def\multright#1{\hbox to size{\vbox {\halign {\rt{##}\cr #1}}\hfill}\par}

\def\today{\ifcase\month\or January\or February\or March\or April\or May\or
      June\or July\or August\or September\or October\or November\or December\fi
      \space\number\day, \number\year}
\def\s{\hbox{\phantom{5}}}      

\def\cm{{\rm\thinspace cm}}

\def\erg{{\rm\thinspace erg}}

\def\K{{\rm\thinspace K}}
\def\keV{{\rm\thinspace keV}}
\def\km{{\rm\thinspace km}}
\def\kpc{{\rm\thinspace kpc}}

\def\Mpc{{\rm\thinspace Mpc}}
\def\Msun{\hbox{$\rm\thinspace M_{\odot}$}}
\def\pc{{\rm\thinspace pc}}

\def\s{{\rm\thinspace s}}

\def\yr{{\rm\thinspace yr}}

\def\cmsq{\hbox{$\cm^2\,$}}

\def\ergpcmsqpspm{\hbox{$\erg\cm^{-2}\s^{-1}\mu{\rm m}^{-1}\,$}}
\def\ergps{\hbox{$\erg\s^{-1}\,$}}

\def\kmps{\hbox{$\km\s^{-1}\,$}}

\def\Msunpyr{\hbox{$\Msun\yr^{-1}\,$}}
\def\pcm{\hbox{$\cm^{-3}\,$}}

\def\kmpspMpc{\hbox{$\kmps\Mpc^{-1}$}}

\def\H2{\hbox{H$_{2}$}}

\def\Pa{Pa$\alpha$}
\def\Ha{H$\alpha$}

\documentclass[usegraphicx]{mn2e}

\usepackage{times}
\usepackage{amssymb}
\usepackage{eso-pic}
\include{defn}

\begin{document}
\hsize=6truein
\title[Integral field spectroscopy of ionized and molecular gas in cool cluster cores]
{Integral field spectroscopy of ionized and molecular gas in cool cluster cores: evidence for cold feedback?
\thanks{Based on observations performed at the European Southern Observatory, Chile (Programme ID: 77.A-0057(A)).}}
\author[R.J.~Wilman et al.]
{\parbox[]{6.in} {R.J.~Wilman$^{1}$, A.C.~Edge$^{2}$ and A.M.~Swinbank$^{2}$\\ \\
\footnotesize
1. School of Physics, University of Melbourne, Victoria 3010, Australia \\
2. Department of Physics, University of Durham, South Rd, Durham, DH1 3LE \\ }}
\maketitle

\begin{abstract}
We present {\em VLT-SINFONI} K-band integral field spectroscopy of the central galaxies in 
the cool core clusters A1664, A2204 and PKS 0745-191, to probe the spatio-kinematic
properties of the Pa$\alpha$ and ro-vibrational \H2~line emission. \smallskip

\noindent
In A1664 the two emission-line velocity systems seen in our previous \Ha~spectroscopy
appear in both \Pa~and \H2~emission, with notable morphological differences. 
The recession velocity of the red component of \Pa~increases linearly with decreasing radius, 
particularly along an 8\kpc~filament aligned with the major axis of the underlying galaxy 
and the cluster X-ray emission. These kinematics are modelled as gravitational free-fall as 
gas cools rapidly out of the hot phase. In A2204 the gas shows 3 or 4 filaments reaching radii 
of 10\kpc, three of which lie towards `ghost bubbles' seen in X-ray imaging by Sanders et al. 
For PKS 0745-191, we confirm the twin-arm morphology in the narrow-band images of Donahue et 
al.; the \Pa~kinematics suggest rotational motion about an axis aligned with the kiloparsec-scale 
radio jet; on nucleus, we find an underlying broad \Pa~component (FWHM 1700\kmps)~and a secondary 
\H2~velocity system redshifted by $+500$\kmps.

\smallskip \noindent
The \H2~v=1-0~S(3)/\Pa~ratio is highest in the most isolated and extended 
regions where it matches the levels in the NGC 1275 filaments as modelled by Ferland 
et al. Regions with much lower ratios highlight active star formation and are often 
kinematically quiescent (FWHM~$<200$\kmps). Our findings suggest that the three clusters may be 
captured in different stages of the `cold feedback' cycle of Pizzolato \& Soker, with
A1664 in a short-lived phase of extreme cooling and star formation prior to an AGN heating event; 
PKS 0745-191 in an outburst state with the AGN accreting from a cool gas disk, and 
A2204 in a later phase in which cool gas is dragged out of the galaxy by the buoyant rise of old radio bubbles.

\end{abstract}


\section{INTRODUCTION}
The past decade has witnessed a flood of new observational insights into the X-ray cooling
flow phenomenon in the cores of galaxy clusters. On the X-ray side, results from {\em XMM-Newton} 
and {\em Chandra} have led to a sharp downward revision in X-ray cooling rates and exposed a strong 
deficit of line emission from gas cooling below one third of the ambient cluster
temperature (see review by Peterson \& Fabian~2006). The implication is that the cooling of
the hot X-ray gas is quenched, with the most likely cause being that episodic outbursts from
radio-loud active galactic nuclei (AGN) somehow heat the gas 
(see review of McNamara \& Nulsen~2007). 

Significant progress has also been made on the discovery and 
characterization of various phases of cool gas and dust in cluster cores. In CO line emission, 
Edge~(2001) reported detections of 16 clusters, consistent with $10^{9}-10^{11.5}$\Msun~of \H2~at 
20--40\K~for a standard CO:\H2~conversion (see also Salom\'{e} \& Combes~2003). Interferometry showed 
further that the CO emission is localised within the central few arcsec of the central cluster galaxy (CCG) (Edge \& Frayer~2003; 
Salom\'{e} \& Combes~2004). The CO emission often coincides with K-band rovibrational emisison from much 
smaller masses ($\sim 10^{5}-10^{6}$\Msun) of \H2~at $\sim 2000$\K~(Edge et al.~2002; Jaffe, 
Bremer \& van der Werf~2001). Similarly, mid-infrared spectroscopy with {\em Spitzer} has revealed 
pure rotational \H2~emission from molecular gas at 300--400\K~in several systems (e.g. Egami et al.~2006a, 
Johnstone et al.~2007). All these molecular emission line components appear to correlate well with the 
strength of H$\alpha$ emission whose prevalence in these systems, often in the form of 
spectacular filamentary systems, has been known for several decades. The most comprehensive spectroscopic
survey for H$\alpha$ emission is that of Crawford et al.~(1999), who showed that in the most luminous systems
the emission is powered by massive star formation. The latter is also the heating mechanism for the excess 
thermal far infrared dust emission discovered by {\em Spitzer MIPS} photometry of a sample of CCGs by 
Egami et al.~(2006b). These results are corroborated by the more extensive {\em Spitzer} photometric 
survey of line-emitting CCGs by Quillen et al.~(2008), revealing excess far-infrared emission
in about half the sample, consistent with star-formation in the known molecular gas reservoirs (O'Dea et al.~2008).

The emerging consensus is that whilst AGN heating successfully offsets most of the cooling in a time-avergaed sense, 
residual cooling to sub-X-ray temperatures persists at levels of, at most, a few per cent of the rates 
previously expected. This gas accumulates in molecular gas reservoirs 
within the central galaxy, giving rise to star formation and associated emission line nebulosity. The physical 
processes at work in these cooled gas reservoirs are, however, not well understood. It is not known whether the star 
formation is steady and continuous or subject to outbursts, perhaps triggered by the central radio-loud AGN or 
interaction with a passing galaxy. The excitation mechanisms of the emission line nebulosity are also uncertain: 
whilst star formation can power the H$\alpha$ emission on energetic grounds, the overall emission line spectrum 
differs markedly from a typical HII region or starburst galaxy, with optical forbidden and rovibrational
\H2~lines being far stronger than observed in such systems (e.g. Wilman et al.~2002). Simulations by 
Ferland et al.~(2008,2009) suggest that the \H2~emission in CCGs may arise from excitation by cosmic 
rays or heating via dissipative magnetohydrodynamic processes in dense gas shielded from external radiation fields. 
Support for this scenario comes from {\em Hubble Space Telescope} observations of the H$\alpha$ filaments in NGC 1275 by 
Fabian et al.~(2008), who showed that the filaments can be supported by threads of magnetic field and thereby 
stabilised against collapse and star formation.

To shed further light on the cool gas systems in cluster cores, we have begun detailed studies
of individual systems using optical and near-infrared integral field spectroscopy. The aim is to map the 
warm ionized gas and hot molecular hydrogen to reveal their spatial distributions, kinematics and excitation 
properties. In the first paper, Wilman, Edge \& Johnstone (2005) presented high-resolution near-infrared 
integral field spectroscopy (IFS) of the nuclear region of NGC 1275 using {\em UIST} on {\em UKIRT}, leading to the discovery of a 50-pc radius disk of hot \H2 and a dynamical measurement of the nuclear black hole mass, the latter in good agreement with the $M_{\rm{BH}}-\sigma$ relation. The existence of this disk has been confirmed by first results from the 
adaptive optics Near-infrared Integral Field Spectrograph {\em (NIFS)} on the {\em Gemini} telescope (McGregor et al.~2007). Observations of NGC 1275 with the {\em Smithsonian Millimetre Array} by Lim et al.~(2008) have shown
how this disk is fed from the inflow of cool molecular gas along filaments extending out to radii of 8\kpc. These filaments coincide with the coolest X-ray plasma between the two radio cavities, and together these observations
demonstrate, for the first time, how the cooling of X-ray gas in a cluster core feeds the central supermassive 
black hole. 
 
In a second study, Wilman, Edge \& Swinbank (2006) presented {\em VLT-VIMOS} optical IFS of four of the most 
H$\alpha$-luminous CCGs known. The full 2-d view revealed
a variety of disturbed morphologies, ranging from the smooth but distorted to the extremely clumpy and disturbed, 
with velocity gradients and splittings of several hundred \kmps~on scales of 20\kpc~or more, and evidence for 
an association between disturbed H$\alpha$ emission and secondary galaxies. The global H$\alpha$ kinematics 
match those of the CO(1-0) emission in single-dish data, suggesting that they arise in the same gas clouds.
The kinematic coupling between the H$\alpha$ and ro-vibrational \H2~emission appears less strong. Despite the 
disturbed kinematics, the ratio of optical forbidden line emission to H$\alpha$ is remarkably constant within and 
between CCGs, implying that the line ratios in these extreme environments are saturating due to 
widespread star formation. Less luminous systems exhibit more spatial variation in line ratios (e.g. 
Hatch et al.~2007), as expected if star formation is more patchy and irregular. 

To build on this work, we
present in this paper near-infrared IFS with {\em VLT-SINFONI} of three CCGs (two from the Wilman et al.~2006 sample), to perform a direct comparison between the Pa$\alpha$ and \H2~emission at high spatial and spectral 
resolution. Compared with our previous studies, {\em VLT-SINFONI} instrument offers a 5-fold improvement in 
spatial sampling over {\em VLT-VIMOS}, and substantially better sensitivity than {\em UKIRT-UIST}. 

\section{OBSERVATIONS AND DATA REDUCTION}
The observations of the CCGs of A1664, A2204 and PKS 0745-191 were taken in service mode in 2006 with 
the SINFONI integral field unit (IFU) (Eisenhauer et al.~2003) on UT4 of the Very Large Telescope (VLT) at the European Southern Observatory in Paranal, Chile. The observation
dates and exposure times are listed in Table 1. The IFU was operated throughout in the $8 \times 8$~\arcsec$^{2}$
field-of-view mode, with 32 slitlets of 0.25\arcsec~width, with a detector pixel scale of 0.125\arcsec~along the slits. The K-band grism spanning a wavelength range of 1.95-2.45\micron~at resolution $\sim 4000$ was employed, resulting
in a dispersion of $2.45 \times 10^{-4}$\micron~per pixel. The instrument was operated without adaptive optics and with an ABBA pattern of 300~s on/off source exposures to enable sky subtraction. The seeing 
was approximately 1\arcsec~throughout. Standard pipeline data reduction was carried out using the {\sc esorex} pipeline.  
In brief, the reduction comprises the following steps: correction for bad pixels, flat fielding, correction for 
geometrical distortion, arc-line-based wavelength calibration, subtraction of sky emission (using ABBA sequences) 
and merging of the individual frames into a single data cube for each object, with sub-pixel dithering enabling re-sampling 
to square pixels of 0.125\arcsec. A refined background subtraction was performed during post-processing in IDL, using a background spectrum synthesised from regions free of CCG line emission in the pipeline-reduced datacubes. All subsequent scientific analysis was also performed in IDL. The spectra were flux calibrated but a telluric standard was not observed, potentially corrupting the spectra outside the 2.08--2.34\micron~window of clean atmospheric transmission. This is for the most part not a problem because most of the emission lines of interest lie in this range. The exception is PKS 0745-191, where Pa$\alpha$ and \H2~v=1-0~S(5) fall just blue-ward of 2.08\micron. In this one target, 
we corrected for this during the analysis by dividing the raw spectra by a scaled version of the standard Mauna Kea atmospheric transmission curve for a comparable spectral resolution (the data were obtained from the UKIRT worldwide 
webpages, and originally produced using the program IRTRANS4). The scaling (i.e. the absorption optical depth) was adjusted in order to flatten out the continuum across the absorption troughs. 

In the analysis we devote one section to each of the three CCGs, followed by some generalised conclusions.
The cosmological parameters $H_{\rm{0}}=70$\kmpspMpc, $\Omega_{\rm{M}}=0.3$ and $\Omega_{\rm{\Lambda}}=0.7$ are 
assumed throughout.

\begin{table*}
\caption{Observation log}
\begin{tabular}{|lllll|}\hline
Target           & Redshift & Observation & On-source exposure & Scale \\
                 &          & dates       & time (s)           & (\kpc~arcsec$^{-1}$) \\ \hline
A1664            & 0.1276   & 2006 Apr 4, May 19,20,30,31           & 7200                   &   2.3     \\  
A2204            & 0.1514   & 2006 Jul 1,4,5,7, Aug 2,3             & 10800                   &  2.6     \\ 
PKS 0745-191     & 0.1028  & 2006 Apr 5,6,20,26,27                 & 10200                   &  1.9     \\ \hline
\end{tabular} \\
\end{table*}

\section{Results on A1664}

A1664 has a redshift of $z=0.1276$ and from the {\em Spitzer} mid-infrared spectral
energy distribution O'Dea et al.~(2008) inferred a star formation rate of 14.6\Msunpyr, compared
with a cooling rate deduced from X-ray spectra of $60^{+20}_{-20}$\Msunpyr. At $\sim 50$\kpc, the 
size of the star formation region implied by the Schmidt law is the second largest in the O'Dea et 
al.~sample (most of the rest are in the 10--20\kpc~range). However, the source is poorly resolved in 
the mid-infrared which would instead imply a size around 15\kpc.
 
A1664 was the most spectacular and complex of the sample of four CCGs in our earlier 
H$\alpha$ study with {\em VLT-VIMOS} (Wilman, Edge \& Swinbank~2006). The H$\alpha$ emission was
characterised by a 31\kpc~filamentary structure passing through the nucleus in a NE-SW direction, an
orientation shared by the larger-scale cluster X-ray emission. Two velocity components were detected 
along the full length of the filament with a splitting of almost 600\kmps~in the nuclear regions. 
A bipolar nuclear morphology in the blue kinematic component suggested that some kind of starburst-driven 
outflow may be responsible for this complexity. However, the presence of a small galaxy just beyond 
the end of the filament (and within 100\kmps~in line-of-sight velocity) led us to favour a scenario in which
this companion cluster galaxy had plunged through the cluster core, disturbed the gas reservoir and triggered
a burst of star formation. The current {\em VLT-SINFONI} IFU data enable us to map this structure at 
higher spatial resolution through the Pa$\alpha$ emission, which is an isolated emission line and thus
free of the line blending seen in the H$\alpha$+[NII] complex.

The Pa$\alpha$ line (rest-wavelength 1.8756\micron) in each 0.125\arcsec~cell was fitted with a gaussian emission line complex atop a flat
continuum over the wavelength range 2.07505--2.13630\micron~(i.e. 250 spectral bins). 
Chi-squared values were computed for fits with zero, one or two velocity components: $\chi^{2}_{0}$, 
$\chi^{2}_{1}$ and $\chi^{2}_{2}$. For $\chi^{2}_{0} - \chi^{2}_{1} > 30$ one component was fitted; 
for $\chi^{2}_{1} - \chi^{2}_{2} > 10$, two components were deemed necessary. A similar procedure was
used to fit the \H2~v=1-0 S(3) emission line (rest-wavelength 1.9576\micron).

\begin{figure}
\includegraphics[width=0.4\textwidth,angle=0]{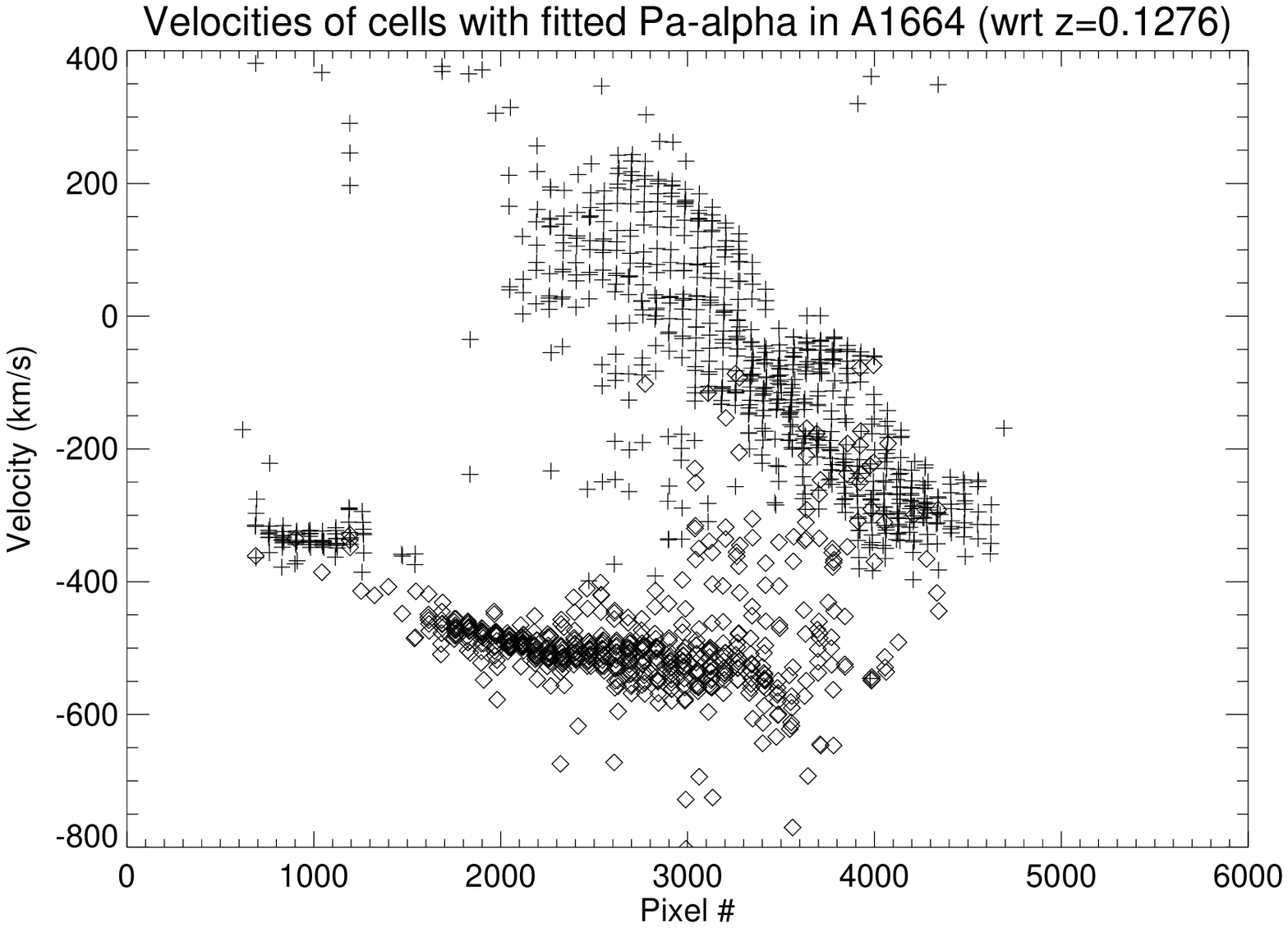}
\includegraphics[width=0.4\textwidth,angle=0]{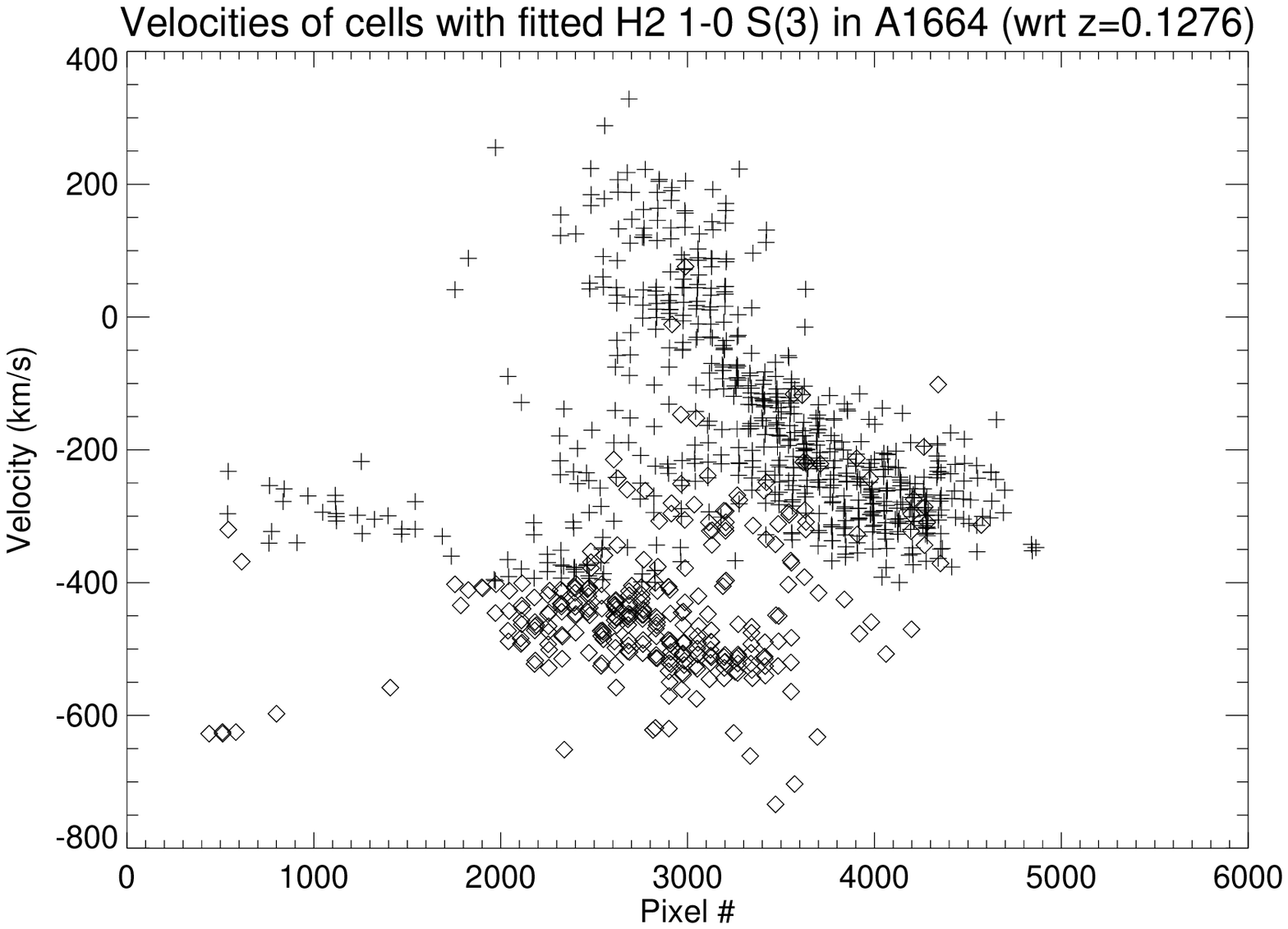}
\caption{\normalsize Velocity distributions for \Pa~and \H2 v=1-0 S(3) in A1664 as a function
of spatial pixel number within the IFU aperture, clearly demonstrating the existence of two 
distinct velocity systems: a red system (denoted by the crosses) and a blue system (diamonds). 
The dividing line between the two is chosen to be --400\kmps.}
\label{fig:veldist}
\end{figure}

\subsection{Gas morphologies and kinematics}
\begin{figure*}
\begin{centering}
\includegraphics[width=5.8cm,angle=0]{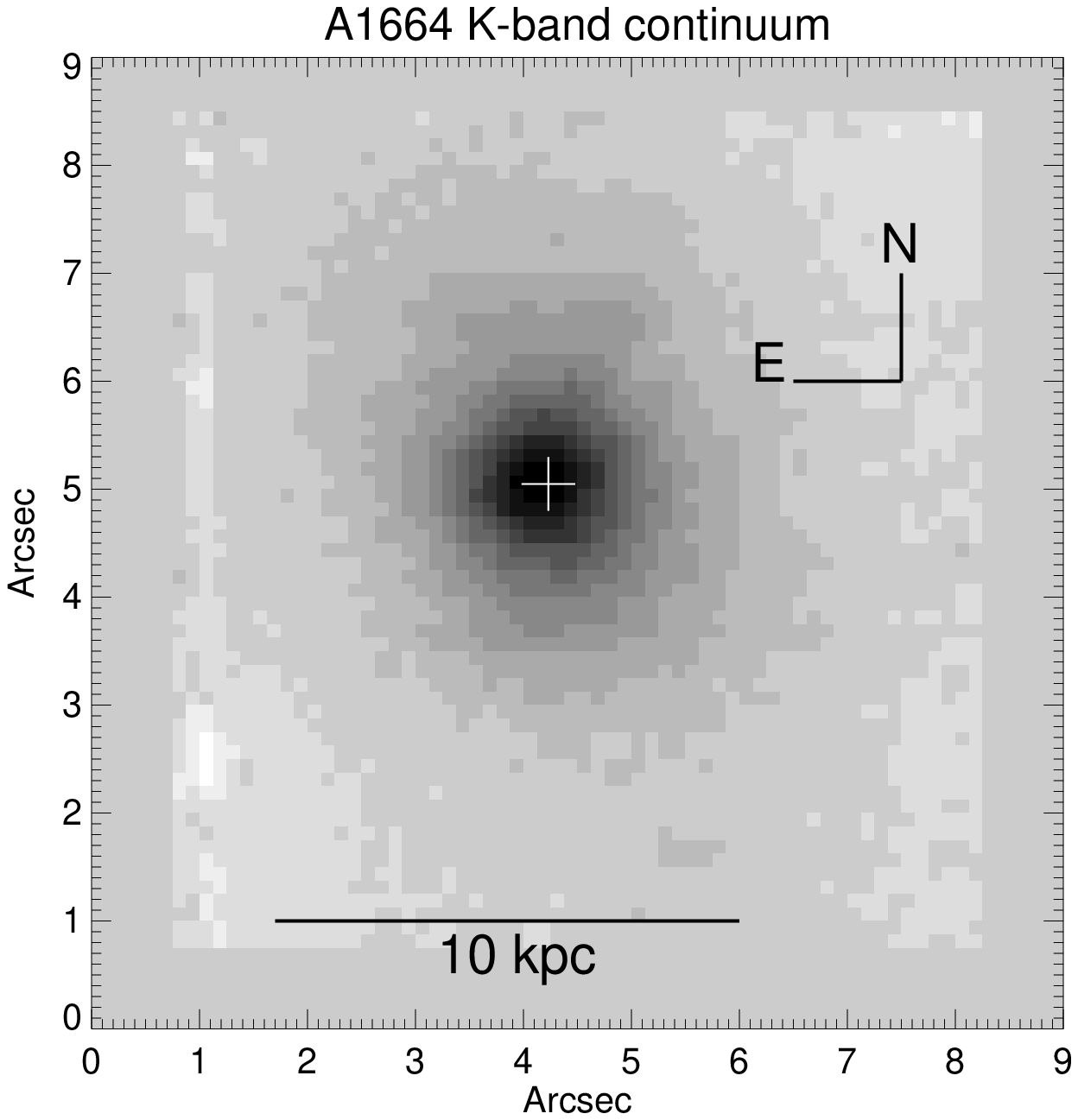}
\includegraphics[width=5.8cm,angle=0]{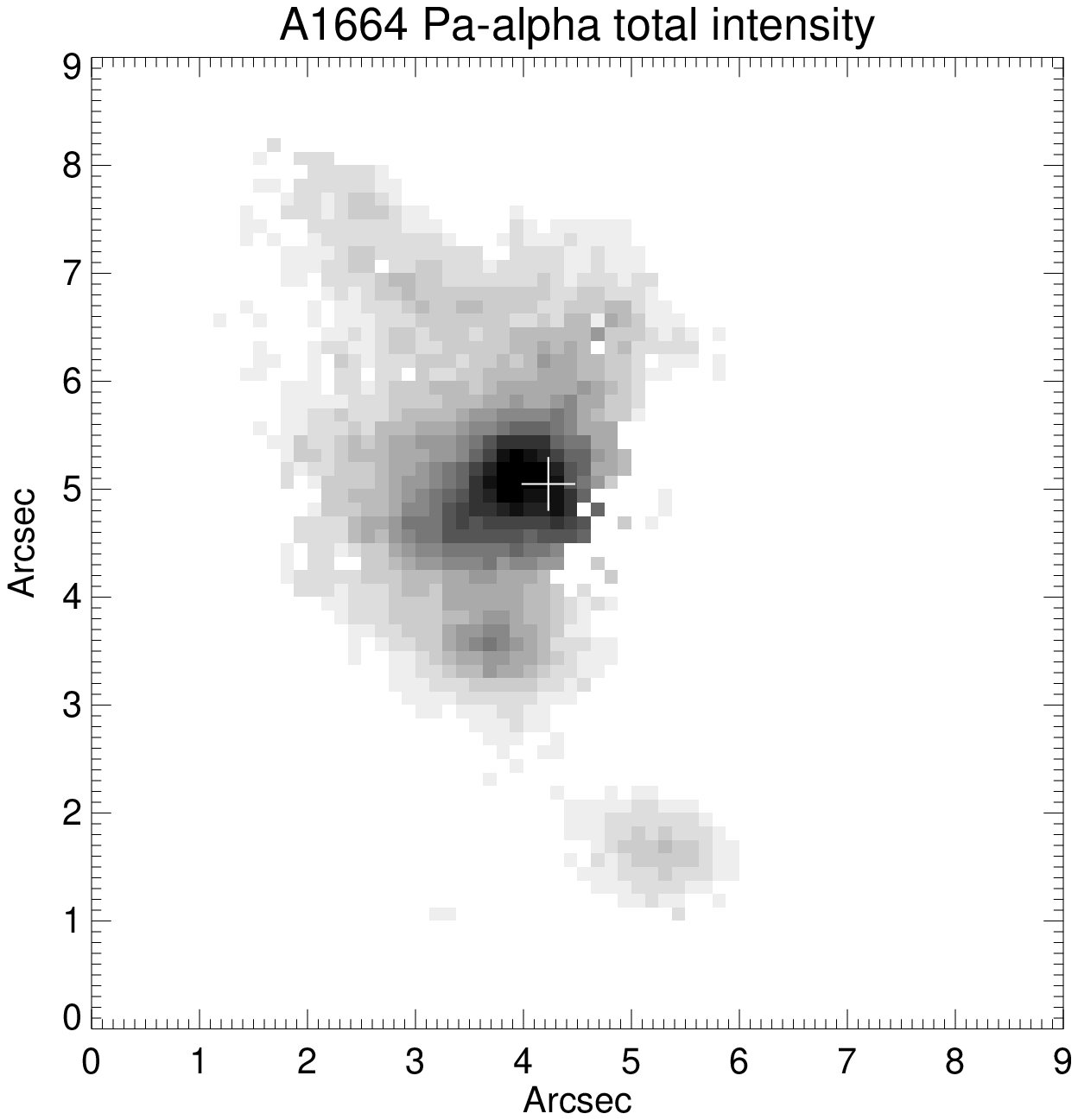}
\includegraphics[width=5.8cm,angle=0]{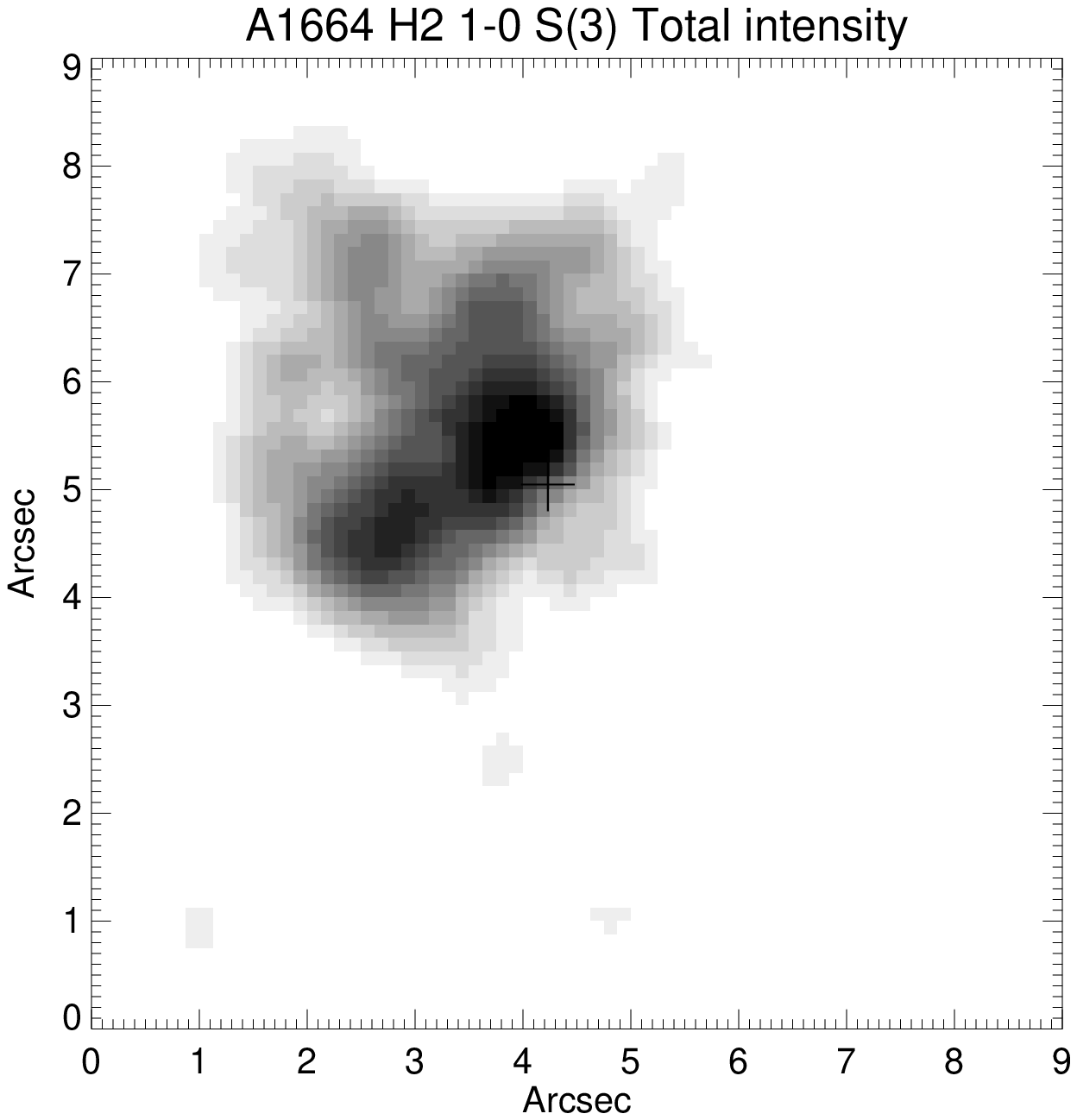}
\includegraphics[width=5.8cm,angle=0]{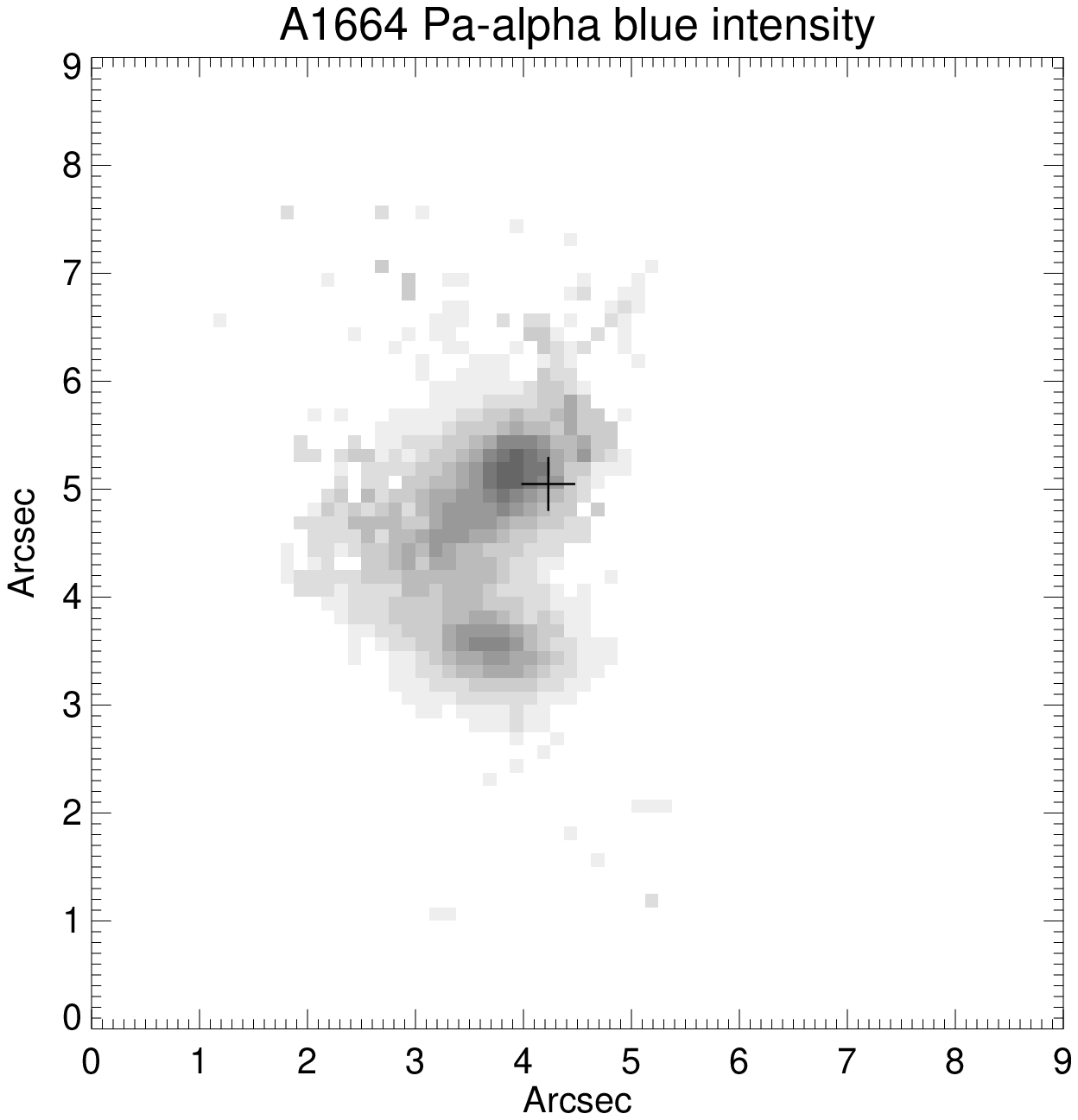}
\includegraphics[width=5.8cm,angle=0]{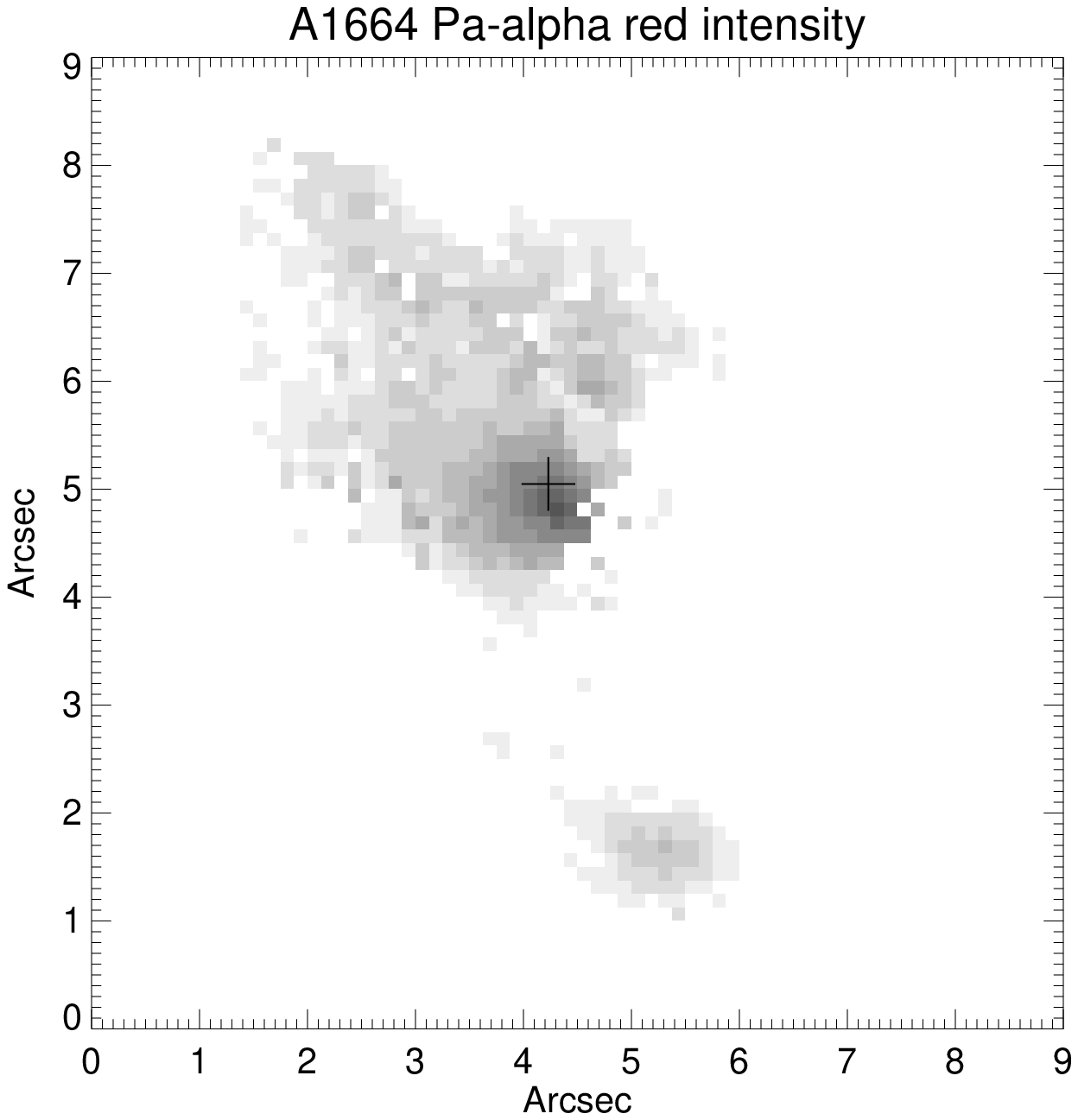}
\includegraphics[width=5.8cm,angle=0]{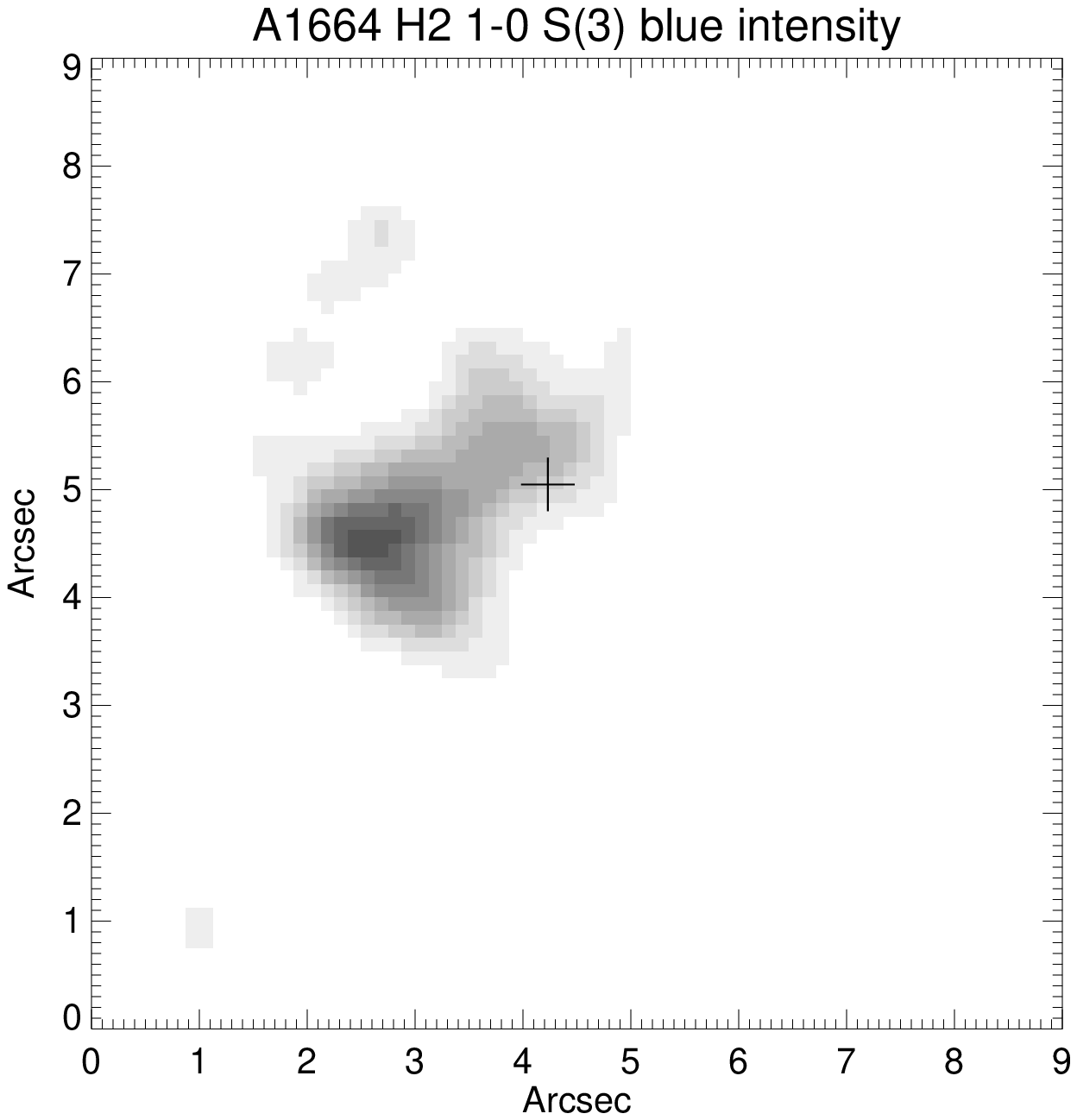}
\includegraphics[width=5.8cm,angle=0]{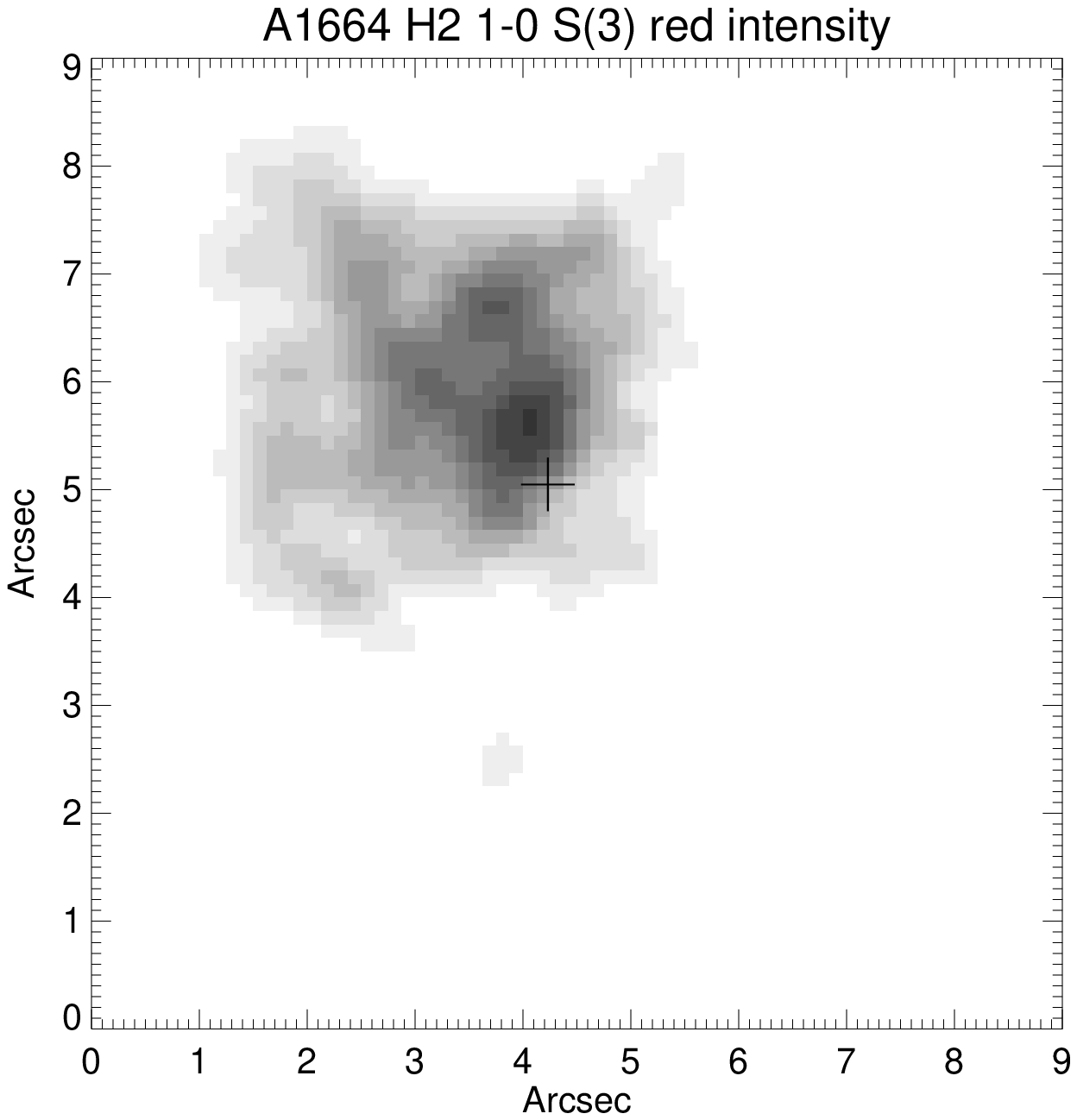}
\caption{Continuum and emission line maps for A1664. The maps for \H2~1-0 S(3) have been
smoothed with a gaussian of FWHM 0.5\arcsec. For each of the two emission lines, the red, blue and total intensity 
maps are plotted on a common greyscale range for ease of comparison. The peak surface brightness of \Pa~maps is 4 times that of the smoothed \H2~v=1-0~S(3) maps. The crosshair in each panel denotes 
the location of the continuum nucleus of the galaxy, as assessed from the K-band image.}
\label{fig:A1664totals}
 \end{centering}
\end{figure*}

An immediate insight into the complex kinematics is shown in the velocity distributions derived
from the emission line fits to Pa$\alpha$ and \H2~v=1-0~S(3) (Fig.~\ref{fig:veldist}). Two 
distinct velocity systems are evident: one spanning --400 to +200\kmps, the other --400 to --600\kmps~(relative to a systemic redshift of $z=0.1276$). Hereafter we refer to these as the `red' and `blue' 
components respectively. The morphologies of these components are shown alongside a K-band continuum map 
in Fig.~\ref{fig:A1664totals}. 

As expected, the \Pa~total intensity map bears a close resemblance to the \Ha~map shown in Wilman et al.~(2006).
Minor differences exist between the morphologies of the `blue' and `red' velocity components in \Pa~and \Ha, but
these are due to the improved separation which is now possible for \Pa~compared with the \Ha+[NII] complex. The
reduced extinction in the near-infrared compared to the optical may also lead to morphological differences 
(Crawford et al.~(1999) derive reddening of $E(B-V) = 0.46$~mag from the Balmer decrement).

\begin{figure*}
\includegraphics[width=0.5\textwidth,angle=90]{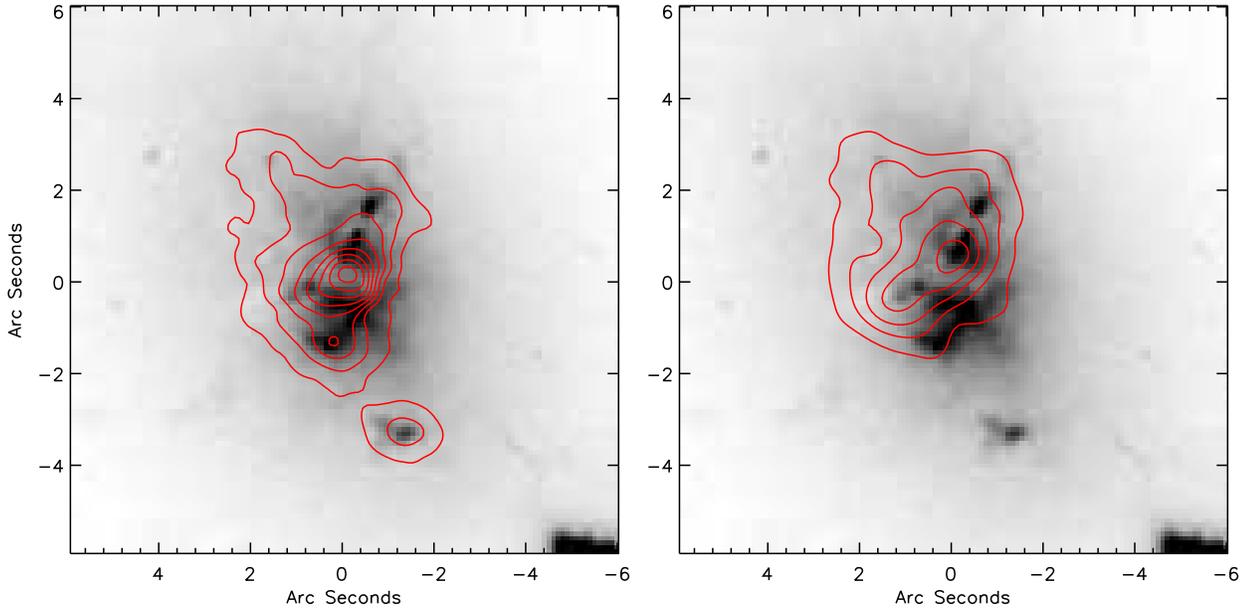}
\caption{\normalsize An HST WFPC2 image of A1664 (F606W filter) overlaid with contours of the emission in \Pa~(left) and \H2~v=1-0~S(3) (right).}
\label{fig:A1664hst}
\end{figure*}

\begin{figure*}
\includegraphics[width=1.0\textwidth,angle=0]{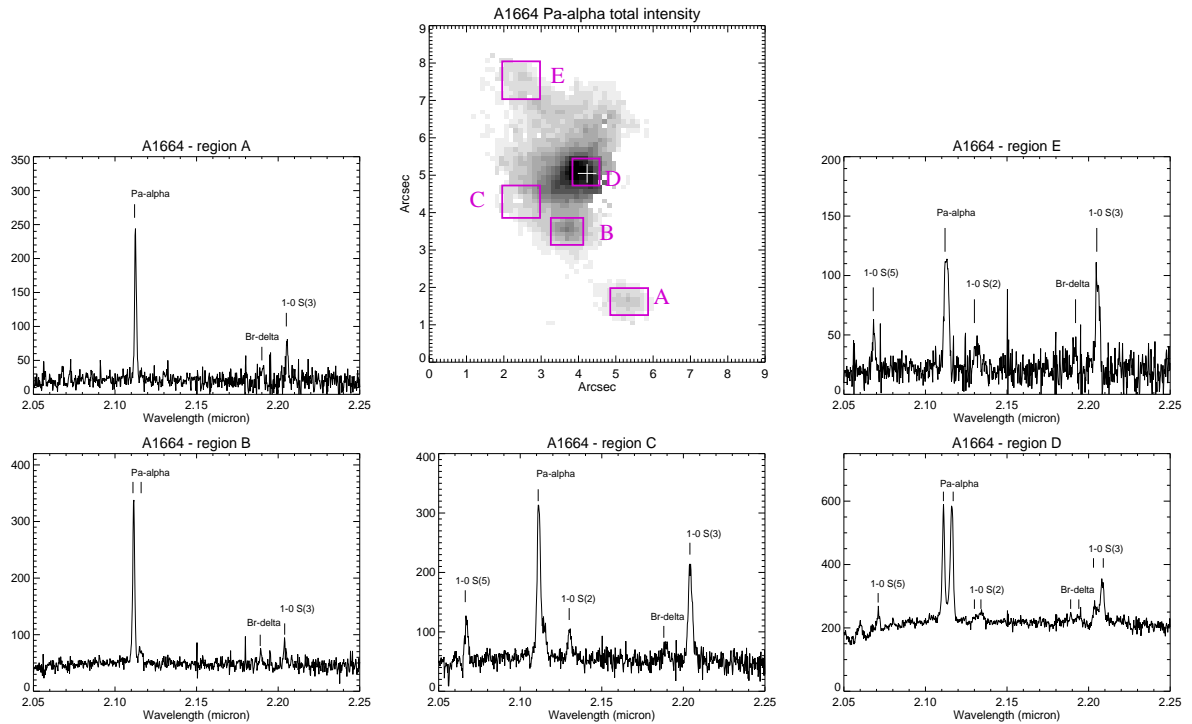}
\caption{\normalsize Spectra of selected regions in A1664 showing the
spatial variations in Pa$\alpha$ and \H2~line ratios. The y-axis units on all spectra are $10^{-15}$\ergpcmsqpspm.}
\label{fig:A1664boxspec}
\end{figure*}

The morphologies in \Pa~and \H2~v=1-0~S(3) are notably different, as was already hinted at by the disparity in 
their line widths in our original K-band long-slit spectrum (Edge et al.~2002). The peaks of all the emission line 
maps are spatially offset from one another and from the K-band continuum nucleus of the galaxy. In Fig.~\ref{fig:A1664hst} 
we show an HST WFPC2 (F606W filter) image of A1664 (private communication, C.P.~O'Dea), overlaid with contours of the 
\Pa~and \H2~v=1-0~S(3) total intensities. This wide V-band image reveals a complex morphology with several bright knots of 
star formation and obscuring dust lanes. The \Pa~emission peaks on the central dust lane and for the most part follows
the distribution of the star-forming knots. The \H2~emission largely avoids the star-forming region to the south of
this dust lane, but exhibits a secondary peak coincident with another prominent dust feature. A full analysis of the
HST image will appear elsewhere. Spectra for a selection of spatial regions are shown in Fig.~\ref{fig:A1664boxspec} 
and confirm the marked variations in \H2~v=1-0~S(3)/\Pa~ratio, to which we will return in section 6.

\begin{figure*}
\begin{centering}
\includegraphics[width=5.8cm,angle=0]{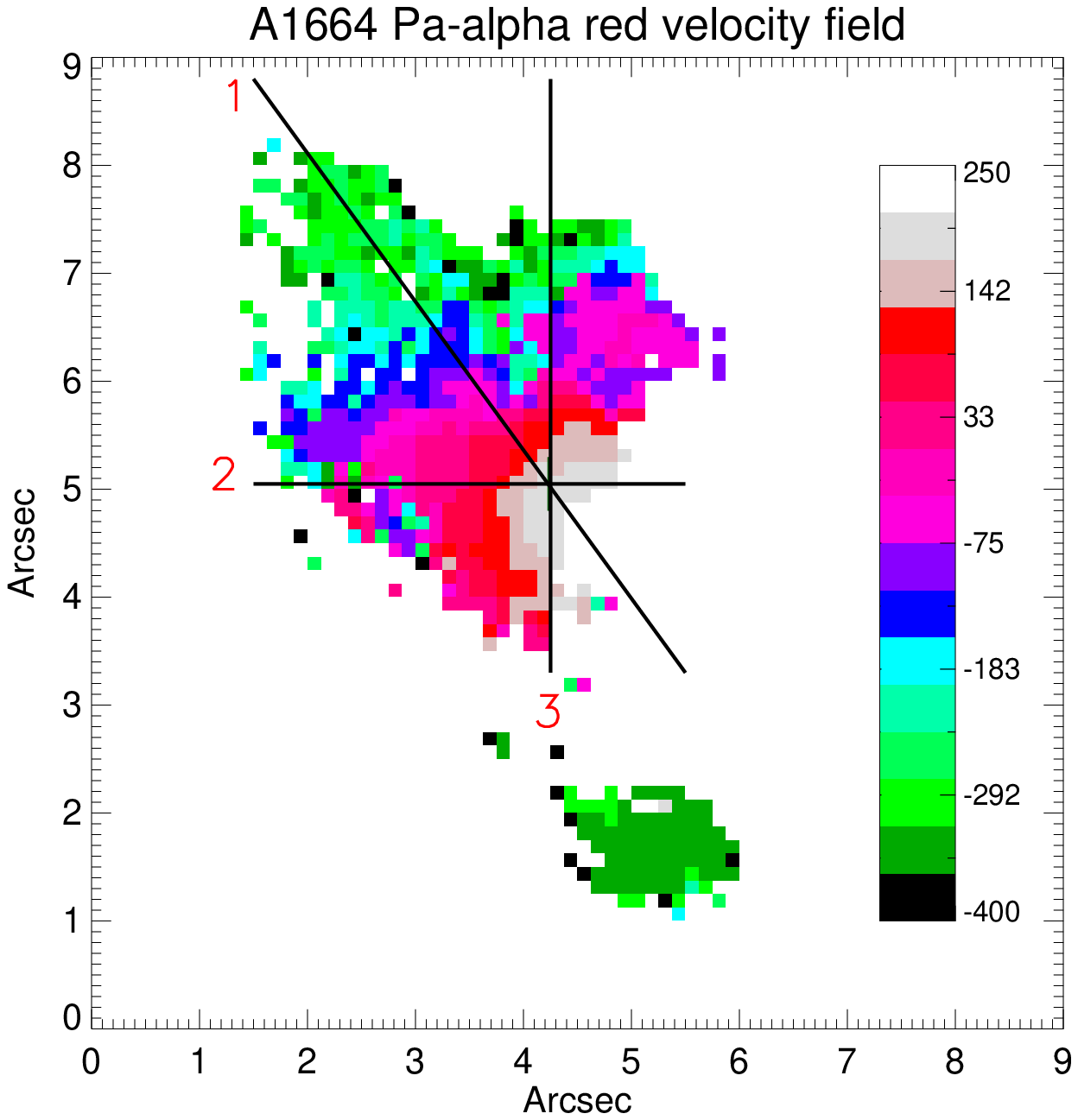}
\includegraphics[width=5.8cm,angle=0]{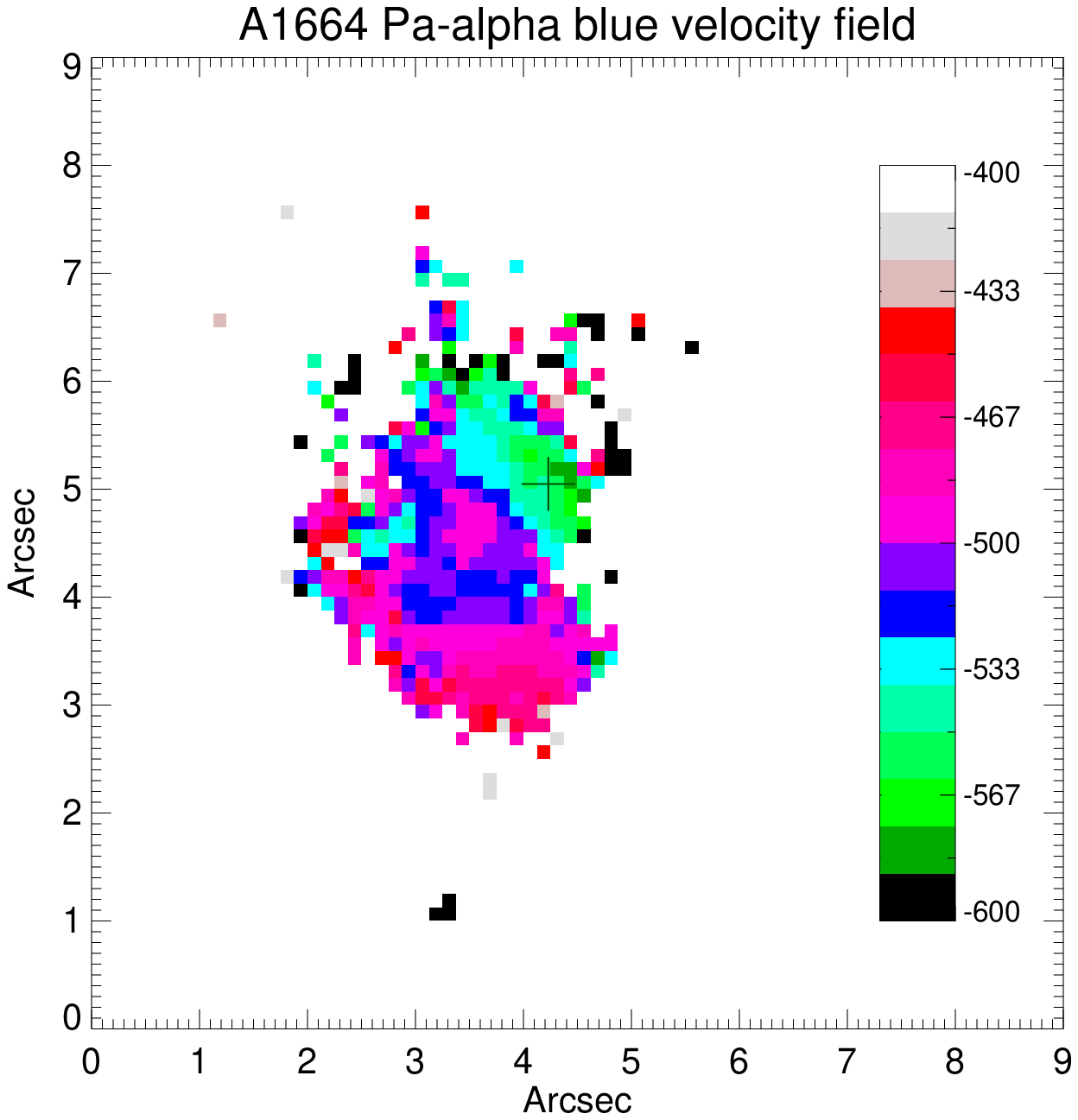}
\includegraphics[width=5.8cm,angle=0]{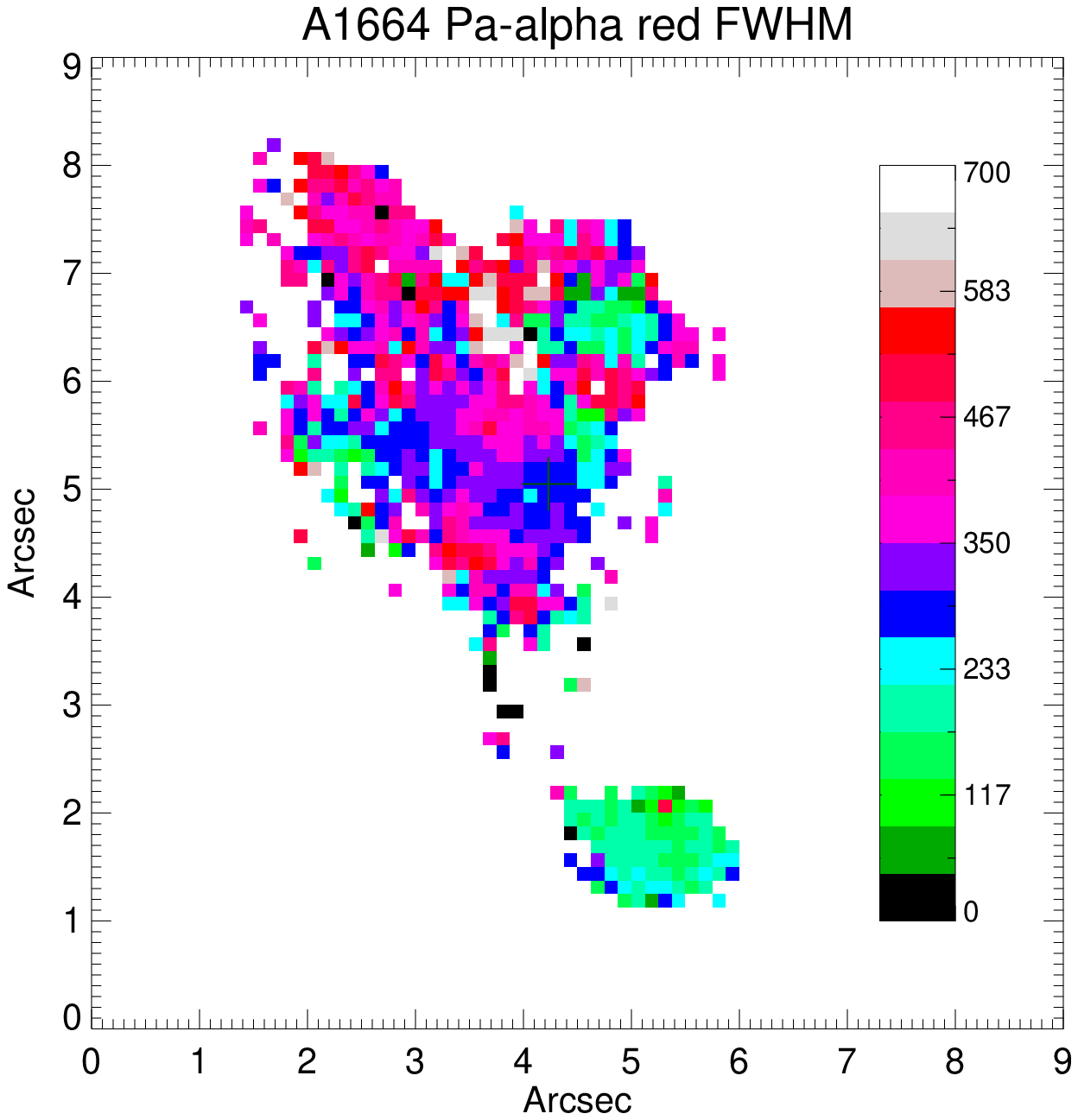}
\includegraphics[width=5.8cm,angle=0]{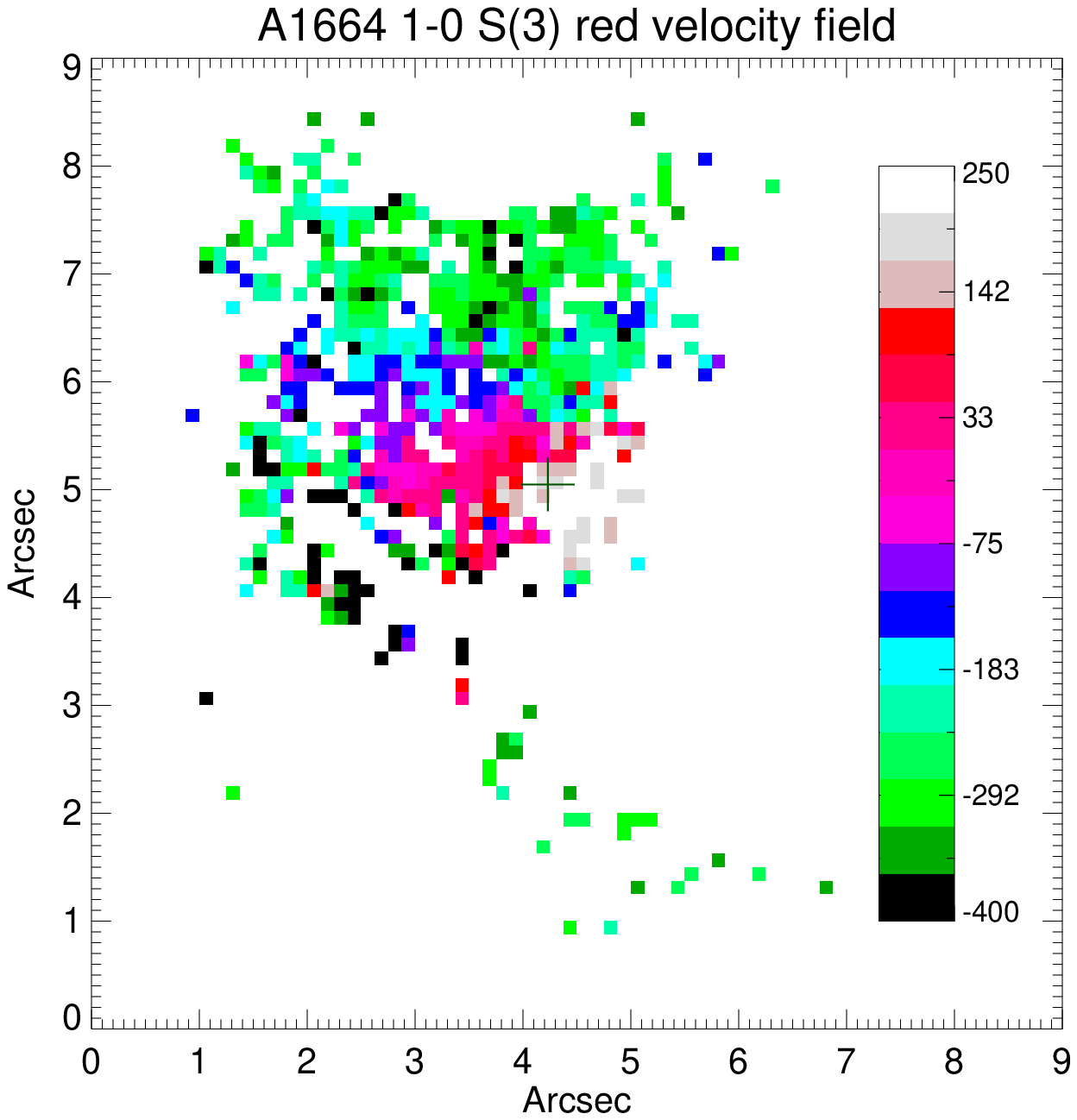}
\includegraphics[width=5.8cm,angle=0]{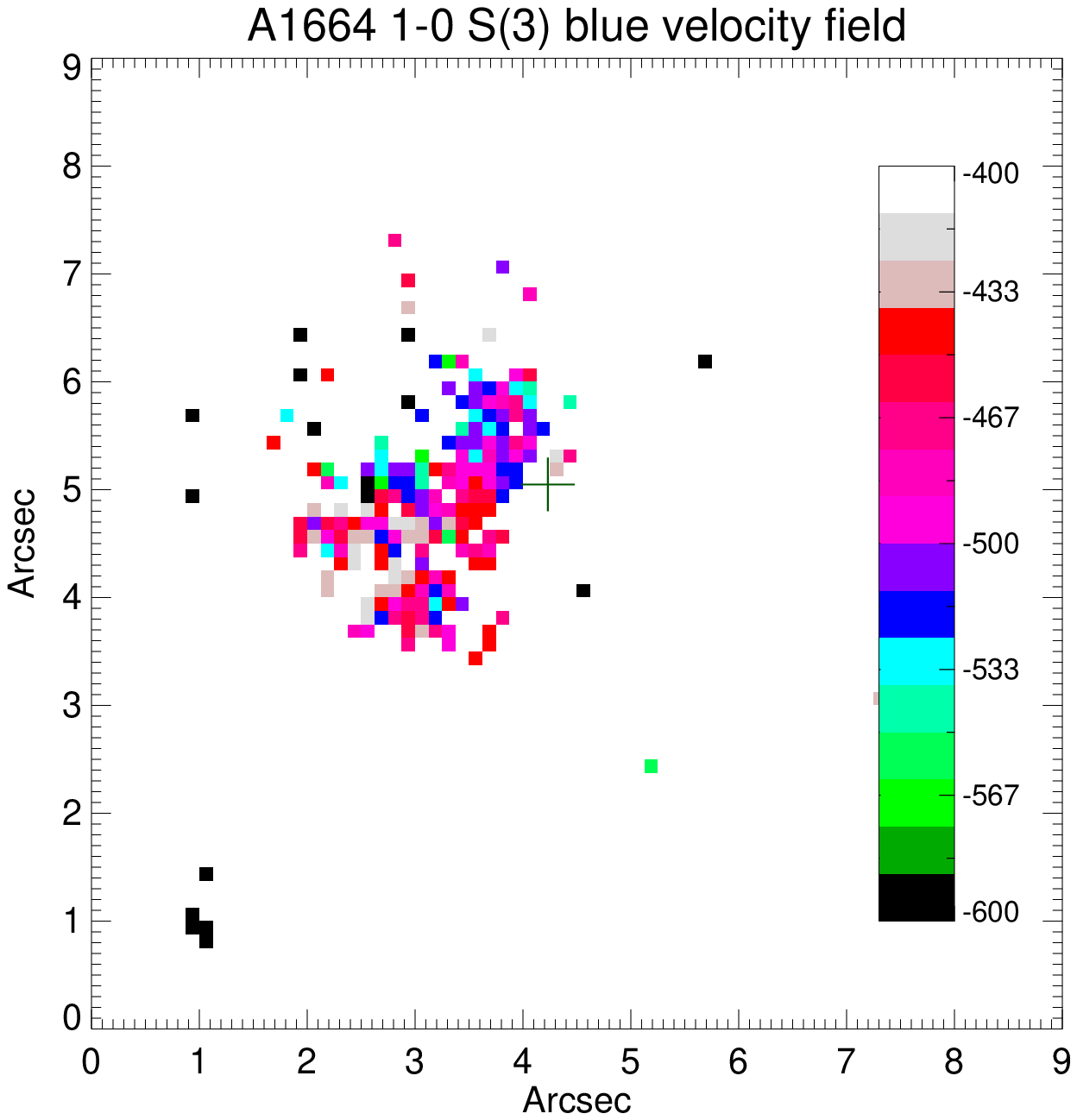}
\includegraphics[width=5.8cm,angle=0]{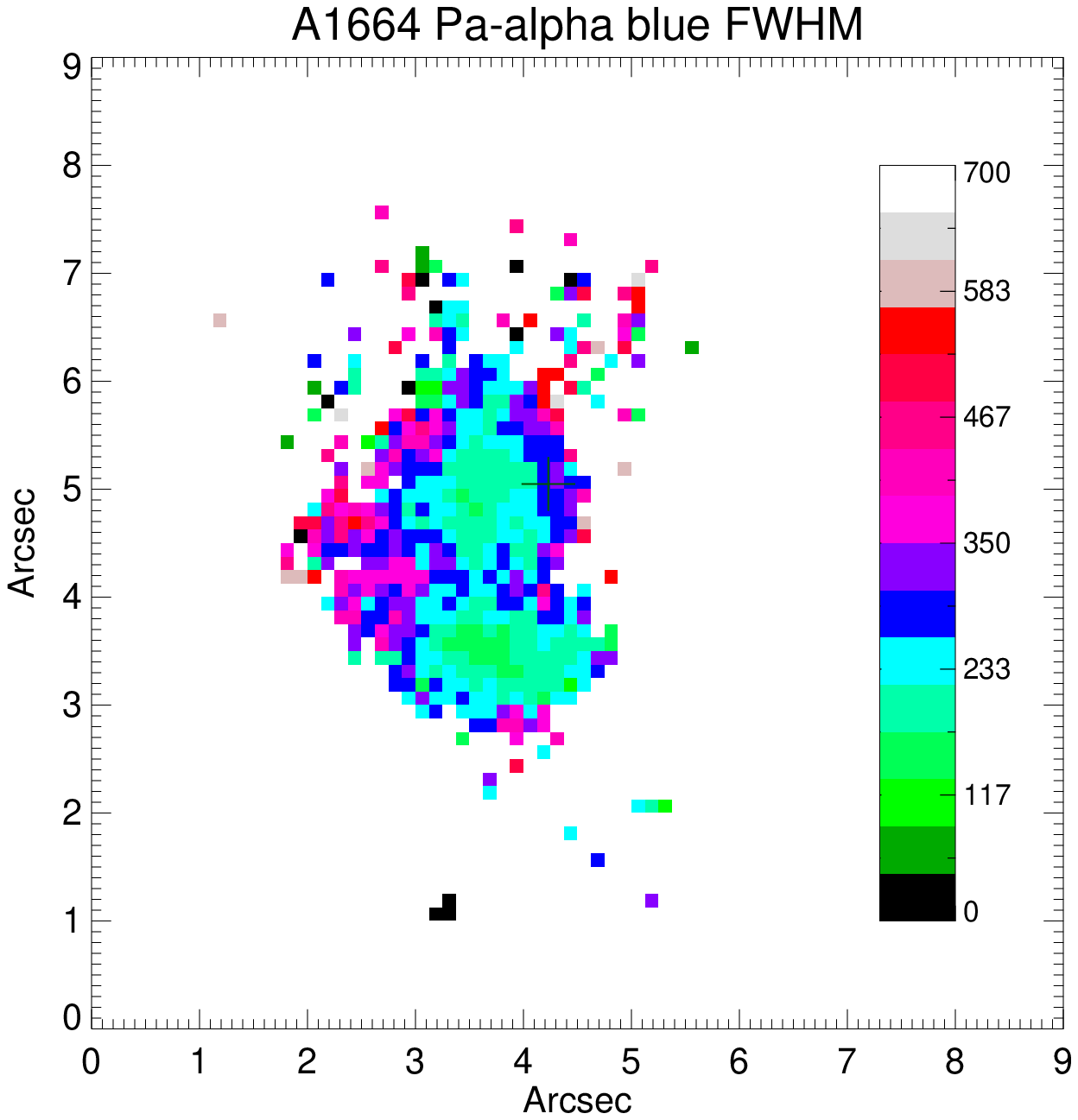}
\caption{Kinematic maps for the red and blue velocity components of \Pa~and \H2~v=1-0~S(3) in A1664. All velocities 
are in \kmps, with reference to a systemic redshift of $z=0.1276$. For reference, the crosshair locates the 
continuum nucleus.}
\label{fig:A1664kinemaps}
 \end{centering}
\end{figure*}

The two-dimensional gas kinematics are shown in Fig.~\ref{fig:A1664kinemaps}. The most coherent large-scale motions
are those of the \Pa~red component, whose recession velocity increases as one moves in towards the nucleus. The 
strongest velocity gradient is along the north-east--south-west direction (`axis 1'), where the FWHM of the emission 
is also elevated with respect to the other parts, at $\sim 400$\kmps. The blue \Pa~component is largely 
quiescent, with FWHM $< 200$\kmps~in the regions with the most coherent velocity structure. The \H2~kinematic maps are
necessarily noisier but the red velocity field resembles that of \Pa. 

Integrating the emission within the SINFONI field-of-view, the total line luminosities are $5.3 \times 10^{41}$\ergps~(\Pa) and
$3.1 \times 10^{41}$\ergps~(\H2~v=1-0~S(3)). These are 5 and 12 times larger, respectively, than the line luminosities 
implied by the fluxes quoted in Edge et al.~(2002) within a 1.22\arcsec-wide north-south slit.

\subsection{Interpretation: inflowing filaments of cooling gas?}
\begin{figure*}
\includegraphics[width=0.40\textwidth,angle=0]{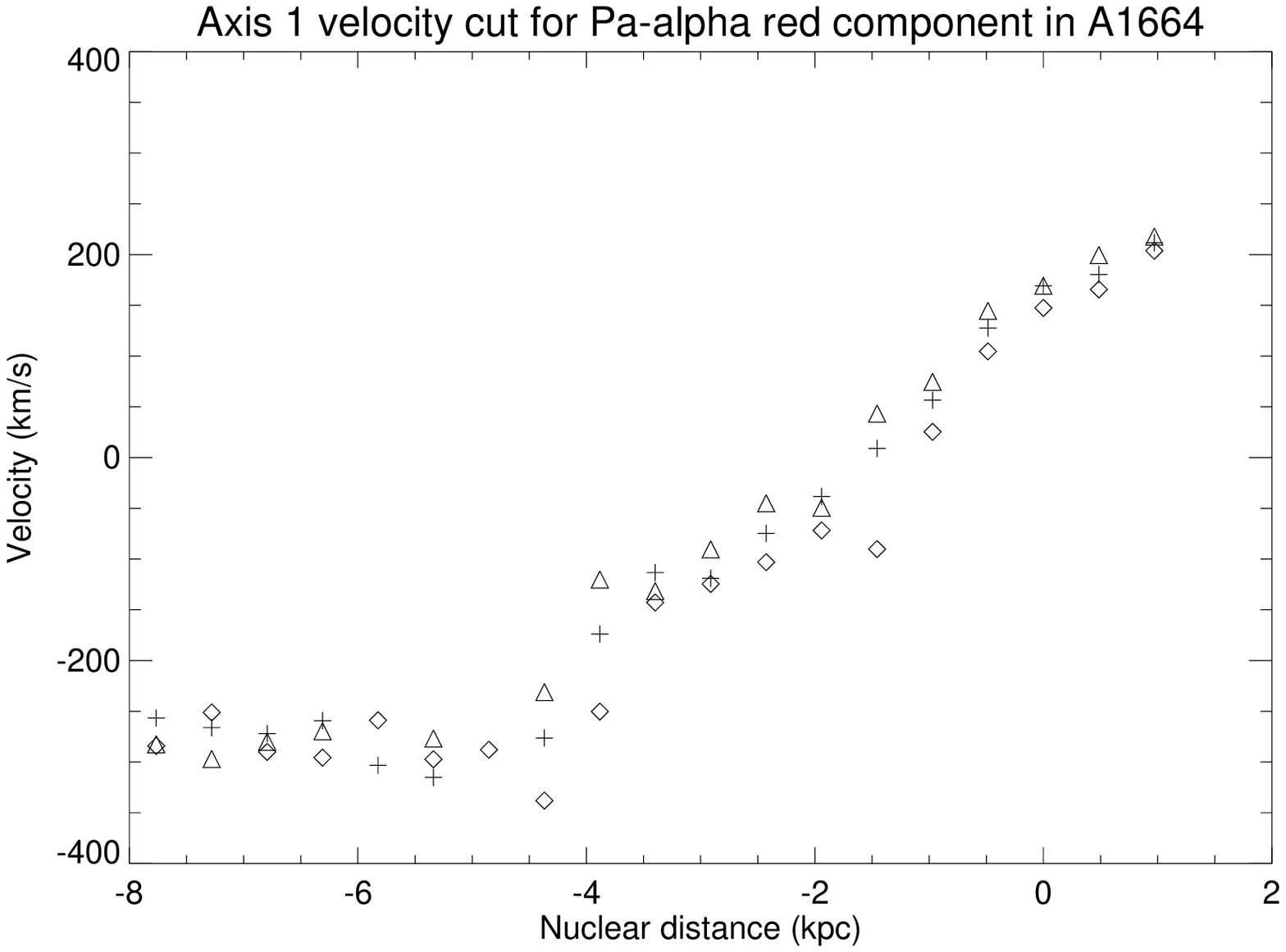}
\includegraphics[width=0.40\textwidth,angle=0]{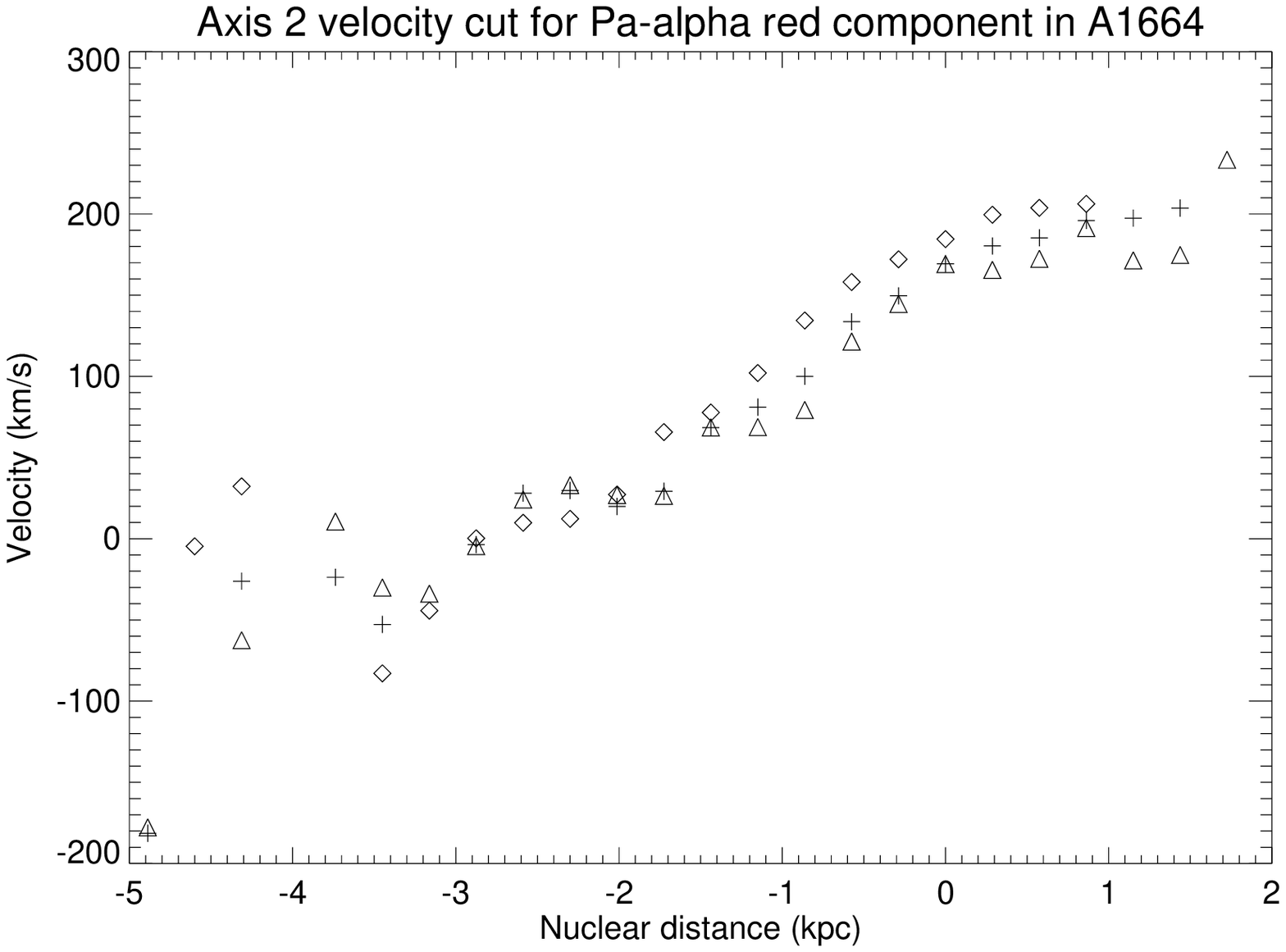}
\includegraphics[width=0.40\textwidth,angle=0]{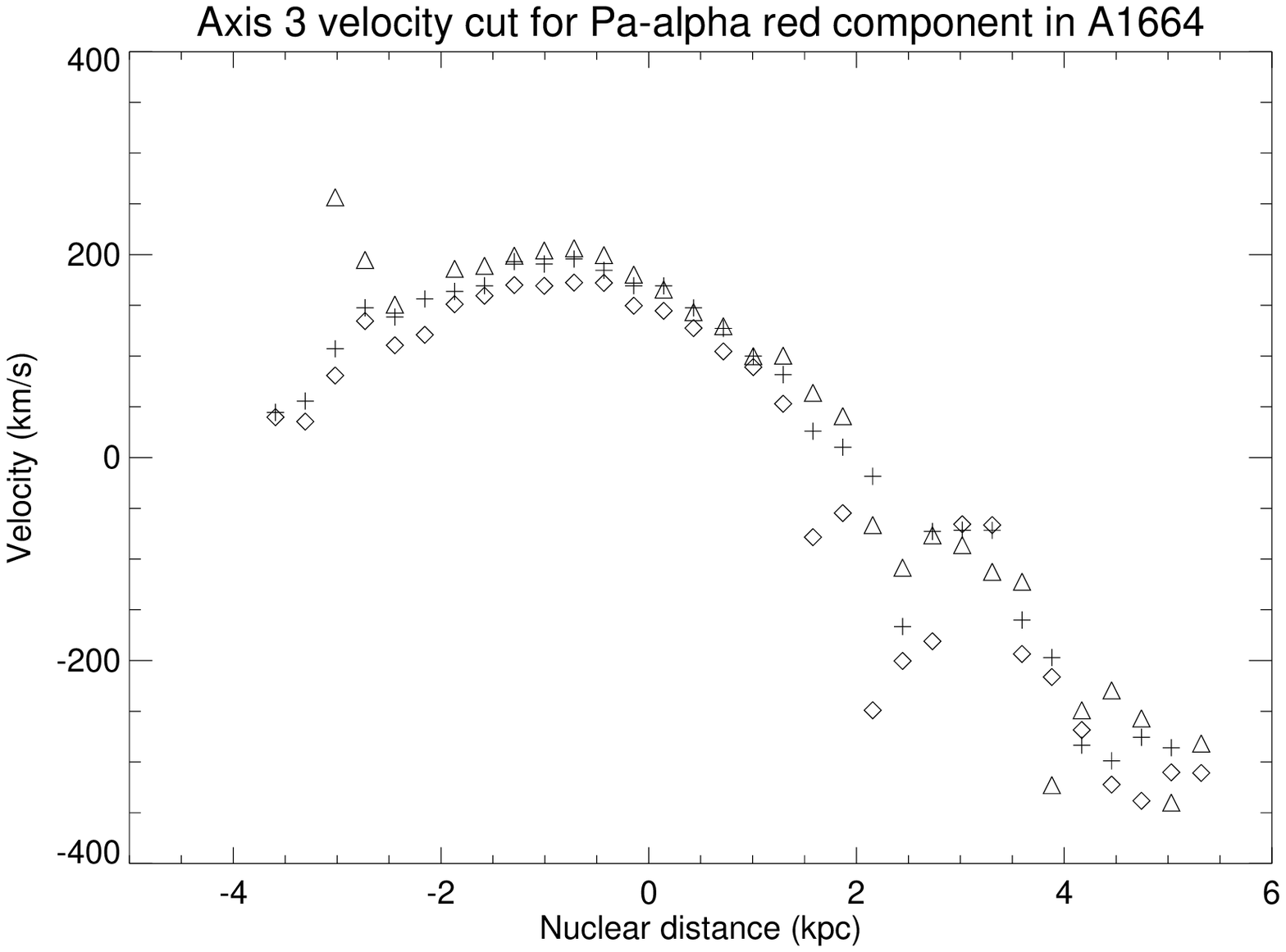}
\caption{\normalsize Velocity cuts for the red velocity component of \Pa~in A1664 as a function of projected 
nuclear distance along the three axes shown in Fig.~\ref{fig:A1664kinemaps}. Negative nuclear distance 
corresponds to the numbered ends of the axes shown in Fig.~\ref{fig:A1664kinemaps}. To give some indication 
of the errors on these velocities, the different symbols refer to three adjacent cells at the same nuclear distance along each axis.}
\label{fig:velcuts}
\end{figure*}

The increasing recession velocity of the red component of \Pa~with decreasing projected radius could be 
interpreted as the accelerated infall of gas from the outer regions of the galaxy towards the nuclear regions. 
For more insight into this, Fig.~\ref{fig:velcuts} shows one-dimensional cuts in this velocity field along 
three separate axes indicated in Fig.~\ref{fig:A1664kinemaps}. These demonstrate that, 
beyond an initial starting radius, the velocity increases linearly with decreasing radius. For axis 1, 
the velocity increases by 500\kmps~over 5\kpc, along axis 2 by 250\kmps~over 3kpc, and along axis 3 by 
400\kmps~over 4\kpc. 

Similar kinematics were observed Lim et al.~(2008) for the CO filaments in NGC 1275 extending out to 8\kpc, 
and in the near-infrared long-slit spectroscopy of Wilman et al.~(2000) for Cygnus A. Lim et al. demonstrated that 
the linear increase of velocity with radius could be modelled as free fall within the gravitational
potential of the galaxy, for a parcel of cool gas released at rest from some initial radius. The CO 
filaments in NGC 1275 are co-spatial with the coolest X-ray gas in the system (orientated roughly 
perpendicular to the radio lobes), suggesting that the filaments represent gas which has cooled out of the hot phase, 
decoupled from hydrostatic equilibrium, and started to free fall inwards.

The soft X-ray emission in A1664 (Fig.~9 of Wilman et al.~2006; Kirkpatrick et al.~2009) is also aligned with the 
inflow of gas along the \Ha~filament (axis 1), and with the major axis of the K-band continuum. 
The alignment of the major axis of the CCG with the cluster X-ray emission in A1664 was noted by 
Allen et al.~(1995). Following Lim et al.~(2008), we model the velocity field by considering free-fall 
in a gravitational potential given by the Hernquist~(1990) profile characteristic of elliptical galaxies: 

\begin{equation}
\rho(r) = \frac{M}{2\pi}\frac{a}{r}\frac{1}{(r+a)^3},
\end{equation}

where $M$ is the total mass of the galaxy and $a$ a scale radius. As shown by Lim et al., the radial velocity 
$v(r)$ of a test particle released from rest at radius $r=r_{\rm{0}}$ is given by:

\begin{equation}
v(r)^{2}= \frac{2GM}{(r+a)} - \frac{2GM}{(r_{\rm{0}} + a)}.
\end{equation}

The parameters $M$ and $a$ are not well-constrained for A1664, so we adopt fiducial values of 
$M = 10^{12}$\Msun~and $a=13.3$\kpc. The latter assumes an effective radius 
$R_{\rm{e}} = 24$\kpc~(the mean of a broad distribution measured by Schneider et al.~(1983) for a CCG sample) 
and the relation $R_{\rm{e}} \simeq 1.8153a$ given by Lim et al. Analysis of the K-band IFU image (Fig.~\ref{fig:A1664totals}) 
and the HST image (Fig.~\ref{fig:A1664hst}, ignoring the disturbed central regions), loosely constraints the major axis 
$R_{\rm{e}}$ to the range 18--27\kpc, consistent with the assumed value.  
 
\begin{figure*}
\begin{centering}
\includegraphics[width=5.8cm,angle=0]{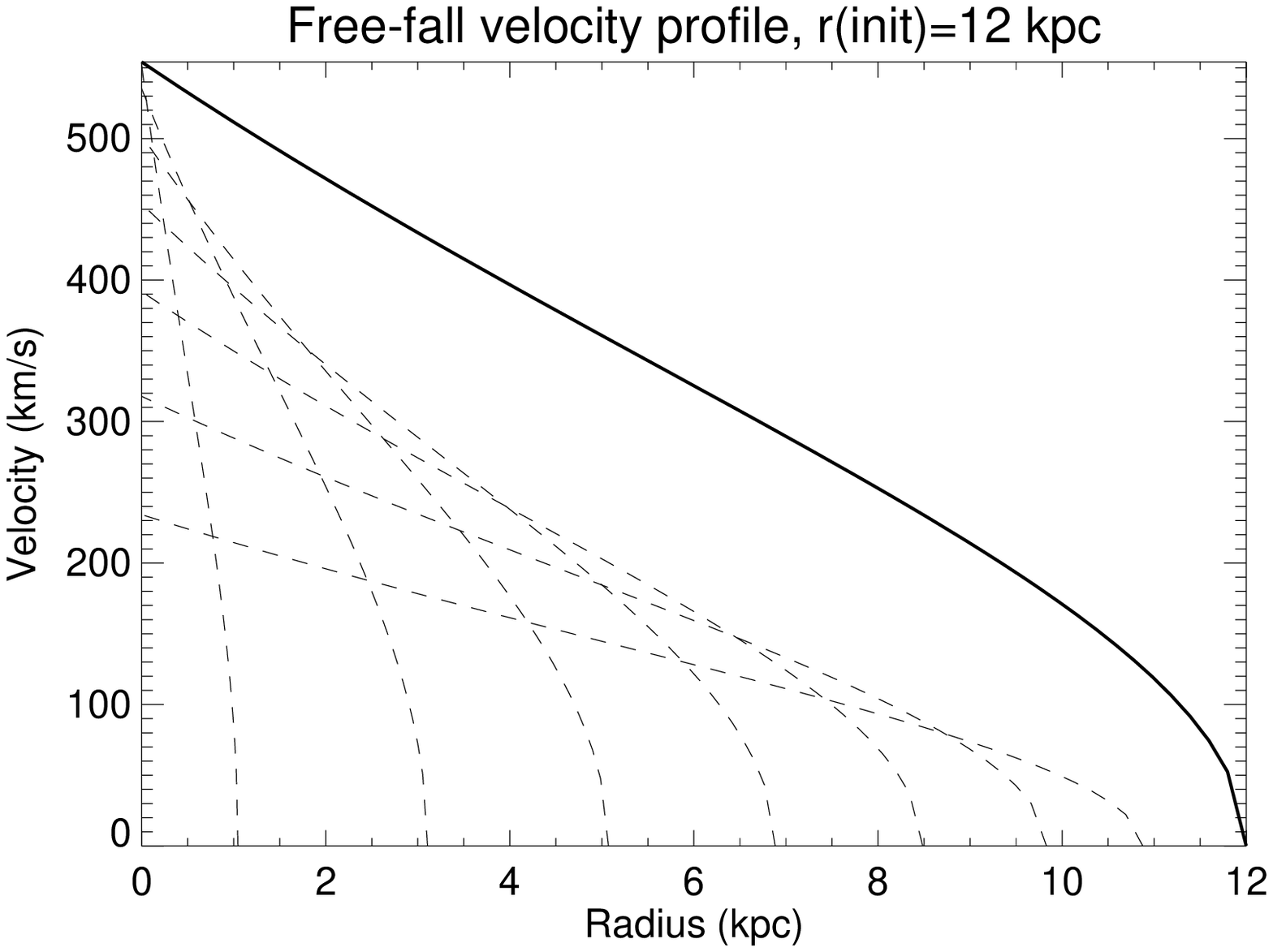}
\includegraphics[width=5.8cm,angle=0]{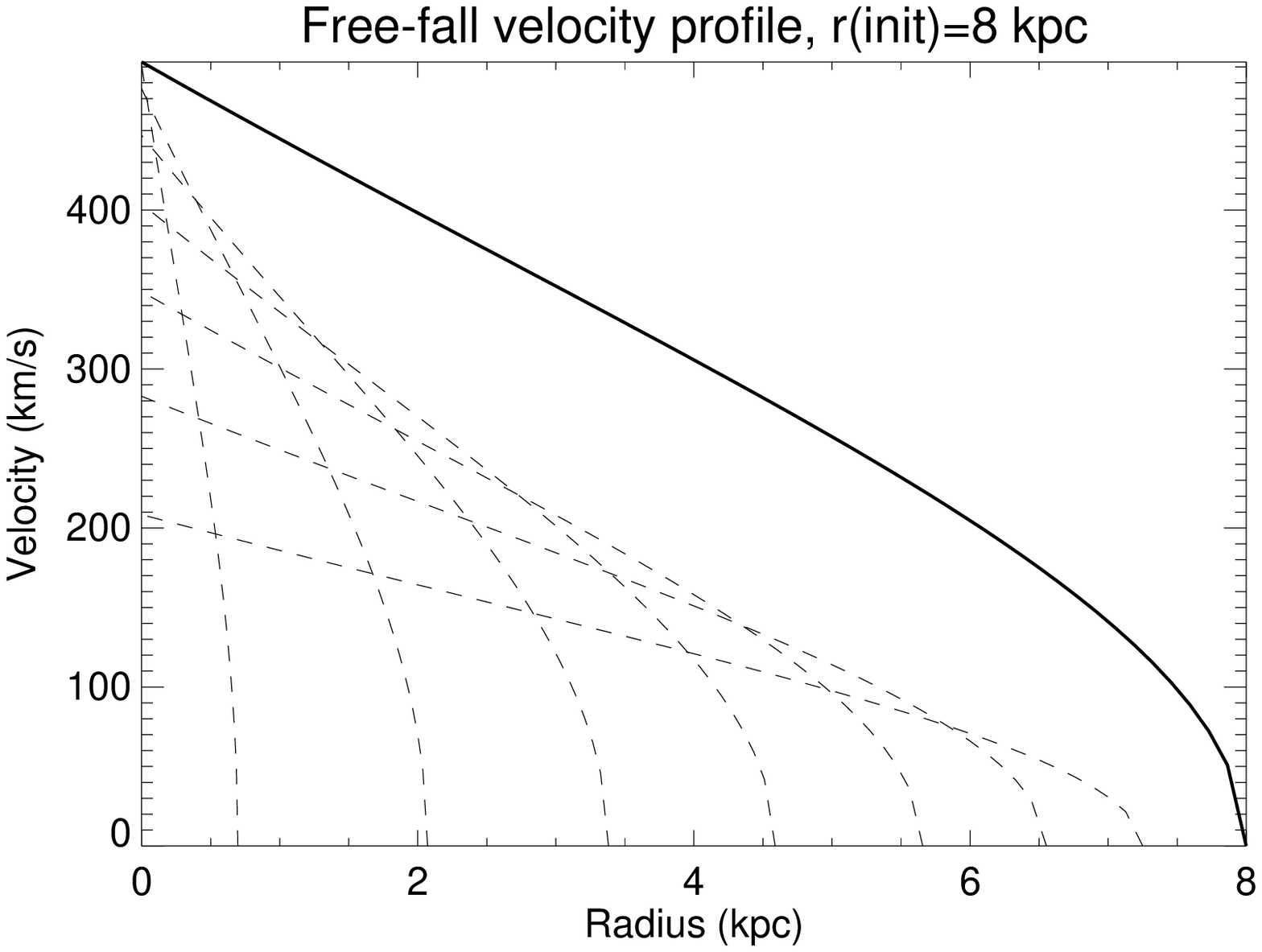}
\includegraphics[width=5.8cm,angle=0]{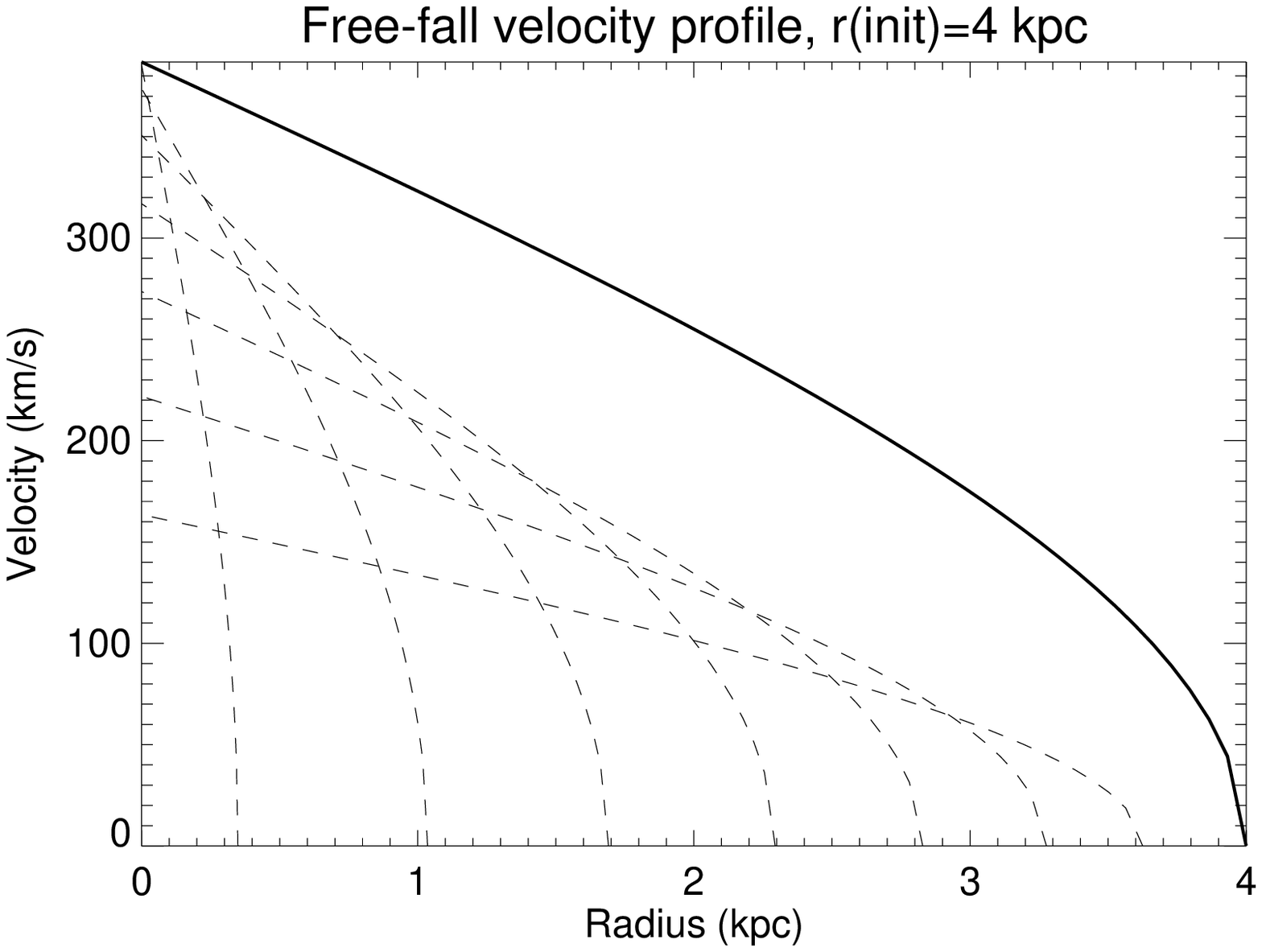}
\caption{For 3 values of the starting radius, the solid lines show the expected radial free-fall 
velocities for a test particle released from rest in a model gravitational potential, as described in 
section 3.2. The dashed lines in each panel show, from right to left, the corresponding {\em line-of-sight} velocities which 
would be observed as a function of {\em projected} radius when these filaments are oriented at angles of 
25, 35, 45, 55, 65, 75, and 85 degrees to the plane of the sky.}
\label{fig:A1664freefall}
 \end{centering}
\end{figure*}

For starting radii $r_{\rm{0}}=12,8$ and 4\kpc, 
Fig.~\ref{fig:A1664freefall} shows the resulting 1-d profiles of line-of-sight velocity versus projected radius 
when the filament is oriented at various angles, $\theta$, to the sky plane. The steepening of the model velocity 
gradient as the filament is oriented closer to the line-of-sight implies that the
observed velocity profiles for axes 1,2 and 3 cannot be modelled by varying the viewing angle for a fixed 
starting radius. Within the context of this model, we find that these gradients can be accounted for with the 
following combinations of parameters $(r_{\rm{0}},\theta)$: (12\kpc, 65$^{\circ}$) for axis 1; (4\kpc, 40$^{\circ}$) 
for axis 2; (8\kpc, 55$^{\circ}$) for axis 3. These are not to be regarded as best fit parameter values, as there are 
clearly degeneracies between the parameters $(r_{\rm{0}},\theta)$ in addition to the uncertainties due to 
the poorly constrained values of $R_{\rm{e}}$ and $M$.

What is not readily explained by this model, however, is the high observed FWHM for the infalling gas along axis 1. 
This may suggest that the filament is observed from a more foreshortened perspective with the infalling gas covering 
a range in solid angle. It may instead arise from turbulence due to hydrodynamic interaction of the clouds 
with the diffuse medium. 

In the absence of better constraints on the parameters of the gravitational potential, this analysis 
serves only to demonstrate that an infall interpretation of the observed velocity field is plausible. 
The observed kinematics may have a different origin, e.g. from an interaction with a secondary cluster galaxy, 
as suggested by Wilman et al.~(2006). A second alternative is that the filamentary \Ha~emission and extended 
`bar-like' X-ray emission (Kirkpatrick et al.~2009) may be due to a `sloshing' motion of the CCG within the 
cluster potential, as invoked to explain the X-ray and \Ha~filament in A1795 (Fabian et al.~2001). In that case, 
however, the \Ha~filament has quiescent kinematics over the bulk of its length with more violent motions only 
seen around the powerful radio source within the CCG (Crawford et al.~2005).

\section{Results on A2204}
In Wilman et al.~(2006), A2204 was found to have an irregular \Ha~morphology consisting of a several
weakly-defined filaments extending from the nuclear regions, possibly coincident with dust features seen
in an {\em Hubble Space Telescope} optical image. The velocity field appeared to comprise 3 clumps of 
emission at velocities of --200, --50 and --100\kmps~(relative to $z=0.1514$), with a FWHM$\simeq 200$\kmps, 
rising to $800\kmps$ in the vicinity of the nucleus. From the {\em VIMOS} data, the secondary galaxy located
$\simeq 10$\kpc~in projection to the south-west of the CCG was measured to have a line-of-sight velocity of $\simeq -100$\kmps~relative to the CCG. It is thus plausible that the two galaxies are interacting. Narrow-band \Ha+[NII] imaging was presented by Jaffe et al.~(2005), which revealed diffuse line emission out to radii of 24\kpc~with a north-west--south-east alignment. 

\begin{figure*}
\begin{centering}
\includegraphics[width=5.8cm,angle=0]{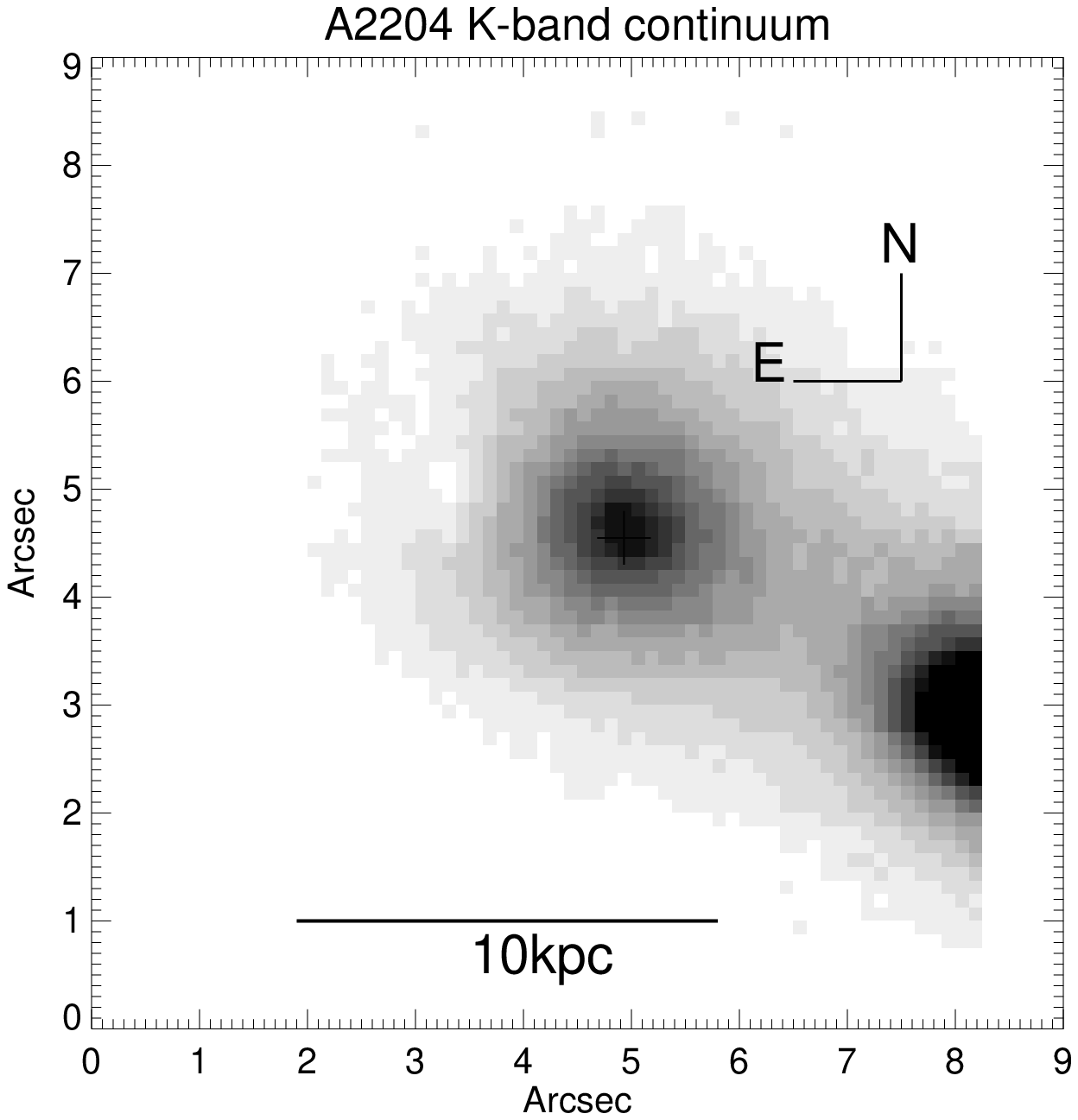}
\includegraphics[width=5.8cm,angle=0]{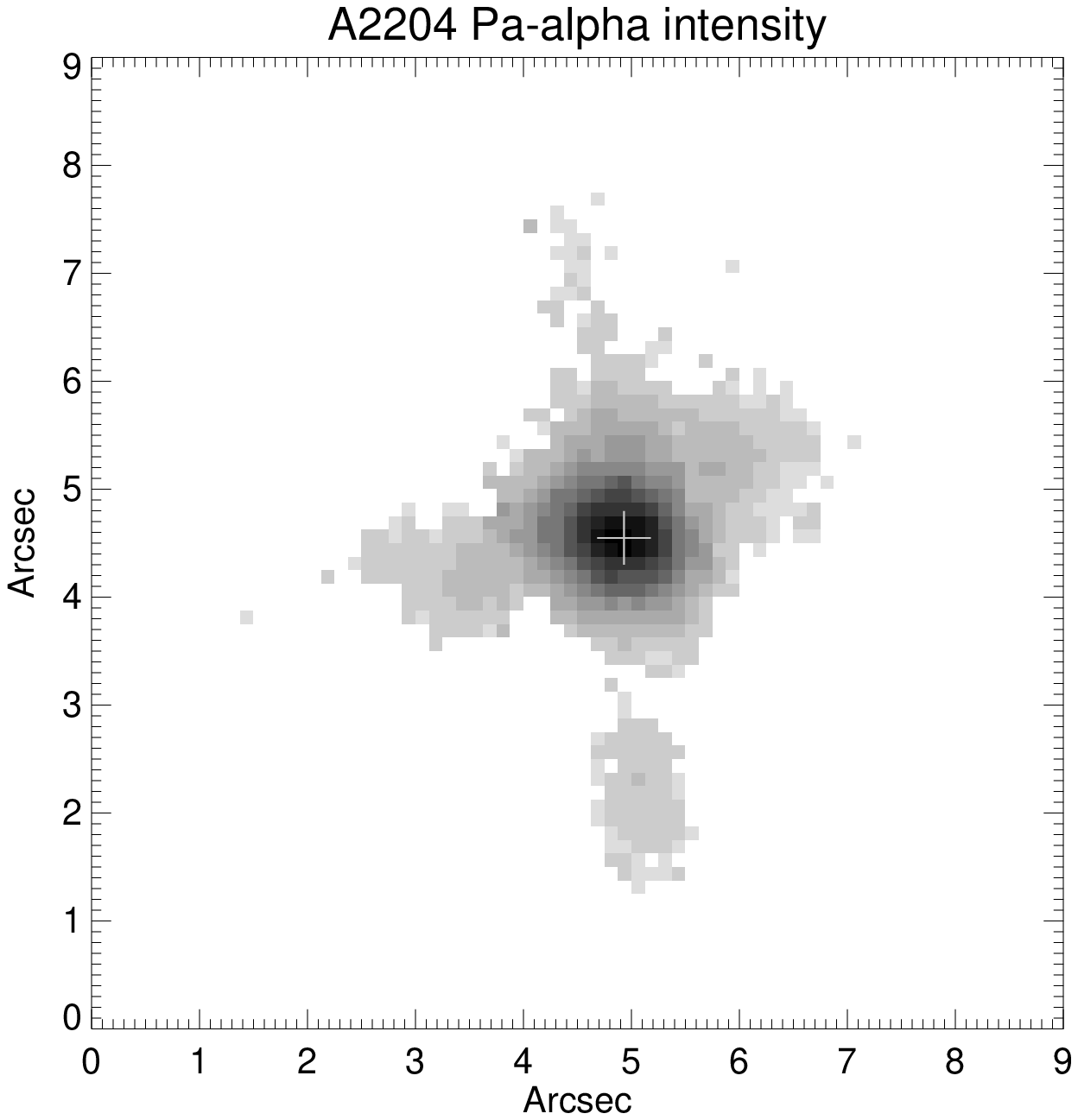}
\includegraphics[width=5.8cm,angle=0]{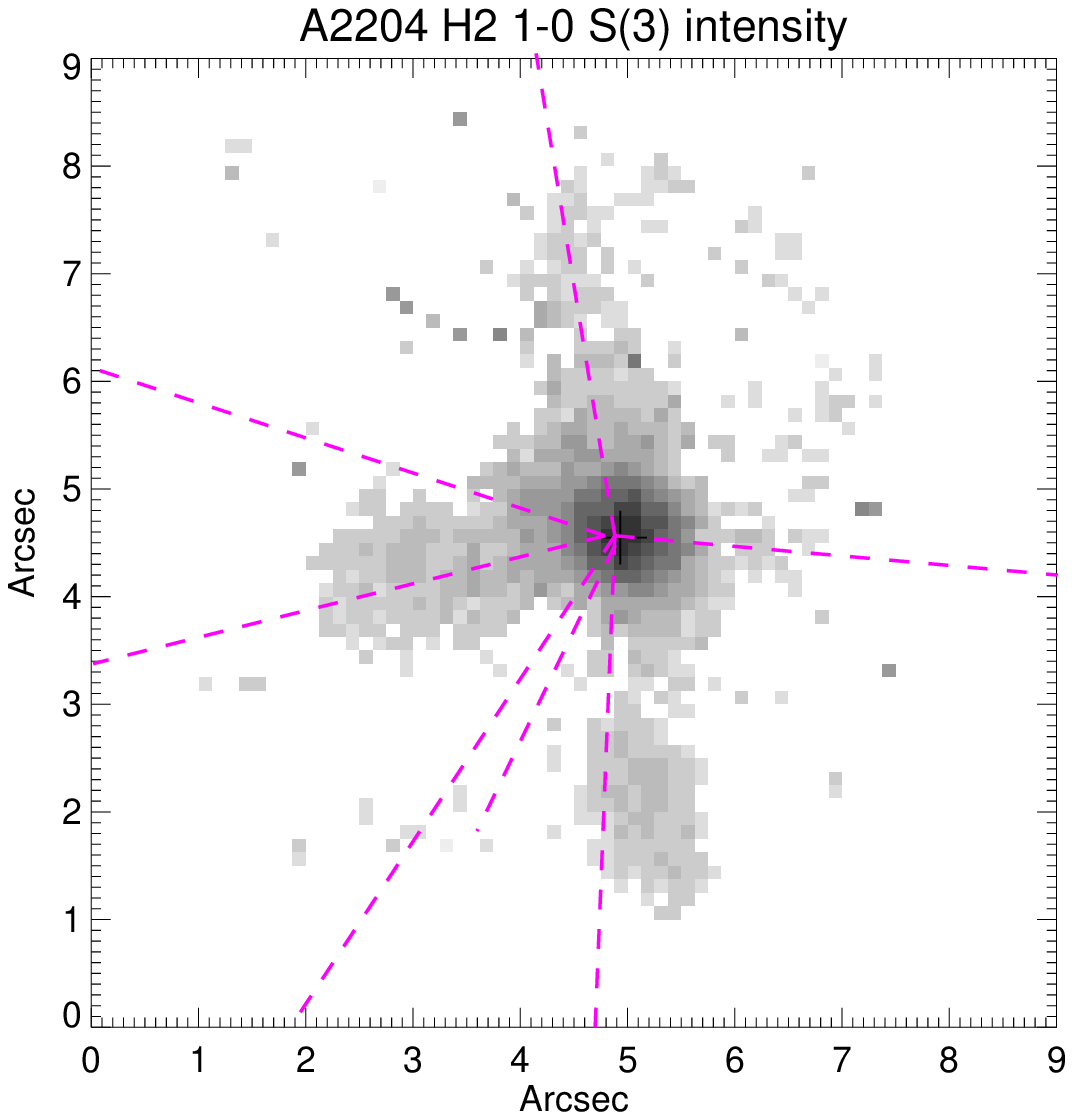}
\caption{Continuum and emission line maps for A2204. For ease of comparison, the crosshair in each panel denotes 
the location of the continuum nucleus of the galaxy, as assessed from the K-band image, and the emission 
line maps use a common greyscale. The galaxy to the south-west is a secondary cluster galaxy with a line-of-sight velocity blue-shifted by 100\kmps~relative to the central
cluster galaxy (see Wilman et al.~2006 for details). The dashed lines on the 1-0~S(3) image show the radial
vectors in the directions of the 7 `ghost bubbles' identified in X-ray imaging by Sanders et al.~(2008) (6 of the bubbles lie at projected radii of 20\kpc, outside the SINFONI field-of-view). }
\label{fig:A2204totals}
 \end{centering}
\end{figure*}

\begin{figure*}
\begin{centering}
\includegraphics[width=5.8cm,angle=0]{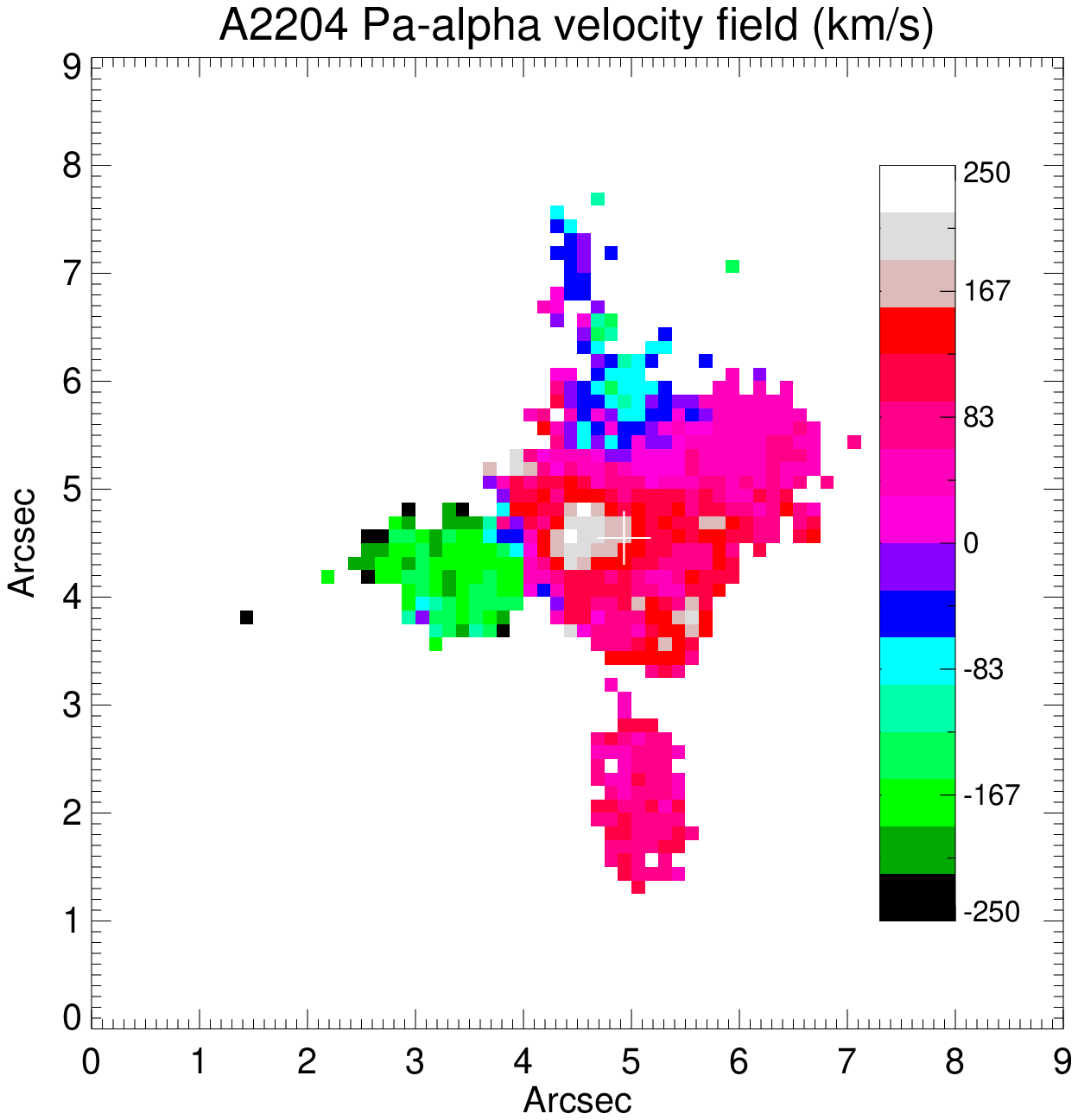}
\includegraphics[width=5.8cm,angle=0]{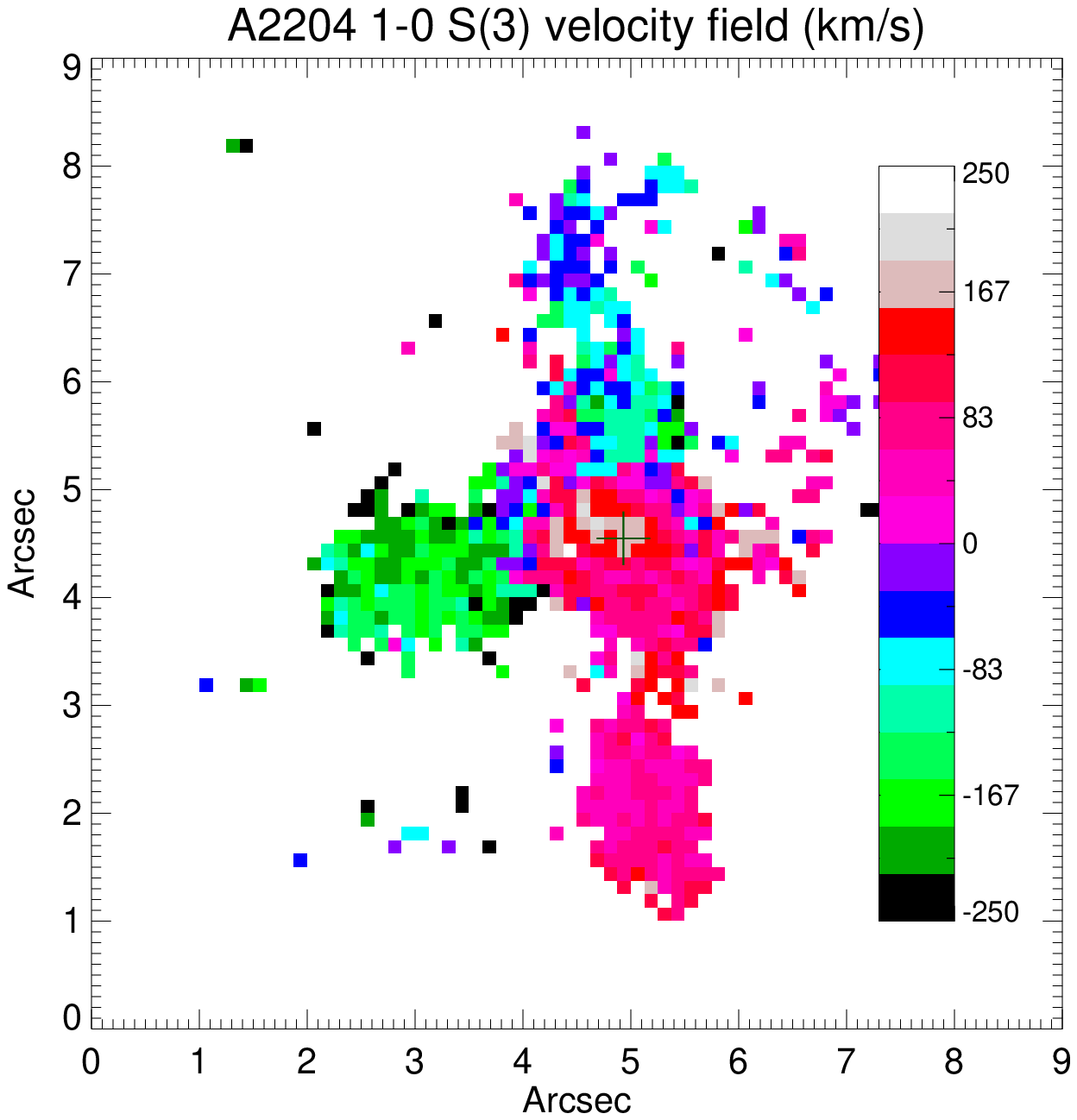}
\includegraphics[width=5.8cm,angle=0]{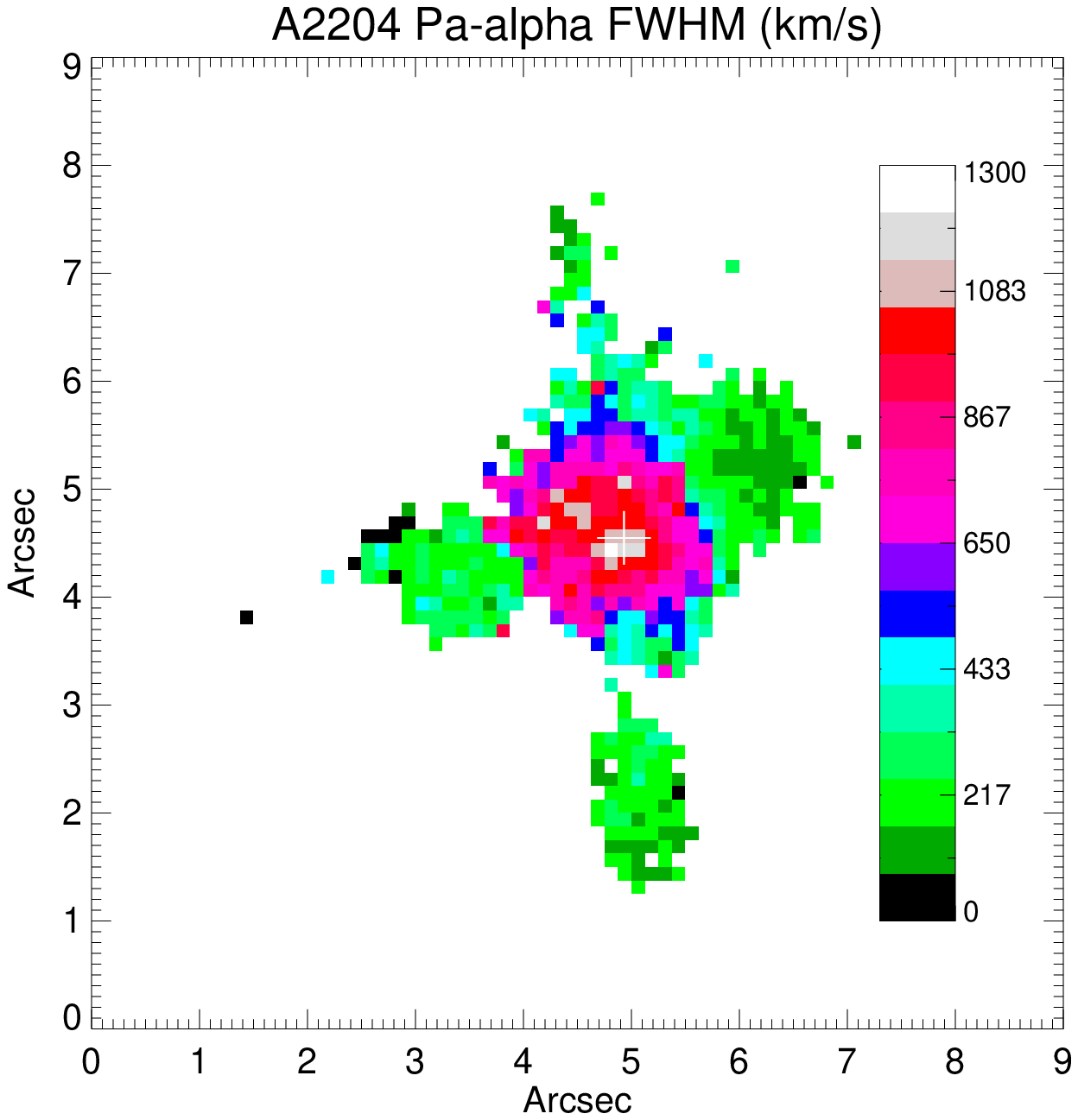}
\includegraphics[width=5.8cm,angle=0]{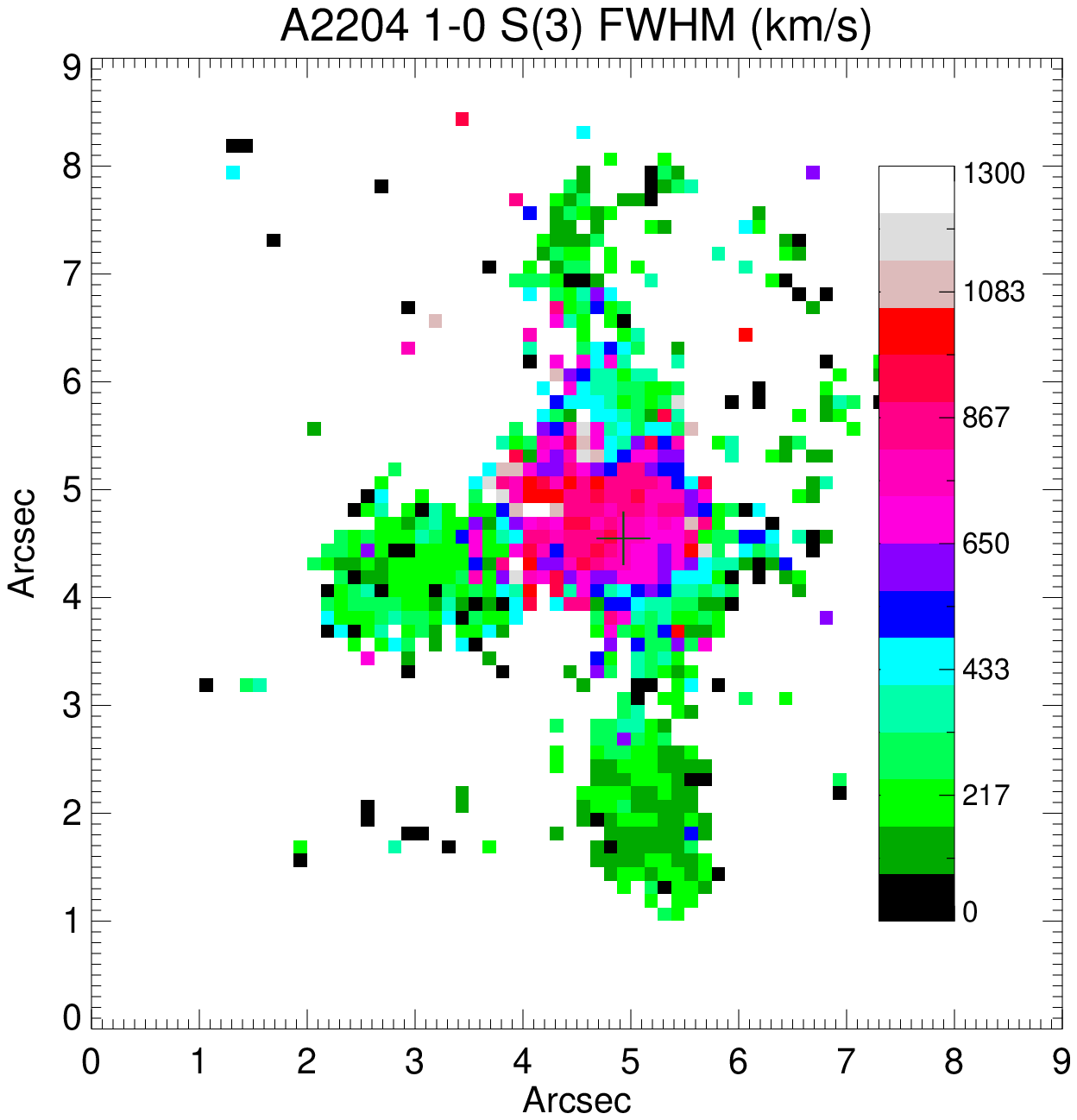}
\caption{Kinematic maps for \Pa~and \H2~v=1-0~S(3) in A2204. All velocities 
are in \kmps, with reference to a systemic redshift of $z=0.1514$. For reference, the crosshair locates the 
continuum nucleus.}
\label{fig:A2204kinemaps}
 \end{centering}
\end{figure*}

\begin{figure*}
\includegraphics[width=1.0\textwidth,angle=0]{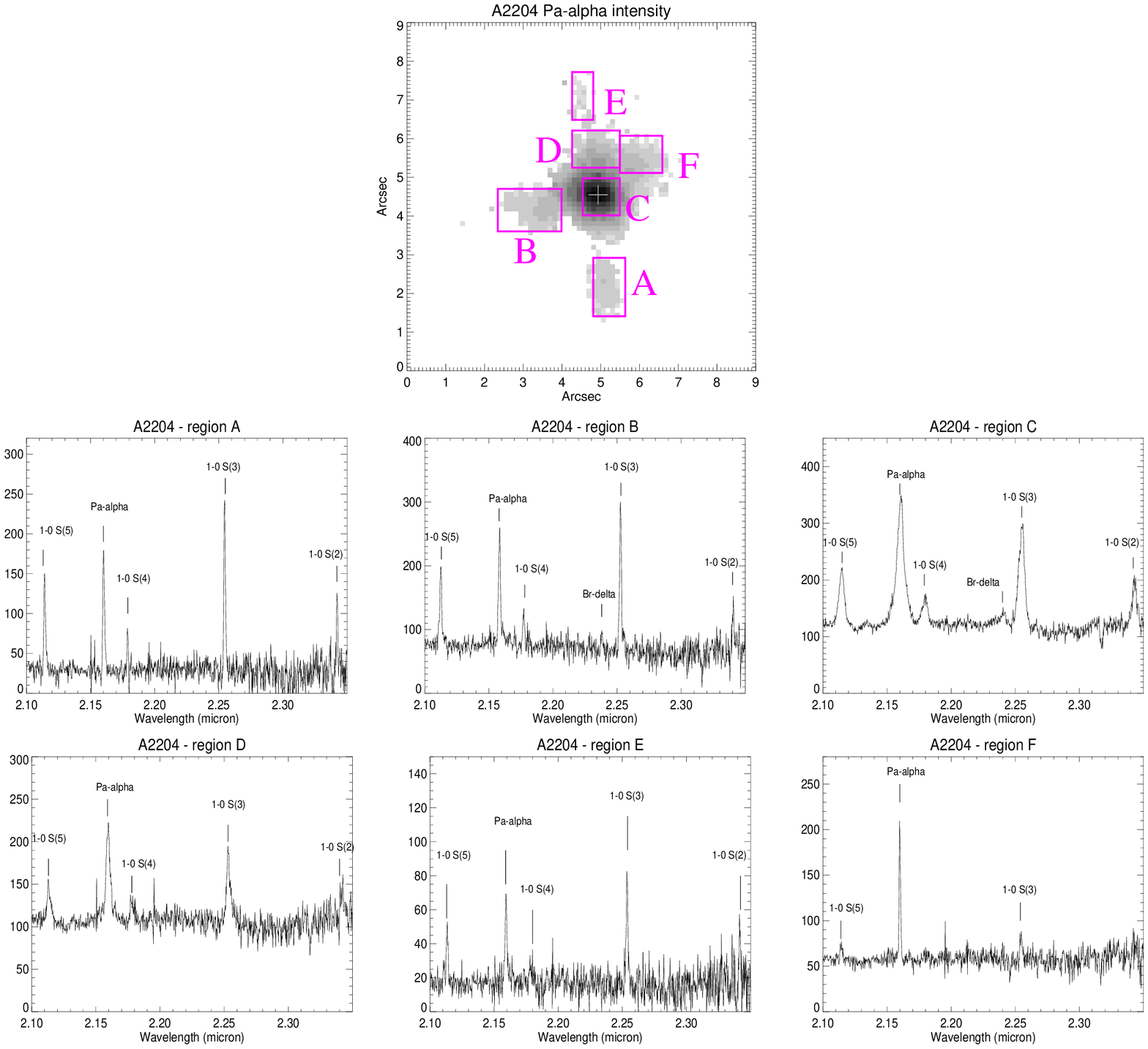}
\caption{\normalsize Spectra of selected regions in A2204. The y-axis units on all spectra are $10^{-15}$\ergpcmsqpspm.}
\label{fig:A2204regionspec}
\end{figure*}

O'Dea et al.~(2008) derived a mid-infrared estimate of the star formation rate of 14.7\Msunpyr. 
The latest X-ray and radio observations are consistent with a cooling
rate of 65\Msunpyr~down to the lowest detectable temperature (Sanders, Fabian \& Taylor~2008). Within 
20\kpc~of the X-ray cluster centroid, Sanders et al. identify between 5 and 7 X-ray surface
brightness depressions, each $\sim 4$\kpc~in size. None of them is associated with the radio emission, 
so they may be `frothy ghost bubbles'. Their estimates of the enthalpy contained within the bubbles fall short
by an order of magnitude of the amount required to offset cooling in the central 100\kpc. 

The present IFS data delineate the gas filaments much more clearly, showing
that they extend roughly perpendicularly from the nuclear regions to the north, south, east, with an additional
protrusion to the west-north-west. The latter feature is absent in \H2~v=1-0~S(3) emission, but the overall 
morphologies
in \Pa~and \H2~v=1-0~S(3) are otherwise very similar. Intriguingly, three of these extended features appear to 
lie along the radial vectors towards `ghost bubbles' identified by Sanders et al. The bouyant rise of these 
bubbles may have dragged the cool line-emitting gas out of the central regions, as seen most spectacularly in 
the Perseus cluster (Fabian et al.~2003).

The kinematic maps in Fig.~\ref{fig:A2204kinemaps} reveal that the extended filaments are 
blue-shifted in velocity from the emission in the main body of the galaxy. This suggests that they may have
been pulled out of the galaxy towards the observer; the velocity gradient appears to change sign at the
tip of the northern filament, suggesting that the outward motion stalls there. Such velocity reversals 
have been observed in the \Ha~filaments in NGC 1275 by Hatch et al.~(2006) and are in qualitative agreement
with the expected flow patterns around rising buoyant bubbles. The line-width of \Pa~rises to 1000\kmps~FWHM on 
nucleus but there is no evidence for the presence of distinct broad and narrow-line components (cf. PKS~0745-191).

Spectra are shown for various regions in Fig.~\ref{fig:A2204regionspec}. As in A1664, significant variations in
\H2/\Pa~line ratio are seen: in extended `filamentary' regions A, B and E, \H2~v=1-0~S(3) is stronger than \Pa, 
whereas in region F \H2~v=1-0~S(3) is extremely weak. The spatially-integrated line luminosities are $3.0 \times 10^{41}$\ergps~(\Pa) and $2.8 \times 10^{41}$\ergps~(\H2~v=1-0~S(3)); these are 2.2 and 3.7 times larger, respectively, than those implied
by the slit fluxes quoted in Edge et al.~(2002).

\section{Results on PKS 0745-191}

\begin{figure*}
\begin{centering}
\includegraphics[width=5.8cm,angle=0]{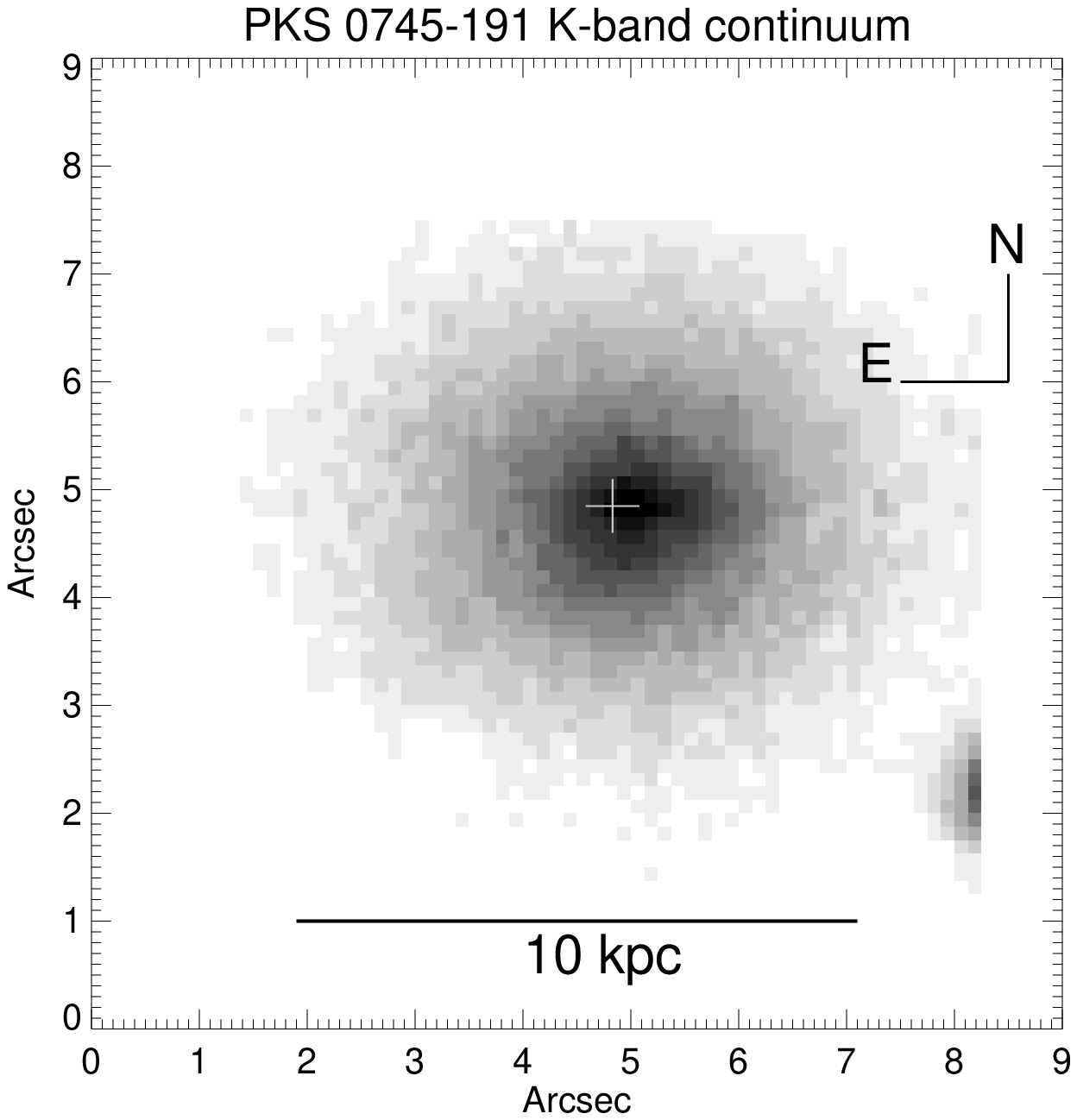}
\includegraphics[width=5.8cm,angle=0]{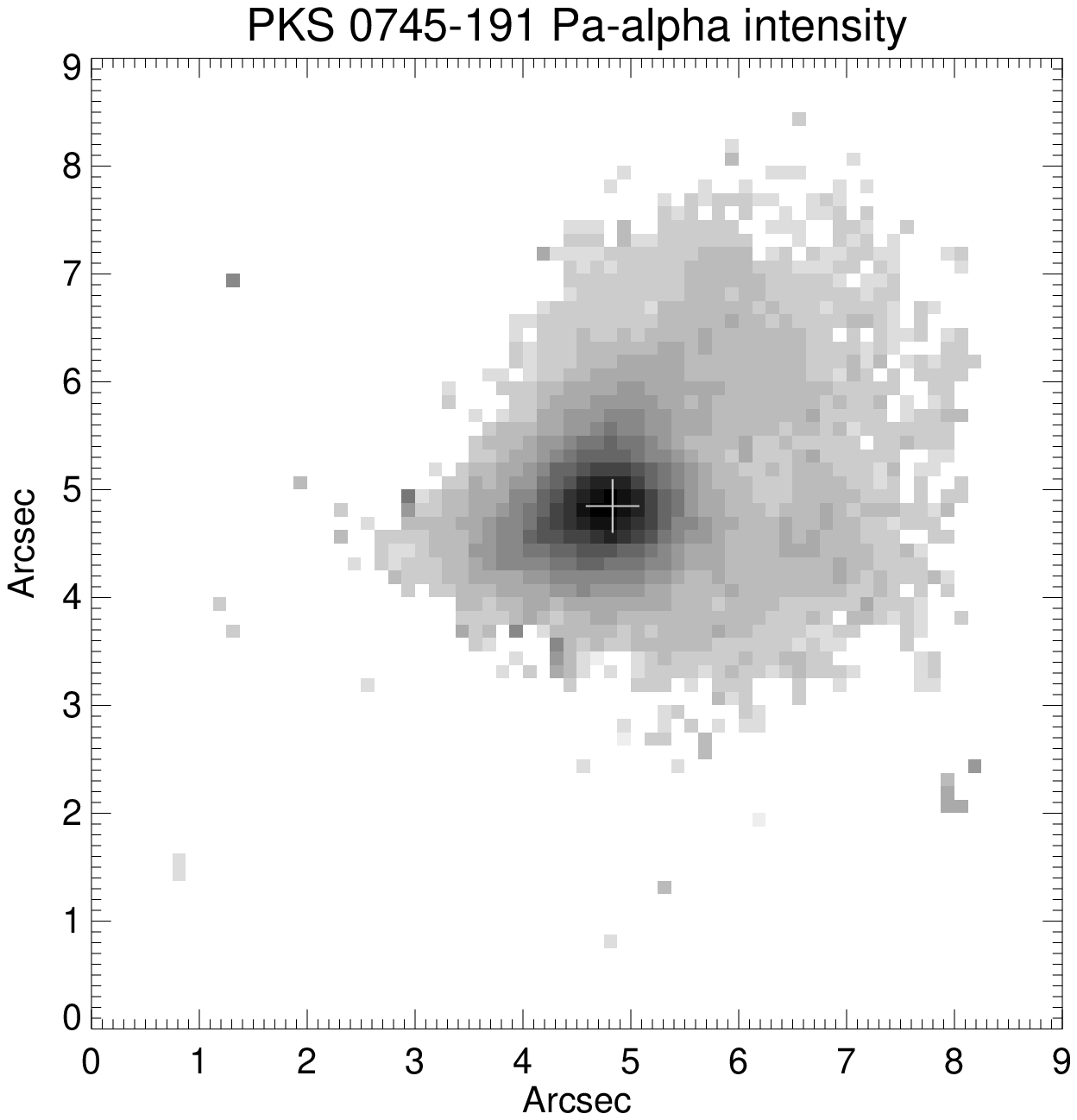}
\includegraphics[width=5.8cm,angle=0]{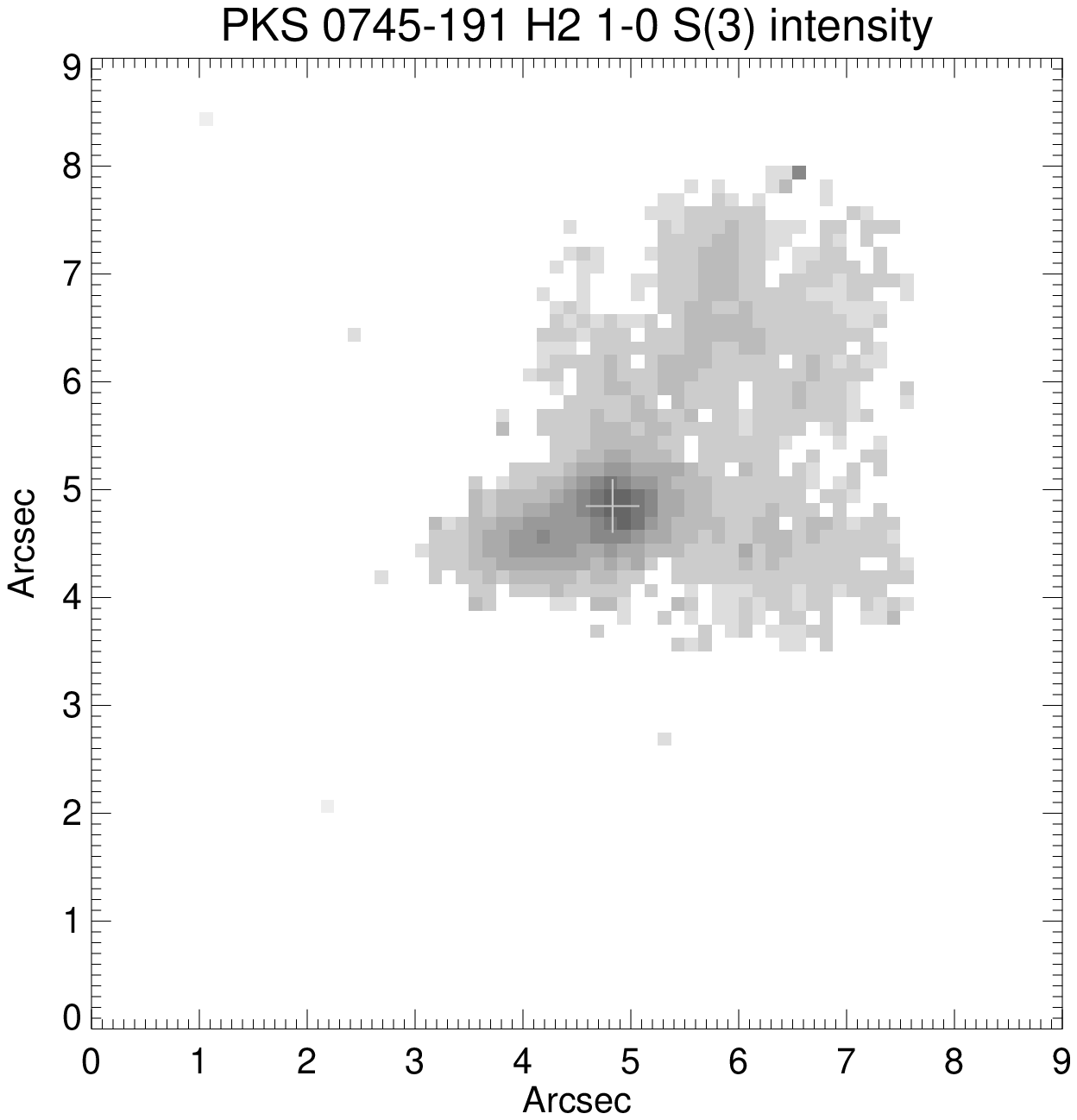}
\caption{Continuum and emission line maps for PKS0745-191. For ease of comparison, the crosshair in each panel 
denotes the location of the continuum nucleus of the galaxy, as assessed from the K-band image, and the emission line maps use a common greyscale.}
\label{fig:PKS0745totals}
 \end{centering}
\end{figure*}

\begin{figure*}
\begin{centering}
\includegraphics[width=5.8cm,angle=0]{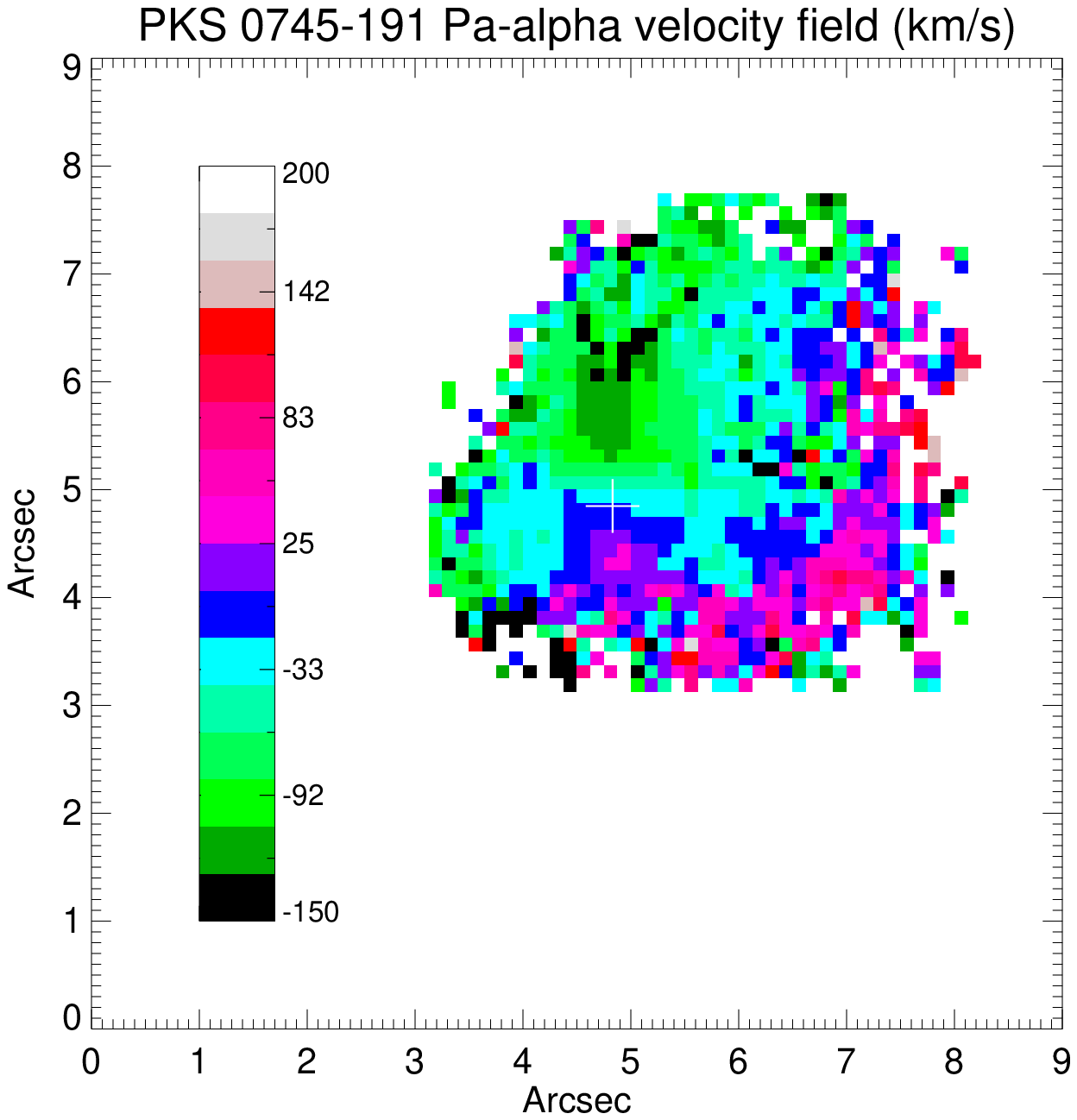}
\includegraphics[width=5.8cm,angle=0]{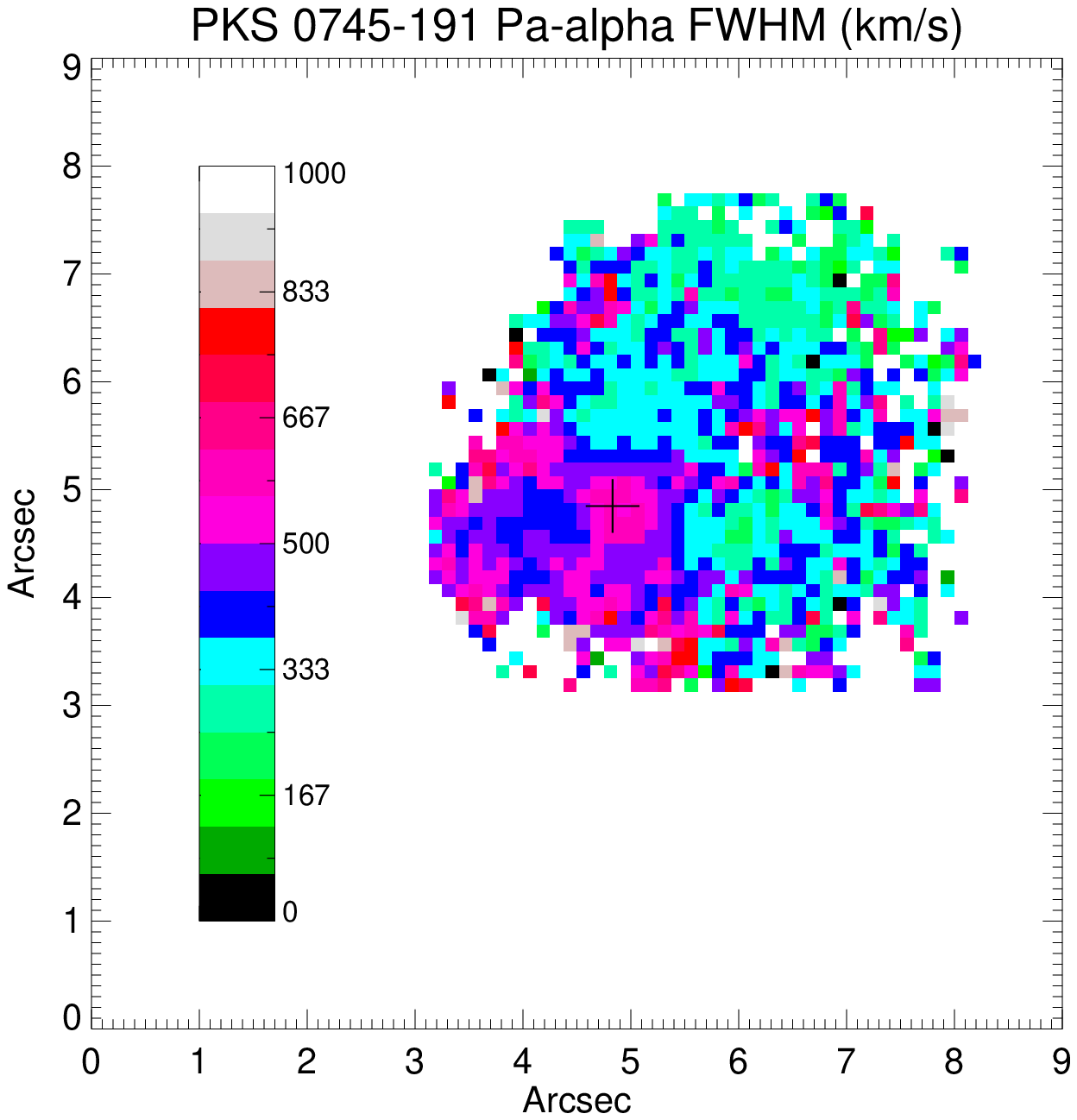}
\includegraphics[width=5.8cm,angle=0]{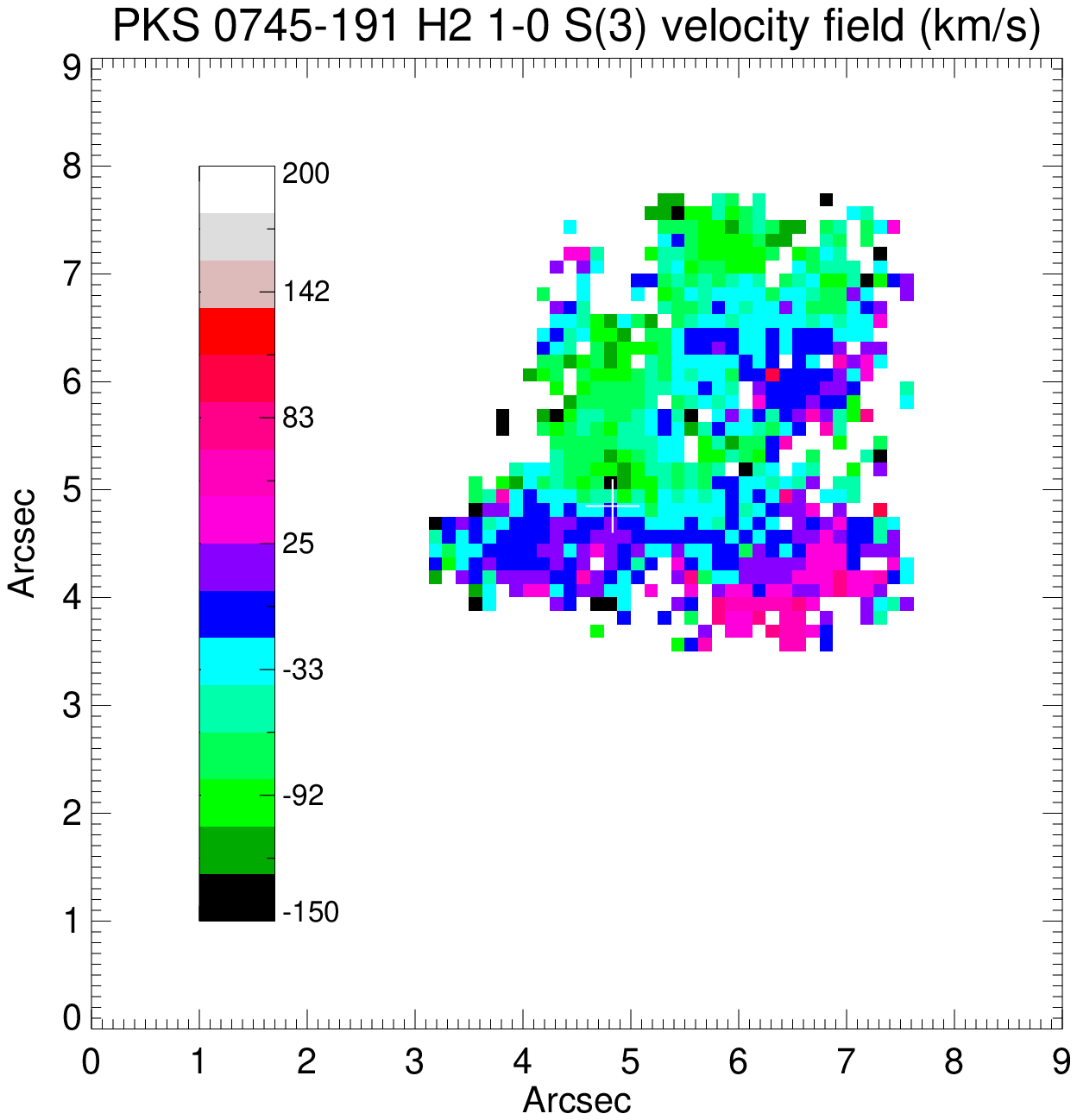}
\includegraphics[width=5.8cm,angle=0]{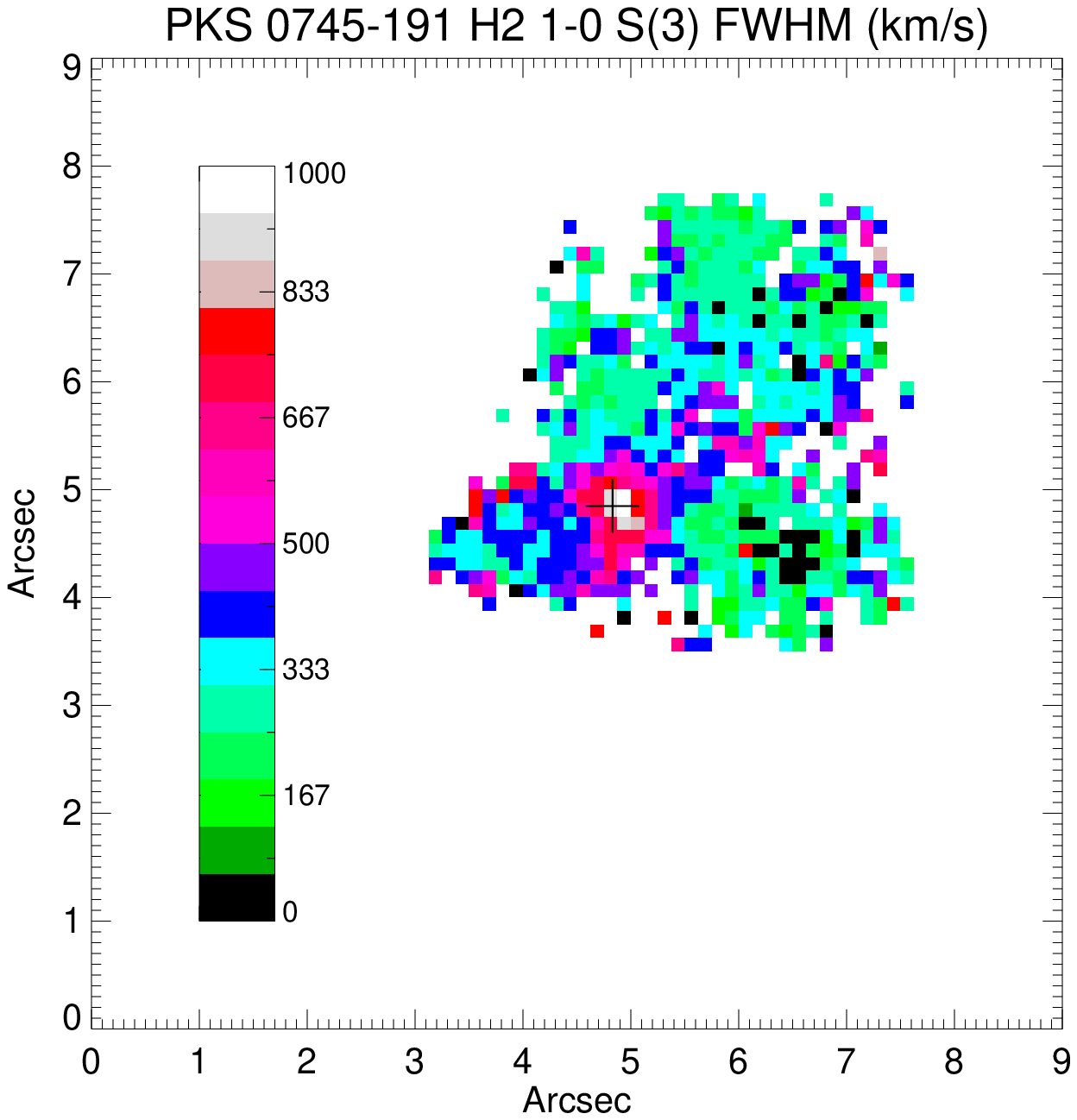}
\includegraphics[width=6.8cm,angle=0]{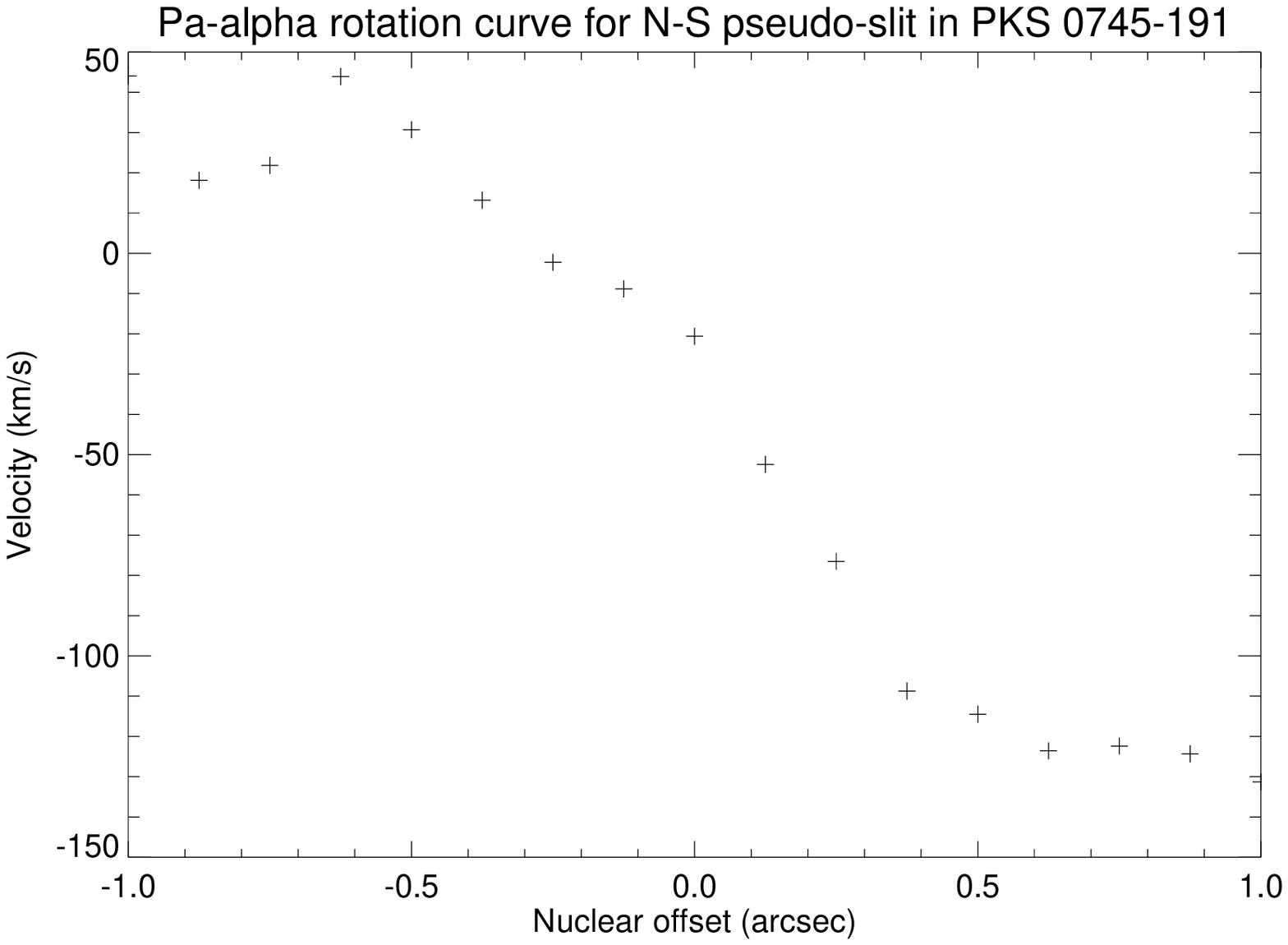}
\caption{Kinematic maps for \Pa~and \H2~v=1-0~S(3) in PKS 0745-191. All velocities 
are in \kmps, with reference to a systemic redshift of $z=0.1028$. For reference, the crosshair locates the 
continuum nucleus. Also shown is a 1-d velocity cut for a 2-pixel (0.25\arcsec) wide pseudo-slit aligned north-south
through the nucleus (negative distance is to the south).}
\label{fig:PKS0745kinemaps}
 \end{centering}
\end{figure*}

\begin{figure*}
\includegraphics[width=1.0\textwidth,angle=0]{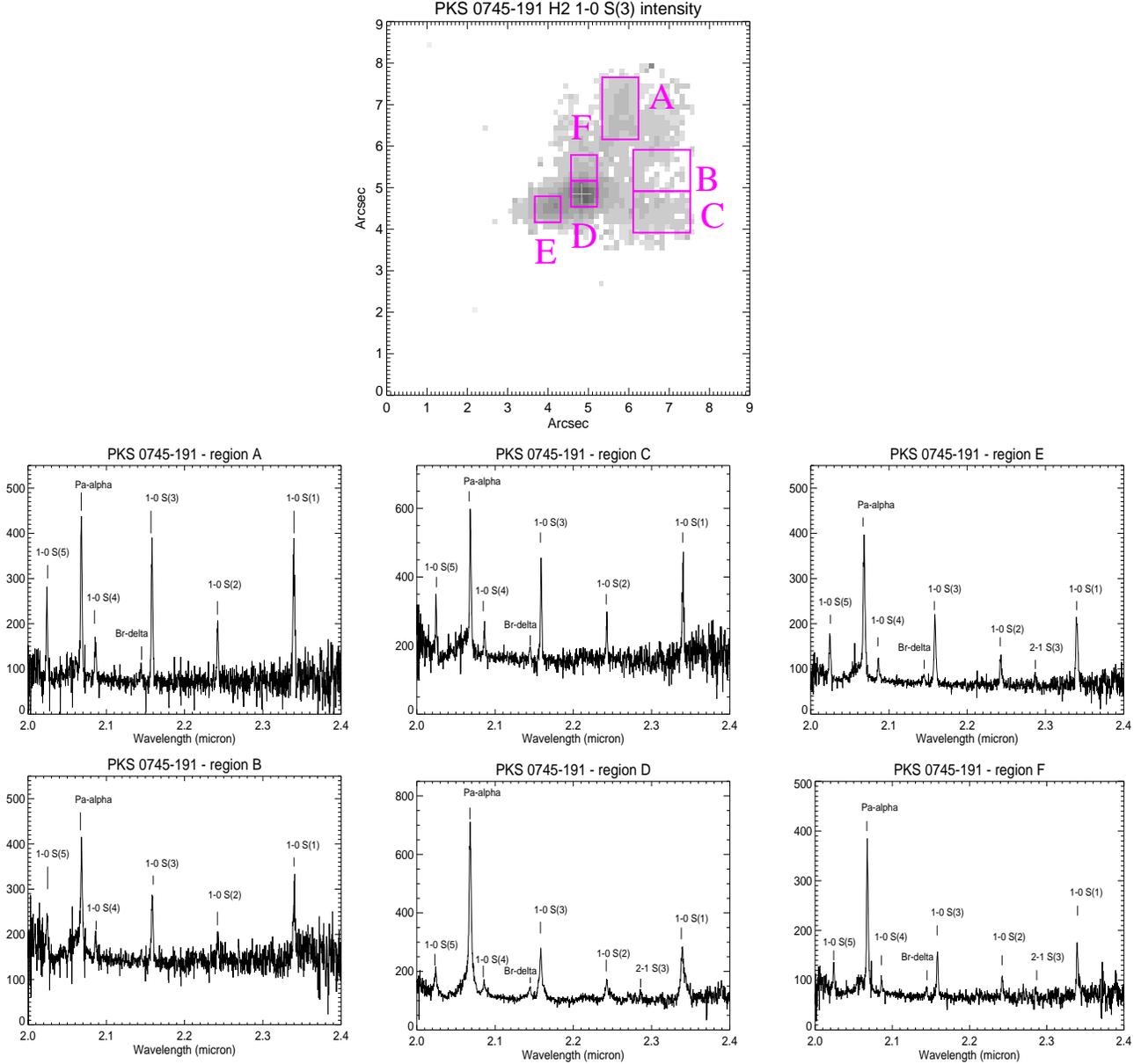}
\caption{\normalsize Spectra of selected regions in PKS 0745-191. The y-axis units on all spectra are $10^{-15}$\ergpcmsqpspm.}
\label{fig:PKS0745regionspec}
\end{figure*}

\begin{figure*}
\includegraphics[width=1.0\textwidth,angle=0]{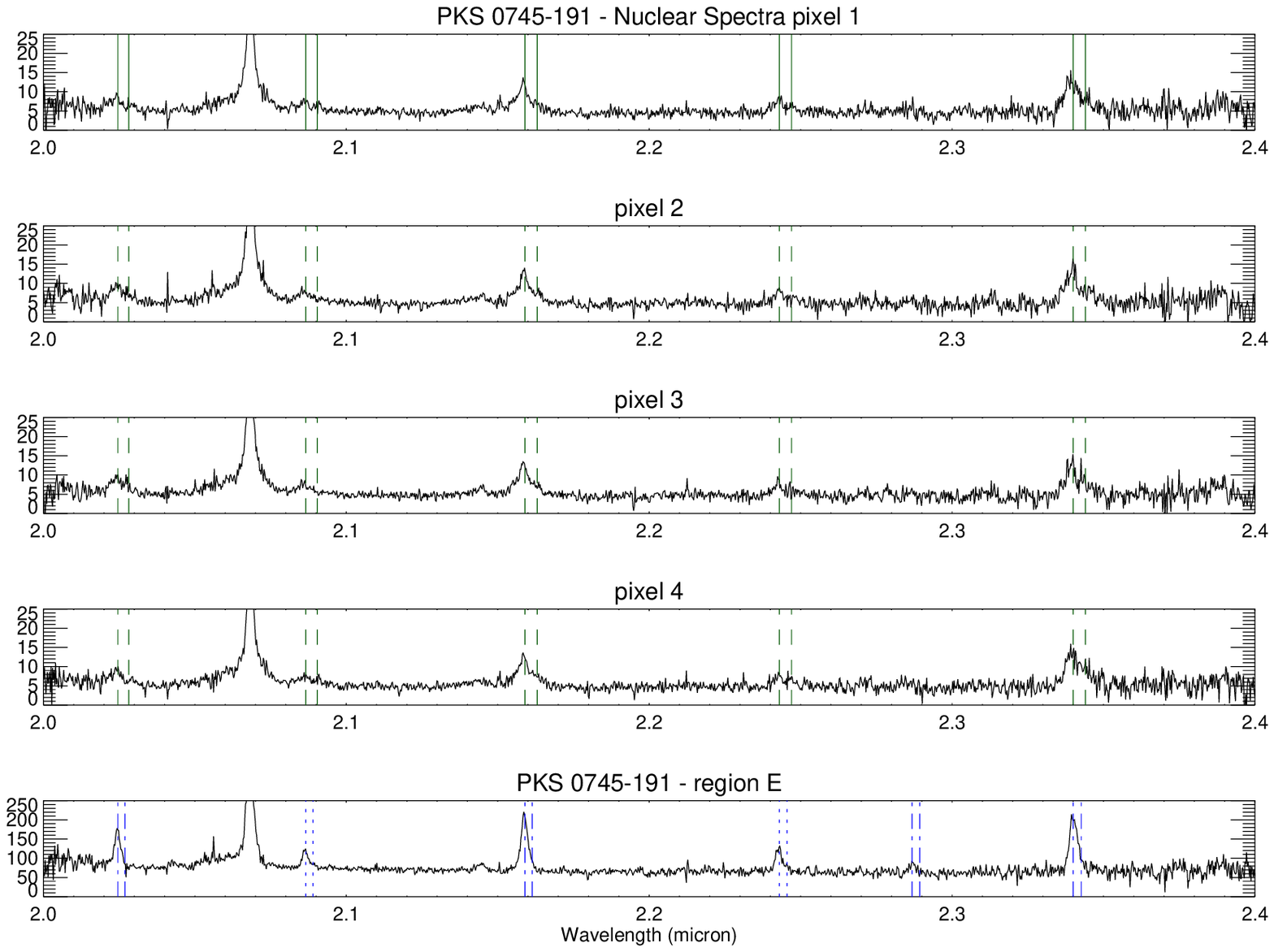}
\caption{\normalsize The top four panels show spectra of the four nuclear pixels in PKS 0745-191 (within the cross-hair in region D of Fig.~\ref{fig:PKS0745regionspec}). The pairs of solid lines
overplotted in the top panel indicate the positions of the two components of the \H2~lines in this 
spectrum; the same lines are overplotted in the next three panels at the same wavelengths. The two
kinematic components are separated by $\simeq 500$\kmps. The bottom panel
shows a blow-up of the spectrum of region E from
Fig.~\ref{fig:PKS0745regionspec}, showing tentative evidence for a red-wing in
\H2~v=1-0~S(2) and S(4) at an offset of $\simeq 330$\kmps (dotted lines). The expected
positions of the two velocity components in the \H2~v=1-0~S(1), S(3), S(5)
and v=2-1~S(3) lines are indicated by the dot-dashed lines.}
\label{fig:PKS0745_NUCH2}
\end{figure*}

Donahue et al.~(2000) presented {\em Hubble Space Telescope} imaging of the central cluster galaxy
of PKS 0745-191 in emission lines of \H2~v=1-0~S(3) and \Ha+[NII]. Our \Pa~and \H2~v=1-0~S(3) 
emission line maps shown in Fig.~\ref{fig:PKS0745totals} are in excellent correspondence with their
observations. The extended line emission is characterised by two roughly-perpendicular, $\simeq 5$\kpc-long,
 arm-like features which protrude on the western side of the nucleus, and by a short spur which extends 
$\simeq 1.5$\kpc~from the nucleus on the eastern side. As in A2204, the \Pa~emission is more strongly nucleated
than the \H2~emission. The current estimates of the star formation and X-ray cooling rates are 17.2\Msunpyr~and 
200$^{+40}_{-30}$\Msunpyr, respectively (O'Dea et al.~2008). 

The H-band continuum imaging of Donahue et al. revealed a putative 
jet-like feature extending about 1\arcsec~from the nucleus in a westerly direction, which they observed
to be coincident with a similar feature in the 2cm~radio image of Baum \& O'Dea~(1991). The feature is 
also marginally present in our K-band image in Fig.~\ref{fig:PKS0745totals}. The 6cm 
radio image of Baum \& O'Dea revealed extended emission but no clear indication of any classical 
jet or lobe features. The negative residuals (after substraction of regular isophotes) in the H-band imaging 
revealed a dust morphology consisting of several `fingers' 1--2\arcsec~in length emanating from the nucleus;
the strongest of these coincides with the spur of line emission to the east-south-east of the nucleus. 

Kinematic maps are shown in Fig.~\ref{fig:PKS0745kinemaps}. They reveal that the two arms of emission exhibit 
distinct velocities approximately 100\kmps~apart. There is evidence for a rotational motion, particularly in
the \Pa~velocity field, about an east-west axis well-aligned with the position angle of the emerging radio jet. The rotation
curve derived from a reconstructed 0.25\arcsec~wide pseudo-slit shows a sharp jump in velocity across the nucleus,
despite the modest (1\arcsec) seeing. This may well be a rotating disk seen close to edge-on, analogous to that observed
in NGC 1275 by Wilman et al.~(2005); future studies under better seeing and with adaptive optics could use it to 
constrain the mass of the black hole. The connection between the two arms of extended emission and this putative disk
is not clear: the two arms could represent cooling material falling into this disk, although
there is no clear velocity gradient along the arms (any motion could be in the plane of the sky, or drag forces may
have caused infalling clouds to reach a terminal velocity); alternatively, the arms may represent ambient material 
pushed aside by the emerging radio source. The spectra for various regions shown in Fig.~\ref{fig:PKS0745regionspec} reveal 
that the \H2/\Pa~line ratio exhibits some spatial variation, but not to the same extent as in A1664 and A2204. 

The spatially-integrated line luminosities are $4.2 \times 10^{41}$\ergps~(\Pa) and $2.0 \times 10^{41}$\ergps~(\H2~v=1-0~S(3)); these are 5.3 and 6 times larger, respectively, than those implied by the slit fluxes quoted in Edge et al.~(2002).

\subsection{Evidence for a second velocity component: a dusty jet-driven outflow?}
The velocity dispersion maps for \Pa~and \H2~v=1-0~S(3) both show peaks within 0.5\arcsec~radius of the 
nucleus. Examination of the binned spectra in Fig.~\ref{fig:PKS0745regionspec} shows that on nucleus 
(region D) the \H2~lines possess a red-shoulder which compromises the fit with a single gaussian. 
Closer inspection of the spectra of the four central 0.125\arcsec~pixels (see Fig.~\ref{fig:PKS0745_NUCH2})
 shows that this is due to a distinct \H2~emission line system offset in velocity by $\simeq 500$\kmps. 
More tentatively, a second component is also apparent in the spectrum of region
E at a relative redshift of $\simeq 330$\kmps. It is most pronounced in the
para-\H2~lines of v=1-0~S(2) and v=1-0~S(4), whilst the ortho-lines of
v=1-0~S(1), S(3) and S(5) have more symmetric profiles. Phenomenologically,
this could arise if the apparent ortho:para ratio in the secondary component
were suppressed significantly below the value of 3 which pertains for local thermodynamic
equilibrium at $T>200$\K. This could occur under UV fluorescent excitation (e.g. Sternberg
\& Neufeld~1999) but in this case we would expect the higher order v=2-1
\H2~lines produced in the radiative cascade to be stronger relative to the
v=1-0 lines. However, this test is inconclusive since the peak in v=2-1~S(3)
coincides with the lower velocity component (with v=2-1~S(3)/v=1-0~S(1)
$\simeq 0.2$) and any redshifted component lies below the noise. 
More plausibly, and consistent with the interpretation of the strong
\H2~filament emission in CCGs (see section 6), the second velocity component could be 
produced in dense, well-shielded gas at 10--100\K~in which the \H2~emission is 
excited by collisions with secondary non-thermal particles, as modelled by 
Ferland et al.~(2008,2009) under the `cosmic ray' case. In such cool gas, the 
LTE \H2~ortho:para ratio would be substantially below 3. The
v=2-1/v=1-0 line ratios would be comparable to the UV
fluorescence case, as set by the balance between radiative cascade and
collisional de-excitation rates (the latter have recently been revised
substantially, as discussed by Ferland et al.~2009). Regardless of the
excitation mechanisms, these secondary velocity components may be related to some form of 
outflow triggered by the emergence of the kiloparsec scale jet visible in the 2cm~radio map.

\begin{figure}
\includegraphics[width=0.45\textwidth,angle=0]{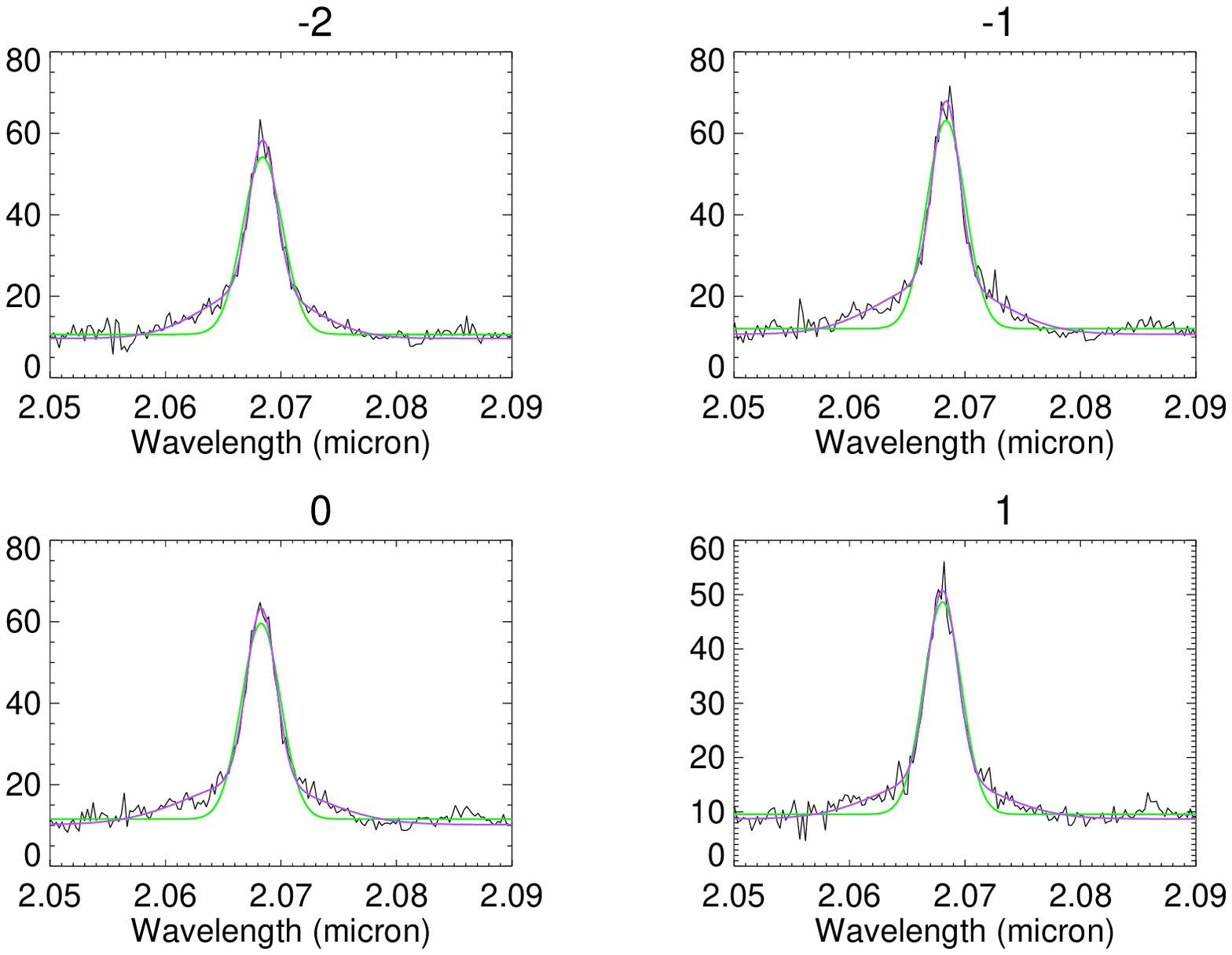}
\caption{\normalsize Profiles of the \Pa~emission line for the central four pixels along the 
north-south pseudo-slit in PKS 0745-191 (from which the rotation curve of Fig.~\ref{fig:PKS0745kinemaps} is derived). 
Fits with single and double gaussians are overlaid, showing the existence of a broad-line component. The 
individual panels are labelled with their distances from the nucleus in 0.125\arcsec~pixels along the pseudo-slit.}
\label{fig:PKS0745PAprof}
\end{figure}

A similar asymmetry is not apparent in the \Pa~profiles, but a closer inspection of the profiles of
this line in the nuclear regions reveals the presence of an underlying broad component centred at the same 
redshift as the narrow component (Fig.~\ref{fig:PKS0745PAprof}). The broad component has a FWHM of 1700\kmps. 
Such a line width falls in the middle of the distribution derived by Landt et al.~(2008) from their near-infrared 
spectral survey of a broad-line AGN sample. The inclusion of the additional broad component clearly improves the
overall fit to the emission line profile, but does not significantly affect the FWHM and velocity fields derived
from the single component fit in Fig.~\ref{fig:PKS0745kinemaps}.

\section{EXCITATION OF THE GAS}
Table~2 lists the \H2~v=1-0~S(3)/\Pa~emission line flux ratio for the labelled spatial regions in each 
galaxy. This particular line ratio was chosen because both lines are within the observed wavelength
range for all three targets and are sufficiently close together to minimise uncertainties arising
from differential flux-calibration errors and extinction.

As already hinted by cruder long-slit spectroscopy (e.g. Edge et al.~2002), the nuclear regions correspond
to local minima in the \H2/\Pa~ratio, most likely due to enhanced \Pa~emission from a photoionized AGN component. 
The regions with the highest \H2/\Pa~ratio are generally isolated features morphologically distinct from the 
contigious body of emission within the galaxy, viz. the extended filaments in A2204 (regions A, B and E);
the extended `arms' in PKS 0745-191; the extremity of the infalling gas stream in A1664 (region E). The 
\H2~v=1-0~S(3)/\Pa~ratios in these regions are extremely high, typically within 30~per cent of unity. The 
latest modelling by Ferland et al.~(2008,2009) accounts for these high values (and other line ratios which 
are anomalous with respect to photoionization predictions) by invoking extra heating from magnetohydrodynamical
waves and/or cosmic-ray-induced non-thermal excitation of the molecular lines. The models invoke emission 
from gas at a range of densities pinned to magnetic fields at a common pressure ($P/k=nT$) matching that of
the diffuse X-ray plasma. The emergent spectrum depends on the assumed power-law density distribution and can be
arbitrarily tuned to match any one line ratio, but for the `Horseshoe filament' in NGC 1275 the best-fitting 
Ferland et al.~(2009) models for the `extra heat' and `cosmic ray' yields \H2~v=1-0~S(3)/\Pa~values of 1.37
and 0.73, respectively. These values bracket the measured ratios in the most extended regions of our galaxies, 
suggesting that the same physical processes are at work.

There are, however, several spatial regions with extremely low \H2~v=1-0~S(3)/\Pa~ratios in the range 0.15--0.3, 
namely regions A, B, C (red velocity component), D (blue) in A1664, region F in A2204, and region F in PKS 0745-191. 
Star formation is most likely the dominant excitation mechanism in these regions. Indeed, region A in A1664 coincides clearly with a knot of star formation emission in the HST image of Fig.~\ref{fig:A1664hst}. To further illustrate the role of star formation, we have made a comparison with the near-infrared spectroscopic survey of a sample of ultraluminous infrared galaxies (ULIRGs) by Dannerbauer et al.~(2005). This unbiased sample of 24 ULIRGS was drawn from those observed with the {\em Infrared Space Observatory} (ISO), and the vast majority 
have mid-infrared and optical spectral classifications of starburst or HII region. Using the published
emission line fluxes in Table 2 of Dannerbauer et al., we compute average \H2~v=1-0~S(3)/\Pa~ratios of
0.19 (median), 0.3 (mean) with a standard deviation 0.46 for the 22 spectral regions with detectable 
\H2~v=1-0~S(3) and \Pa~(some of the ULIRGs are split into independent spatial regions). A further 8 of their spectra
exhibit \Pa~but no detectable \H2~v=1-0~S(3) and in some other cases the latter line may blend with [SiVI]1.957; the 
average \H2~v=1-0~S(3)/\Pa~ratios quoted above are thus likely to over-estimate the true value, but are nevertheless 
comparable with the lowest values measured in our CCGs. 

\begin{figure*}
\includegraphics[width=5.8cm,angle=0]{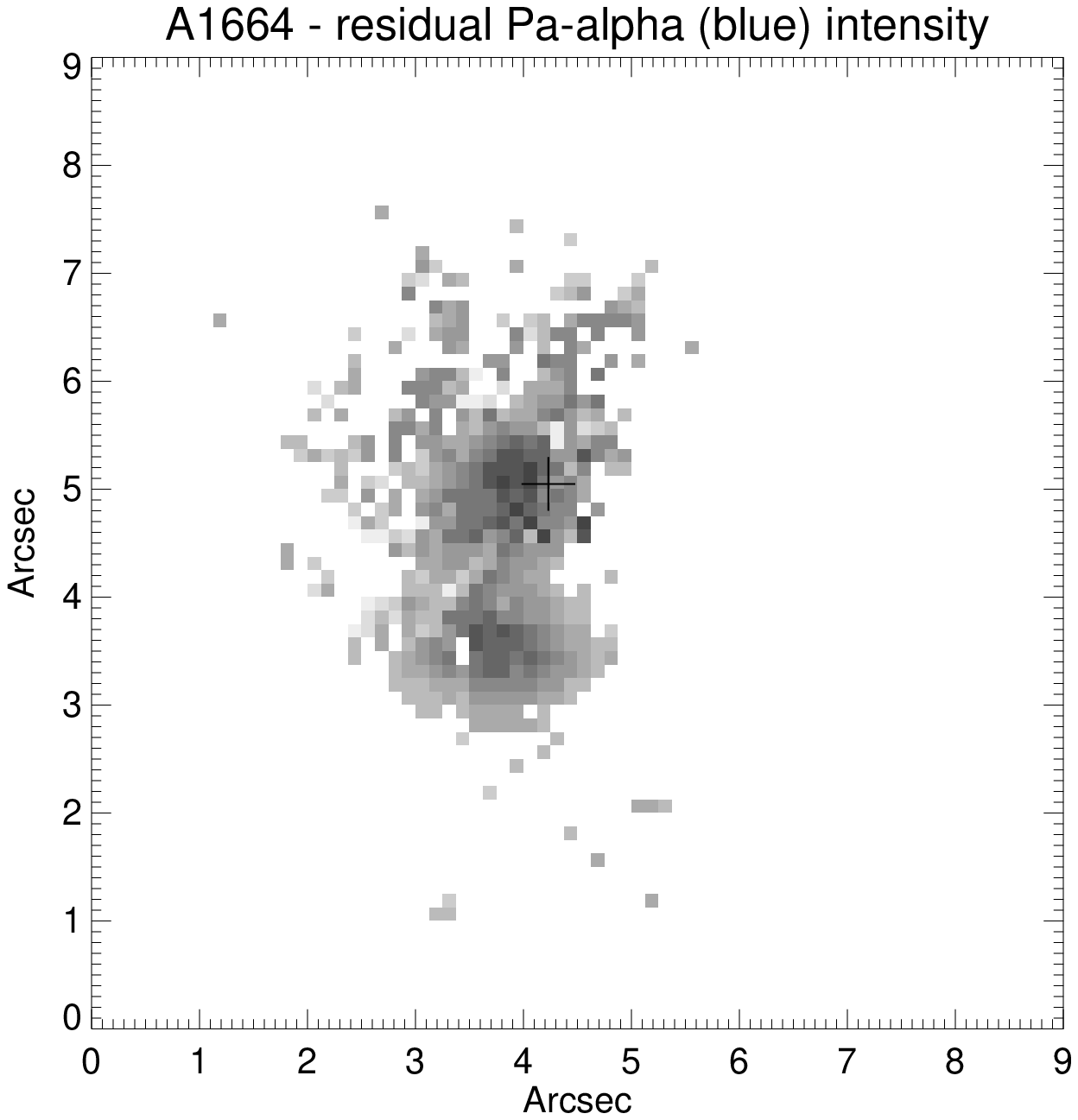}
\includegraphics[width=5.8cm,angle=0]{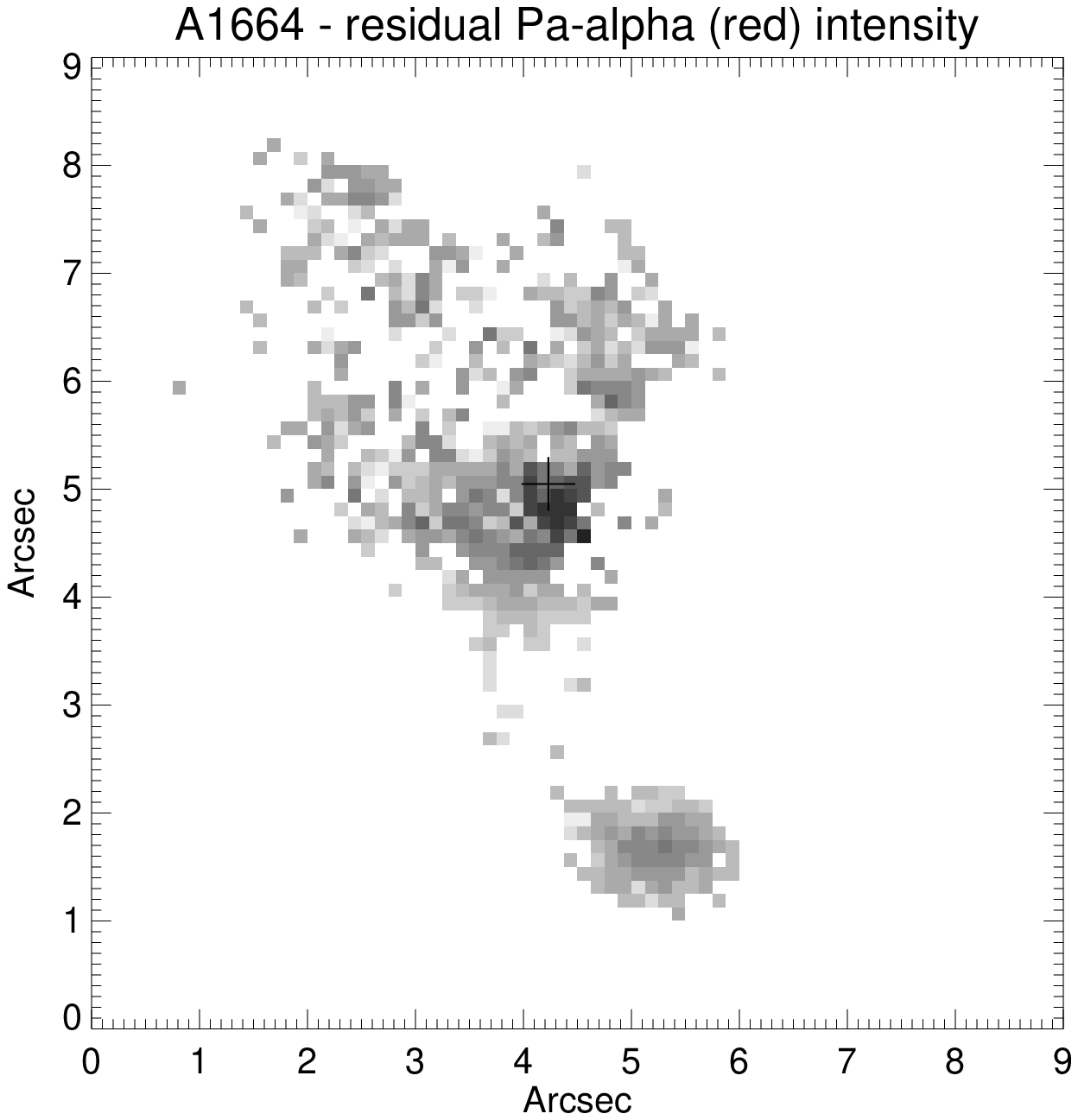}
\includegraphics[width=5.8cm,angle=0]{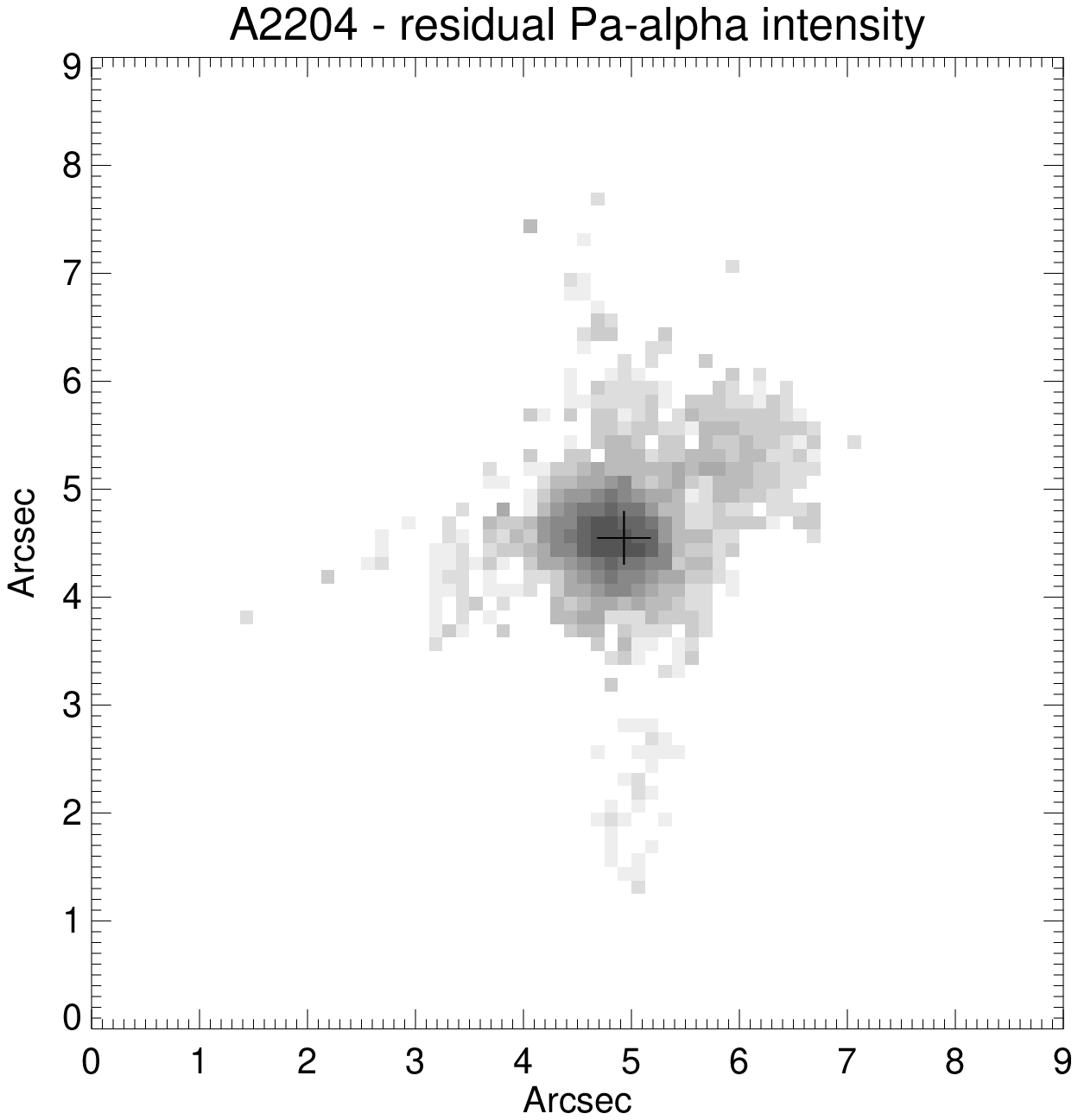}
\includegraphics[width=5.8cm,angle=0]{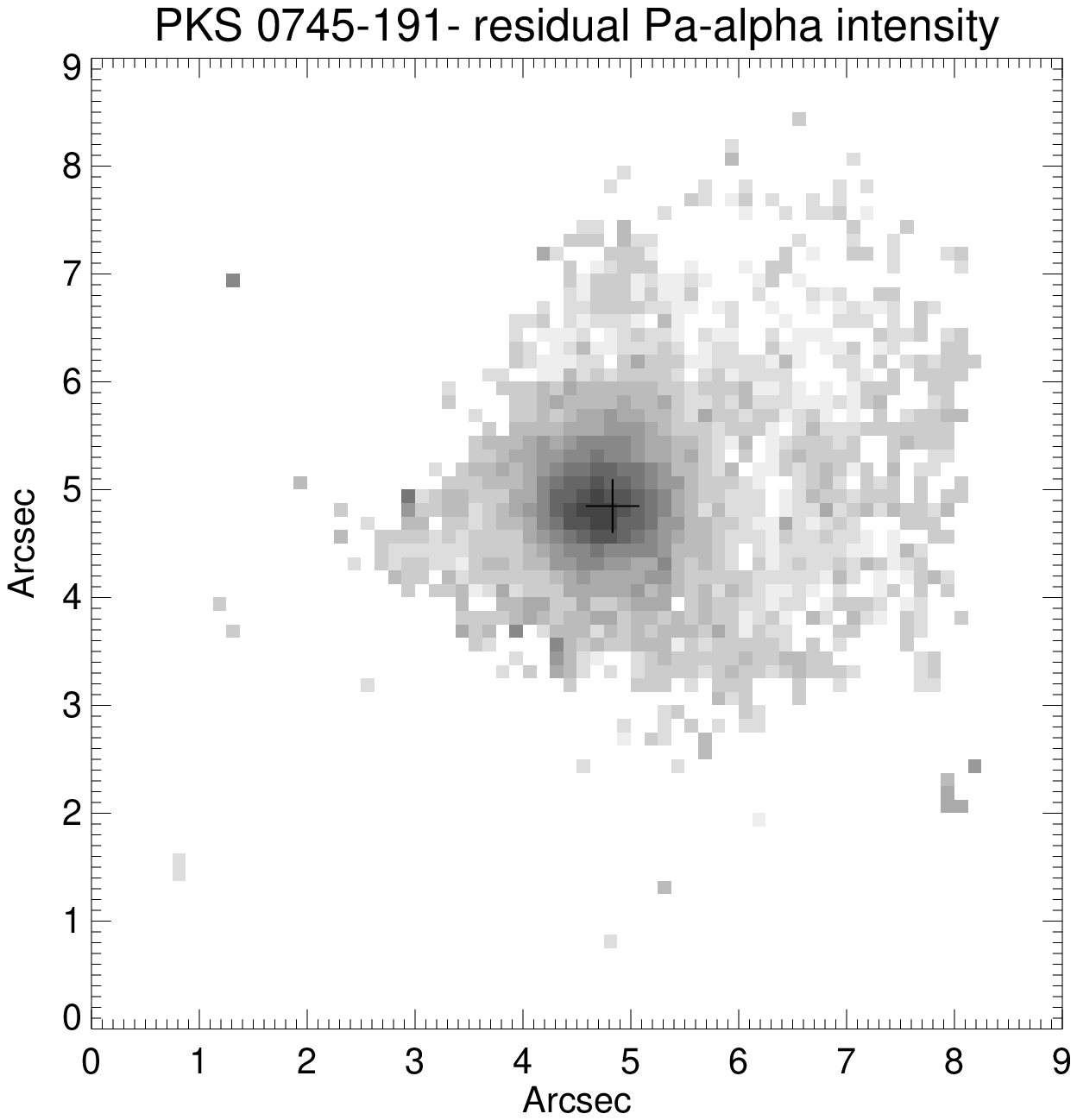}
\caption{\normalsize Residual \Pa~images formed by subtracting a scaled version of the \H2~v=1-0~S(3) image 
from that of \Pa. The residual image should emphasise the regions where the \Pa~emission is due to star formation, 
as described in section 6.}
\label{fig:SF}
\end{figure*}

Regions with \H2~v=1-0~S(3)/\Pa~intermediate between the above extremes are likely to reflect a blend 
of star formation and non-photoionization `filament-excitation'. A crude insight into the physical location 
of the star formation within the CCG can be obtained by substracting $1/f_{\rm{FIL}}$ times the \H2~v=1-0~S(3) image from the \Pa~image, 
such that the residual \Pa~emission mainly reflects the contribution from star formation. 
The value of $f_{\rm{FIL}}$ is set by the maximum observed value of \H2~v=1-0~S(3)/\Pa~for each galaxy, which we take
to be characteristic of the `filament-excitation', i.e. $f_{\rm{FIL}}=1.3$ for A2204, $0.75$ for A1664 and $0.86$ for PKS 0745-191. The residual \Pa~intensity maps are shown in Fig.~\ref{fig:SF}. For A2204, the star formation is confined to the main body of the galaxy with
an additional clump 4\kpc~to the north-west; for A1664, there appear to be five main clumps of star-formation: two circumnuclear 
components $\simeq 1$\kpc~either side of nucleus (one in the red velocity component, one in the blue), a 
second component $\simeq 3.7$\kpc~south of the nucleus in the blue velocity component, region A (see Fig.~\ref{fig:A1664boxspec}), and an
area $\simeq 3.5$\kpc~to the north north-west of the nucleus. These features mostly correspond with the star-forming regions visible in the 
HST image (Fig.~\ref{fig:A1664hst}), except for the two circumnuclear regions which are obscured by the dust lane. The 
infalling gas stream (along axis 1 in Fig.~\ref{fig:A1664kinemaps}) is relatively devoid of star formation. With reference to the 
kinematic maps, most of these regions correspond to gas with the lowest velocity dispersion ($FWHM < 200$\kmps). 
For PKS 0745-191, the residual image emphasises the main body of the galaxy and the area between the `arms' of 
\H2~emission. These residuals are not a pure star formation tracer, since some of the nuclear 
emission with depressed \H2~v=1-0~S(3)/\Pa~may arise from AGN-related photoionization. When the previously-quoted 
total \Pa~luminosities of the three CCGs are converted to \Ha~luminosities assuming case B recombination (\Ha/\Pa$=8.45$), and 
then to a star formation rate (SFR) using the \Ha-SFR relation of Kennicutt~(1998) [SFR(\Msunpyr) = L(\Ha)/$1.26 \times 
10^{41}$\ergps], we obtain the following SFRs: 36\Msunpyr~(A1664), 20\Msunpyr~(A2204) and 28\Msunpyr~(PKS 0745-191). These values 
exceed the infrared measurements of O'Dea et al.~(2008) of 14.6, 14.7 and 17.2\Msunpyr, respectively. If we correct for the 
contribution of filament excitation to the \Pa~emission (assuming \H2~v=1-0~S(3)/\Pa = 0.15 for star formation and the above values of $f_{\rm{FIL}}$ for each CCG), the implied SFRs become: 11 (A1664), 6 (A2204) and 15\Msunpyr (PKS 0745-191).  A correction for dust extinction of 
$A_{\rm{V}} = 2-6$~mag towards the \Pa-emitting star-forming regions (the range estimated from a local sample of luminous infrared galaxies 
by Alonso-Herrero et al.~2006) would boost the SFRs by 30--100~per cent, and the infrared and \Pa~estimates would then be more comparable.

These substantial variations in \H2~v=1-0~S(3)/\Pa~contrast with the constancy 
of the ratio of optical forbidden line emission to \Ha~in A1664 and A2204, as reported by Wilman et al.~(2006). 
In the above scenario where the observed line emission reflects a spatially-varying blend of photoionization due to 
star formation (and possibly an AGN) and non-photoionization `filament' excitation, this implies that the optical line 
ratios for the two types of emission are very similar, whilst the infrared \H2~v=1-0~S(3)/\Pa~ratios are manifestly quite
different. In both cases, the optical forbidden lines and the bulk of the HI lines are produced in gas in a relatively narrow
range of parameter space with $T \sim 10^{4}$\K~and density $n \sim 10^{2.5}$\pcm. In the `cosmic ray' case of filament excitation 
favoured by Ferland et al.~(2009), the volume emissivities of the HI and \H2~lines extend deep into the much cooler, denser and 
neutral parts of the clouds, with a secondary peak in \Ha~emissivity at 100--1000\K~due to collisional excitation by fast electrons 
and a secondary peak in ro-vibrational \H2~emission at 10--100\K. However, the volume occupied by this dense gas in their model is much lower 
and it contributes negligibly to the observed optical line emission, which is dominated by the partially ionized gas.
Moreover, Ferland et al. show that the internal extinction across a 0.3\pc~thread (several hundred of which compose 
each 70\pc-wide filament in NGC 1275) is $A_{\rm{V}} \sim 17$~mag, arising mostly within the dense molecular regions. 
This implies that the observed optical emission is dominated by the low-extinction partially ionized gas. In contrast, the K-band extinction is 
$A_{\rm{K}} \sim 0.1A_{\rm{V}}$, enabling the \Pa~and \H2~lines to probe much deeper; the broad emissivity distributions 
of these lines renders the observed \H2~v=1-0~S(3)/\Pa~ratio sensitive to the details of the gas density distribution, 
leading to spatial variations in \H2~v=1-0~S(3)/\Pa~over and above those caused by a varying mix of `filament' and star formation 
(photoionization) emission.

\begin{table*}
\caption{Spatially-resolved emission line ratios within the CCGs ($\dagger$ indicates the nuclear region).}
\begin{tabular}{|lll|}\hline
Galaxy           & Region & \H2~v=1-0~S(3)/\Pa \\ \hline
A1664            & A & 0.28 $\pm$ 0.025 \\ 
                 & B (blue) & 0.16 $\pm$ 0.017 \\ 
                 & B (red) & 0.19 $\pm$ 0.08 \\ 
                 & C (blue) & 0.75 $\pm$ 0.042\\ 
                 & C (red) & 0.17 $\pm$ 0.086 \\ 
                 & D $\dagger$ (blue) & 0.23 $\pm$ 0.026\\ 
                 & D $\dagger$ (red) & 0.41 $\pm$ 0.020 \\ 
                 & E       & 0.72 $\pm$ 0.057 \\ 
A2204            & A & 1.29 $\pm$ 0.09 \\
                 & B & 1.23 $\pm$ 0.075\\
                 & C $\dagger$ & 0.69 $\pm$ 0.018 \\
                 & D & 0.81 $\pm$ 0.056 \\
                 & E & 1.25 $\pm$ 0.14 \\
                 & F & 0.32 $\pm$ 0.054 \\
PKS 0745-191     & A & 0.86 $\pm$ 0.037 \\
                 & B & 0.57 $\pm$ 0.039 \\
                 & C & 0.62 $\pm$ 0.028 \\
                 & D $\dagger$ & 0.36 $\pm$ 0.01 \\
                 & E & 0.47 $\pm$ 0.015 \\ 
                 & F & 0.31 $\pm$ 0.017 \\ \hline
\end{tabular} \\
\end{table*}

\section{DISCUSSION}
Recent analyses by Rafferty et al.~(2008) and Cavagnolo et al.~(2008) have demonstrated the existence of a threshold
for the onset of star formation and radio-loud AGN activity in CCGs when the central X-ray 
cooling time drops below $5 \times 10^{8}$\yr~(equivalent to a central entropy threshold of 30\keV~\cmsq). This 
underscores the role of AGN heating as the ultimate regulator of X-ray cooling in these environments. The existence of a
common (sharp) threshold in cooling time for star formation and AGN activity suggests that both are a direct consequence 
of X-ray cooling. In the light of this, we discuss the current {\em VLT-SINFONI} observations and the prospects for 
IFS studies of larger samples of CCGs.

In A1664, the evidence for the gravitational free-fall of cooled \Pa~and \H2-emitting gas from radii 
$\sim 10$\kpc~suggests that the system is observed at an early stage in this particular cooling/heating 
feedback cycle. Freshly cooled gas rains down on the CCG and gives rise to extensive star formation within 
the molecular cloud lifetime of $\sim 10^{7}$\yr~(Mouschovias et al.~2006), which is comparable to the free-fall timescale 
from 10\kpc. The AGN has either not been fuelled yet or its activity is still insufficient to arrest the X-ray cooling. 
Some of the infalling gas will eventually fuel the AGN directly if its angular momentum is low, otherwise accretion may 
proceed via a disk (e.g. as observed in NGC 1275; Wilman et al.~2005). The central radio source is indeed 
weak with the unresolved NVSS 1.4~GHz detection of 36mJy implying a monochromatic power of 
$\nu L_{\rm{\nu}} \simeq 2 \times 10^{40}$\ergps, placing it at the lower end of the radio power distribution for
CCGs below the cooling time (entropy) threshold in Cavagnolo et al. These observations of free-falling cool gas and 
a weak AGN are consistent with the `cold feedback' model of Pizzolato \& Soker~(2005). In this model, blobs of gas 
overdense by at least a critical factor $\delta_{\rm{c}} \simeq 2$ relative to the ambient medium initially cool slowly and flow 
steadily inwards over a period of $\sim 10^{8}$\yr~since the last AGN heating event. Over the next $\sim 10^{7}$\yr~they 
enter a free-fall plunge into the CCG over the final $\sim 10$\kpc~of radius as they cool catastrophically, contract in size 
and drag forces become negligible. Some of this cool gas may form stars and some may fuel the AGN, leading to the next major 
heating event and the repetition of the cycle. A1664 may well be captured in this final phase with an another AGN heating
event due within $\sim 10^{7}$\yr. A {\em Chandra} study of A1664 by Kirkpatrick et al.~(2009) confirms that the system is in a
low-state of AGN activity insufficient to offset the X-ray cooling, which occurs at a rate comparable to the star formation rate. 
The alignment between the elongated X-ray bar (possibly a collapsed cavity system; Kirkpatrick et al.) and the \Ha~filament 
further strengthens the hypothesis that the star formation is a direct result of the X-ray cooling and not triggered by interaction 
with a secondary galaxy as suggested by Wilman et al.~(2006).

Not surprisingly, neither of the other two CCGs in our sample is in this short-lived pre-outburst state with cool gas 
free-falling onto the CCG. With its kiloparsec-scale radio jet and higher power ($\nu L_{\rm{\nu}} \simeq 9 \times 10^{41}$\ergps~from the NVSS 
1.4~GHz detection), PKS 0745-191 is arguably observed shortly after an AGN outburst with the central black hole still being
fed from the gas disk observed in the extended \Pa~emission; on nucleus, the broad-line \Pa~component and secondary \H2~velocity
system point to active gas accretion and a possible outflow. A2204, with its `ghost cavities' and $\sim 10$\kpc-scale radio 
source (Sanders et al.~2008), is probably observed at a somewhat later post-outburst epoch. As argued by Rafferty et al., star formation within the cool gas reservoir can persist for more than 
$10^{8}$\yr~and exhibit detectable signatures over several feedback cycles and will thus not appear synchronized with the 
latter. Future IFU observations of a larger sample could identify more of these `pre-outburst' systems, by for example
targeting the subset of clusters in the Cavagnolo et al.~sample which have central cooling times below the critical threshold but 
weak (i.e. currently undetected) radio emission. Studies of the multiphase cool gas in such systems could shed light on the 
black hole fuelling mechanisms and further constrain the `cold feedback' model.

\section*{ACKNOWLEDGMENTS}
RJW thanks the University of Melbourne for an honorary fellowship. We thank Chris O'Dea for permission to use their proprietory
HST image of A1664, Brian McNamara for communicating a pre-print of their
{\em Chandra} study of A1664 (Kirkpatrick et al.), and the referee for a
prompt and valuable report.

{}


\begin{thebibliography}{}
\bibitem[]{} Allen S.W., Fabian  A.C., Edge A.C., B\"{o}hringer H., White D.A., 1995, MNRAS, 275, 741
\bibitem[]{} Alonso-Herrero A., Rieke G.H., Rieke M.J., Colina L., Perez-Gonzalez P.G., Ryder S.D., 2006, ApJ, 650, 835
\bibitem[]{} Baum S.A., O'Dea C.P., 1991, MNRAS, 250, 737
\bibitem[]{} Dannerbauer H., Rigopoulou D., Lutz D., Genzel R., Sturm E., Moorwood A.F.M., 2005, A\&A, 441, 999
\bibitem[]{} Edge A.C., 2001, MNRAS, 328, 762
\bibitem[]{} Edge A.C., Wilman R.J., Johnstone R.M., Crawford C.S., Fabian A.C., Allen S.W., 2002, MNRAS, 337, 49
\bibitem[]{} Edge A.C., Frayer D.T., 2003, ApJ, 594, L13
\bibitem[]{} Cavagnolo K.W., Donahue M., Voit G.M., Sun M.,2008, ApJ, 683, L107
\bibitem[]{} Crawford C.S., Allen S.W., Ebeling H., Edge A.C., Fabian A.C., 1999, MNRAS, 306, 857
\bibitem[]{} Crawford C.S., Sanders J.S., Fabian A.C., 2005, MNRAS, 361, 17
\bibitem[]{} Donahue M., Mack J., Voit G.M., Sparks W., Elston R., Maloney P.R., 2000, ApJ, 545, 670
\bibitem[]{} Egami E., Rieke G.H., Fadda D., Hines D.C., 2006a, ApJ, 652, L21
\bibitem[]{} Egami E., et al., 2006b, ApJ, 647, 992
\bibitem[]{} Eisenhauer F., et al., 2003, SPIE, 4841, 1548
\bibitem[]{} Fabian A.C., Sanders J.S., Ettori S., Taylor G.B., Allen S.W., Crawford C.S., Iwasawa K., Johnstone R.M., 2001, MNRAS, 321, L33
\bibitem[]{} Fabian A.C., Sanders J.S., Crawford C.S., Conselice C.J., Gallagher J.S. III, Wyse R.F.G., 2003, MNRAS, 344, L48 
\bibitem[]{} Fabian A.C., Johnstone R.M., Sanders J.S., Conselice C.J., Crawford C.S., Gallagher J.S. III, Zweibel E., 2008, Nature, 454, 968
\bibitem[]{} Ferland G.J., Fabian A.C., Hatch N.A., Johnstone R.M., Porter R.L., van Hoof P.A.M., Williams R.J.R., 2008, MNRAS, 386, L72
\bibitem[]{} Ferland G.J., Fabian A.C., Hatch N.A., Johnstone R.M., Porter R.L., van Hoof P.A.M., Williams R.J.R., 2009, MNRAS, 392, 1475
\bibitem[]{} Hatch N.A., Crawford C.S., Johnstone R.M., Fabian A.C., 2006, MNRAS, 367, 433 
\bibitem[]{} Hatch N.A., Crawford C.S., Fabian A.C., 2007, MNRAS, 380, 33 
\bibitem[]{} Hernquist L., 1990, ApJ, 356, 359
\bibitem[]{} Jaffe W., Bremer M.N., van der Werf P.P., 2001, MNRAS, 324, 443
\bibitem[]{} Jaffe W., Bremer M.N., Baker K., 2005, MNRAS, 360, 748
\bibitem[]{} Johnstone R.M., Hatch N.A., Ferland G.J., Fabian A.C., Crawford C.S., Wilman R.J., 2007, MNRAS, 382, 1246
\bibitem[]{} Kennicutt R.C., Jr., 1998, ApJ, 498, 541 
\bibitem[]{} Kirkpatrick C.C., McNamara B.R., Rafferty D.A., Nulsen P.E.J., Birzan L., Kazemzadeh F., Wise M.W., Gitti M., 2009, submitted to ApJ
\bibitem[]{} Landt H., Bentz M.C., Ward M.J., Elvis M., Peterson B.M., Korista K.T., Karovska M., 2008, ApJS, 174, 282
\bibitem[]{} Lim J., YiPing A., Dinh-V-Trung, 2008, ApJ, 672, 252
\bibitem[]{} McGregor P., Dopita M., Sutherland P., Beck T., Storchi-Bergmann T., 2007, ApSS, 311, 223 
\bibitem[]{} McNamara B.R, Nulsen P.E.J., 2007, ARA\&A, 45, 117
\bibitem[]{} Mouschovias M., Tassis K., Kunz M.W., 2006, ApJ, 646, 1043
\bibitem[]{} O'Dea C.P., et al., 2008, ApJ, 681, 1035 
\bibitem[]{} Peterson J.R., Fabian A.C., 2006, PhR, 427, 1
\bibitem[]{} Pizzolato F., Soker N., 2005, ApJ, 632, 821
\bibitem[]{} Quillen A.C., et al., 2008, ApJS, 176, 39
\bibitem[]{} Rafferty D.A., McNamara B.R., Nulsen P.E.J., 2008, ApJ, 687, 899
\bibitem[]{} Salom\'{e} P., Combes F., 2003, A\&A, 412, 657
\bibitem[]{} Salom\'{e} P., Combes F., 2004, A\&A, 415, L1
\bibitem[]{} Sanders J.S., Fabian A.C., Taylor G.B., 2008, accepted by MNRAS (arXiv0811.0743)
\bibitem[]{} Schneider D.P., Gunn J.E., Hoessel J.G., 1983, 268, 476 
\bibitem[]{} Sternberg A., Neufeld D.A., 1999, ApJ, 516, 371
\bibitem[]{} Wilman R.J., Edge A.C., Johnstone R.M., Crawford C.S., Fabian A.C., 2000, MNRAS, 318, 1232 
\bibitem[]{} Wilman R.J., Edge A.C., Johnstone R.M., Fabian A.C., Allen S.W., Crawford C.S., 2002, MNRAS, 337, 63
\bibitem[]{} Wilman R.J., Edge A.C., Johnstone R.M., 2005, MNRAS, 359, 755
\bibitem[]{} Wilman R.J., Edge A.C., Swinbank A.M., 2006, MNRAS, 371, 93

\end{thebibliography}
\end{document}